\documentclass[12pt]{iopart}
\usepackage{color}
\usepackage{graphicx}
\usepackage{graphicx,epsfig}
\usepackage{tabularx}
\usepackage{epstopdf}
\usepackage{axodraw4j}
\usepackage{pstricks}
\usepackage{amssymb}
\usepackage{simplewick}
\usepackage{verbatim} 
\newcommand{\Rmnum}[1]{\expandafter\@slowromancap\romannumeral #1@}

\def\cs2{c_{s}^{2}}

 \def\be   {\begin{equation}}   \def\ee   {\end{equation}}
 \def\ba   {\begin{array}}      \def\ea   {\end{array}}
 \def\bea  {\begin{eqnarray}}   \def\eea  {\end{eqnarray}}
 \def\bean {\begin{eqnarray*}}  \def\eean {\end{eqnarray*}}

\begin{document}

\begin{flushleft}
Universit\`{a} degli Studi di Milano Bicocca, Dipartimento di Fisica ``G. Occhialini''
\end{flushleft}

\begin{flushleft}
\textit{Dottorato di Ricerca in Fisica ed Astronomia, XXIII ciclo}
\end{flushleft}
\vspace{6cm}
\title{Effective Field Theory for Inflation}
\vspace{10cm}

\begin{flushright}
{ \large \textbf{Doctoral Thesis of:} Matteo Fasiello}
\end{flushright}
\vspace{1mm}
\begin{flushright}
{ \large \textbf{Supervisors:} Prof. Claudio Destri and Prof. Sabino Matarrese}
\end{flushright}
\newpage
\section*{}


\newpage

\tableofcontents

\newpage
\begin{flushright}
\textbf{\textit{To Emanuela}}
\end{flushright}
\newpage
\section{Acknowledgments}
I would like to thank my advisor at Milano Bicocca, Claudio Destri, for allowing me absolute freedom to work with the Padova Group while at the same time keeping a benevolent, informed and supportive eye on my progress.
\newline
\newline
I am extremely happy to thank Sabino Matarrese, my advisor in Padova, for a whole lot of reasons: for being always willing to bet on me thus instilling confidence and enthusiasm, for being cheerful and encouraging all the time and, most of all, for deeply caring about his students.
\\
\noindent It is a great pleasure to thank Nicola Bartolo for being the most enjoyable person one could hope to work with, for teaching me very many things, for being always kind, cheerful and supportive; for constantly sharing his ideas in extremely stimulating discussions.\\
\noindent I am delighted to thank Toni Riotto for being always available even from a distance, for his being constantly supportive, for sharing many extremely useful insights on physics and for being a source of important advice on many other matters.

\noindent I am really grateful for the time I spent in Padova in such a stimulating, nice and warm environment Nicola, Sabino and Toni have created. I feel truly privileged to have seen first hand the very rare positive combination they represent for their students.
\newline
\newline
I also wish to thank the experts whose suggestions and comments have helped improve the work presented in this thesis: Xingang Chen, Paolo Creminelli, Leonardo Senatore as well as the three anonymous referees of \cite{b,t,rn}.

\newpage
\section*{*}
\newpage

\section*{Introduction}
Inflation \cite{guth,toni1} stands as the central paradigm of modern cosmology. It earned its place by automatically solving the so-called flatness, horizon and monopoles problems of standard Big-Bang cosmology. It further provides a mechanism for the generation of density perturbations in the early Universe, which lead to Large Scale Structures (LSS) \cite{lss1}-\cite{lss5} in the distribution of galaxies and temperature anisotropies in the Cosmic Microwave Background (CMB) \cite{smoot92}-\cite{wmap5}.\\
Over the years, many inflationary models have been put forward that account for the spectrum of primordial perturbations. These perturbations are to a good approximation Gaussian as they are well-described by the leading quadratic terms in the inflationary Lagrangian. On the other hand, studying higher order interaction terms is crucial in characterizing the various inflationary mechanisms and in pursuing a deeper understanding of the inflationary dynamics. With the advent of new generation experiments (the launch of the Planck satellite~\cite{Pl,Mand} and the continued analysis of WMAP data~\cite{kom}), which provide improved sensitivity to deviations from Gaussian statistics, such a theoretical investigation is being timely complemented by observations and is rendered all the more urgent and important.\\ 
In studying primordial non-Gaussianites \cite{reviewk,bisp,bispectrum,trispectrum} one aims at handling quantities that are conveniently related to observables and that, at the same time, have an immediate expression in terms of higher order operators in the inflationary action. This is the case for n-point functions of the (scalar) degree of freedom that drives inflation. Starting with the three-point function, these correlators give an explicit measure of the deviation from Gaussianity and depend directly on the interactions in that they would vanish if we were to truncate the Lagrangian at quadratic order in perturbations. Althought in this work we will be concerned only with higher order correlators at tree level, it is important to point out that interactions give also a (small) correction to the power spectrum of curvature perturbations once loop calculations \cite{loop} are considered, so that too is a measure of non-Gaussianity.\\
The process of characterizing the various inflationary mechanisms according to their non-Gaussian properties can be schematically outlined. First, one determines if the amplitude of the n-point function, or, more conveniently, its Fourier transform, is large enough to be detectable. In the affirmative case, it is instructive to proceed and study the complete dependence of the correlator on the external momenta, the so called \textit{shape-function}. It is in fact crucial to know in what type of momenta configuration the shape-function peaks. Indeed, in analyzing the data one must know beforehand if a specific shape-function is expected so that the appropriate corresponding non-Gaussianity \textit{estimator} can be built. Not doing so would result in a system which is essentially blind to a whole family of inflationary models (or, in an effective field theory approach, a family of interaction operators) which might well be there and be important.\\
In this work we aim at differentiating among the many inflationary mechanisms trough the study of their non-Gaussianities. As a consequence, it seems natural to look for an approach that captures all the general features of an inflationary theory and reduces to specific models in appropriate regimes. We specify here that we are limiting our analysis to the case of inflation driven by a single scalar degree of freedom and therefore will not discuss vectors as has recently been done, for example, in the context of statistical anisotropies. As one might suspect, an approach that allows for a unifying perspective on inflation can be found in the realms of effective field theories.\\
A precise prescription based on these ideas has been put forward in \cite{eft08}, whose approach we will closely follow in this work. In \cite{eft08} the authors give, subjected to mild caveats, the most general action for inflation driven by a single scalar degree of freedom. The Lagrangian for this theory is written down in detail and it turns out one is able to re-obtain most specific inflationary models by switching on or off appropriate coefficients driving various interaction operators in the action. The hope is to be able to pin down the specific interaction terms generating distinctive features in the bispectrum and trispectrum of curvature perturbations. This could result in observational bounds on the value of the coefficients (we generically call them $\bf{M}_n$) driving the various interactions at third and fourth order \footnote{On the other hand, as we will see, this approach proves itself useful already at second order in perturbations.}.
The power of the effective field theory approach is quite manifest in that, in principle, it allows these coefficients
considerable more freedom than what they are granted in any specific inflationary model. In fact, by being for the most part free parameters (a couple of these coefficients are to obey some inequalities if one wants, as we do, the generalized speed of sound to
be smaller than unity \footnote{One should also mention the general requirement that stems from working within the regime of validity of perturbation theory: namely the $\bf{M}_n$ coefficients are to be smaller than the mass of the underlying theory.}), the $\bf{M}_n$' s allow for the description of known interactions with relative weights which would otherwise be fixed. By employing effective field theory then one enlarges the region of the parameters space than can be spanned. Besides that, in the complete effective Lagrangian some of the $\bf{M}_n$ coefficients multiply (\textit{extrinsic curvature-generated}) operators that are sometimes neglected in the literature but should in principle be studied as, in fact, their contribution to higher order correlators can be relevant and this effectively increases the dimension of the parameters space of the theory.\\
Quite interestingly, the formalism of \cite{eft08} also sheds new light on effects due to symmetries in the action: for example, one can clearly see how a reduced speed of sound often automatically results in an enhanced non-Gaussianity. One more advantage that comes with employing the proposed setup is of calculational nature: in the so called \textit{decoupling regime} (which implies working in a specific energy range) the dynamics of the metric decouples from the one of the scalar that drives inflation thus rendering the Lagrangian itself and the higher order correlators much easier to handle and calculate. This mechanism is very reminiscent of what happens in standard quantum field theory and goes under the name of equivalence theorem.\\
In this work we aim to integrate and summarize the investigations presented in \cite{b,t,rn}. The common ground these papers share relies on the use of effective field theory methods within a general effort to characterize as many inflationary models as possible according to the non-Gaussian properties they exhibit. These properties take the form of various predictions for cosmological observables such as the power spectrum, its tilt, its running; the bispectrum amplitude, $f_{NL}$, its running; the trispectrum amplitude.\\
As mentioned, already at second order in perturbations the effective theory approach reveals interesting facts. In \cite{b,rn}, making full use of the freedom on the $\bf{M}_n$'s, we are able to write a very general quadratic Lagrangian which captures the quadratic theory of the entire class of the so called generalized slow-roll theories (often referred to as $P(X,\phi)$-model) and also covers models of inflation for which terms with more than one derivative acting on the scalar driving inflation are allowed (e.g. Ghost inflation \cite{ghost}). From the general quadratic action we move to the equation of motion: we solve it by imposing the Bunch-Davies vacuum condition and by requiring to re-obtain the known solutions for specific inflationary mechanisms in the corresponding limits. The resulting power spectrum 
is now a function of five, rather than just three, independent coefficients. This in turn means that its tilt depends on the usual three slow-roll parameters but also on two additional coefficients, two generalized slow-roll parameters. These results are presented at leading order and, in particular cases, at next-to-leading order in generalized slow-roll approximation.\\
We then turn our attention to non-Gaussianities (NG), starting with the bispectrum. It is important to stress here that we will consistently study mostly models that generate large, detectable NG and, in doing so, it will become clear that there is an important region of the parameters space of the effective theory where the contribution of (often neglected) extrinsic curvature-generated \footnote{In the approach of \cite{eft08} the gauge choice identifies a preferred slicing of spacetime; the extrinsic curvature tensor $K_{\mu \nu}$ describes the geometry of such slicings.} interaction terms is important and, possibly, leading. We start from the most general third-order effective action that originates from the prescription of \cite{eft08}. Employing the IN-IN formalism \cite{in-in1, in-in3, in-in2, w-qccc}, a thorough analysis of the various terms contribution to the amplitude, $f_{NL}$, and shape of the three-point correlator is performed. 
One immediately reproduces results of the current literature and re-discovers known features such as the fact that large non-Gaussianities may be generated if the sound speed (now a generalization thereof) is much smaller than unity. On the other hand, a number of noteworthy novelties arise which are due precisely to the effect of extrinsic curvature-generated terms. Upon requiring a small generalized speed of sound and using at full the freedom on those $\bf{M}_n$ coefficients that first appear at third order in perturbations, one finds that the leading contribution to $f_{NL}$ can indeed come from curvature contributions and there's more: the shape-function that some of these terms generate has peculiar, distinctive features. In $P(X, \phi)$-models of inflation, but also in Ghost inflation, the typical shape-function generated by a single leading interaction term in the cubic Lagrangian will peak in the so called \textit{equilateral} configuration (that is, in Fourier space, all three external momenta are equal $k_1=k_2=k_3$). Interestingly, a number of curvature terms we study generate a shape-function that peaks in the \textit{flat} configuration ($k_1=2 k_2=2 k_3$). Such a configuration is quite uncommon for single-field models of inflation and cannot be obtained without considering extrinsic curvature-generated terms in the action unless one relaxes the Bunch-Davies vacuum condition on the wavefunction \cite{holman} or considers linear combinations of third-order operators \cite{ssz05}. Our study then enlarges the classes of single-field inflationary models whose bispectrum signature may consist in a flat shape-function and makes this feature somewhat more natural as a flat shape now independently originates from several interaction terms within the B-D vacuum condition for the wavefunction.\\ Prompted by the $f_{NL}$-related findings concerning extrinsic-curvature interactions, we investigated the corresponding contribution to the running of $f_{NL}$ itself. This quantity is expected to be small, roughly of the order of the generalized slow-roll parameters. The actual calculation shows that the contribution from curvature terms might be the leading one for the running and that it can be larger than the slow-roll parameters since here the quantities involved are both (quite constrained) $\bf{M}_n$ coefficients which appear in the quadratic Lagrangian and $\bf{M}_n$'s that first show up in the cubic action.\\
Following \cite{t}, we present what follows quite naturally after the investigations briefly outlined above, namely a study of the trispectrum generated by the effective action up to fourth order. Here again, the novelties are to be found in the effects of the extrinsic curvature-generated interaction terms. These can    provide the dominant contribution to the four-point function and present very distinctive patterns in the form of the trispectrum shape-function.\\
The powerful and very convenient setup that we are going to thoroughly describe below is subjected to some specific limitations. We will discuss them as they arise along the presentation.\\
The paper is organized as follows. In the first section we briefly review the setup of \cite{eft08}. In \textit{Section 2} we proceed with the analysis of the quadratic Lagrangian for the perturbations. We solve the corresponding equation of motion and calculate the resulting spectrum, its tilt and running. In \textit{Section 3} we take on cubic interactions obtaining the amplitudes and shape-functions for the leading contributions to the  bispectrum of curvature perturbations. We also calculate the running of $f_{NL}$ in a number of particularly interesting cases. \textit{Section 4} is dedicated to the trispectrum: here most of the possible quartic interactions are considered and special attention is devoted to operators that are invariant under two specific symmetries recently introduced in the literature. In the \textit{Conclusions} we summarize the significance of our results and comment on further work. In the various \textit{Appendices} we present some explicit calculations which have been omitted from the main text for the sake of simplicity.

\section{The effective action for single-field inflation (up to third order)}
Our goal will eventually be to write down the complete theory of single-field models of inflation  up to fourth order in perturbations. 
We will follow the  effective theory approach first introduced in  Ref.  \cite{eft08} of which we now give an outline.\\
The single scalar field $\phi$ which one assumes to be responsible for inflation is splitted as usual in an unperturbed part, the background, plus a fluctuating one:
\be
\phi(\vec x, t)=\phi_0 (t)+ \delta \phi(\vec x, t).
\ee 
At this stage it is essential to underline the gauge choice that is made as it is not the most common one found in the literature. In \cite{eft08} and in the following one works in the comoving (or unitary) gauge for which $\delta \phi =0$  (see also \cite{luty}), the scalar degree of freedom is now \textit{hidden} in the metric. Once this choice is implemented, the Lagrangian will no more be invariant under full spacetime diffeomorphisms (diffs) but only under spatial reparametrizations.
This is the starting point to write the most general  space diffs invariant Lagrangian at the desidered order in perturbation theory in an effective theory approach.
In \cite{eft08} the authors prove that, once an approximate shift-simmetry is required, their second and third order action is the most general one (see also  \cite{w-e} for an interesting perspective on the most general effective Lagrangian for inflation). One can then use the so called  \textit{Stueckelberg trick} to restore full spacetime reparametrization invariance. As a by-product of this latter procedure, the degree of freedom hidden in the metric shows up again as a scalar field. 

Let us start from the general theory before full spacetime reparametrization invariance has been restored: the most general space diffs-invariant action in unitary gauge can be schematically written as \cite{eft08}:
\be
S = \int d^4 x \sqrt{-g} \; F(R_{\mu\nu\rho\sigma}, g^{00}, K_{\mu\nu},\nabla_\mu,t),\label{unper1}
\ee
where $K_{\mu\nu}$ is the extrinsic curvature tensor on which we will soon elaborate more and the ``0'' components of the metric tensor $g^{\mu\nu}$ are free indices.  Considering fluctuations around a FRW background amounts to studying the following action:
\begin{eqnarray}
S &=& \int d^4x \sqrt{-g} \; \Big[\frac12 M_{\rm Pl}^2 R + M_{\rm Pl}^2  \dot H  g^{00} - M_{\rm Pl}^2 \Big(3 H^2 +\dot H\Big) + \nonumber \\  && 
\sum_{n\geq 2} F^{(n)}(g^{00}+1,\delta K_{\mu\nu}, \delta R_{\mu\nu\rho\sigma};\nabla_\mu;t)\Big],
\label{genspacei1}
\end{eqnarray}
where the  $F^{(n)}$ functions contain fluctuations which are at least quadratic.
The next step is to   restore full spacetime reparametrization invariance. To see how it works, we  borrow a simple example from \cite{eft08} and consider the following sample action terms
\be
\int d^4x\;  \sqrt{-g} \left[A(t)+B(t)g^{00}(x)\right] \ \label{ex01}.
\ee
We are interested in time reparametrization $t\rightarrow t+\xi_{0}(\vec x,t); \quad \vec x \rightarrow \vec x $, under which the above action (after a simple variable redefinition) reads
\be
\fl \int d^4x\;  \sqrt{- g(x)} \left[A( t-\xi_0(x))+B(t-\xi_0(x)) \frac{\partial (t-\xi_0(x))}{\partial x^\mu}\frac{\partial (t-\xi_0(x))}{\partial x^\nu} g^{\mu\nu}(x)\right].\label{ex02}
\ee
Upon promoting $\xi_0$ to a field, $\xi_0(x)=-\pi(x)$ and requiring the following gauge transformation rule $\pi(x)\rightarrow \pi(x)-\xi_0(x)$ on $\pi$,  the above action is invariant under full spacetime diffeomorphisms. The  scalar degree of freedom $\pi$ makes its 
appearance in  the time dependence of the $A,B$ coefficients and in the transformed metric.
More into details, under time reparametrization the metric  $g^{\mu\nu}$ transforms as follows:
\begin{eqnarray}
 \widetilde g^{\alpha \beta} = \frac{\partial \widetilde g^{\alpha}}{\partial x^\mu} \frac{\partial \widetilde g^{\beta}}{\partial x^\nu} g^{\mu\nu}\, , 
\end{eqnarray}
which implies
\begin{eqnarray}
g^{ij}\rightarrow g^{ij}; \quad g^{0i}\rightarrow (1+\dot\pi)g^{0i} +g^{ij}\partial_j \pi; \label{metric} 
\\ \nonumber
g^{00}\rightarrow (1+\dot\pi)^2 g^{00} +2(1+\dot\pi)g^{0i}\partial_i \pi+ g^{ij}\partial_i \pi \partial_j \pi \label{ex03}.
\end{eqnarray}
This procedure has been borrowed, conceptually unchanged, from standard gauge theory: a Goldstone boson which transforms non linearly under the gauge transformation provides the longitudinal component of  a massive   gauge boson. At sufficiently high energy 
such  Goldstone boson  becomes the only relevant 
 degree of freedom. This is the so-called equivalence theorem. The same is true for our case: for sufficiently high energy
 the mixing with gravity becomes irrelevant and the scalar $\pi$ becomes the only relevant mode in the dynamics. This is the
 so-called decoupling regime. 
Let us clarify this concept with a simple example. Consider the following contribution, taken from  Eq.~(\ref{genspacei1})
\be
M_{\rm Pl}^{2} \dot H g^{00} \rightarrow M_{\rm Pl}^{2} \dot H ((1+\dot\pi)^2 g^{00} +2(1+\dot\pi)g^{0i}\partial_i \pi+ g^{ij}\partial_i \pi \partial_j \pi).
\ee
We focus on  the quadratic part of the first term in the above equation. Upon canonical normalization, $\pi_{c}=M_{\rm Pl}{\dot H}^{1/2}\pi$ and $ g^{00}_{c}=M_{\rm Pl}g^{00}$, one gets

\be
M_{\rm Pl}^{2} \dot H(  g^{00} + 2 \dot \pi  g^{00} + {\dot\pi}^2 g^{00} )=   {\dot\pi_{c}}^2 + 2 {\dot H}^{1/2}\dot\pi_{c} g^{00}_{c}  +M_{\rm Pl} \dot H g^{00}_{c}. \label{dec}
\ee
Consider the second term of Eq.~(\ref{dec}) which mixes gravity with the scalar. Since $\dot \pi_{c}\sim E \pi_c$, at energies higher than $\sim {\dot H}^{1/2}$ the term  ${\dot\pi_c}^2$ dominates the dynamics. This turns out to be true in general: the number of derivatives (which in Fourier mode would basically give an energy-dependent coefficient in front of $\pi$) is higher in  terms containing only $\pi$'s  than in the mixed terms and therefore there exists an energy threshold above which the scalar decouples from gravity. Since in explicitating the $F^{(2)}$ 
term in Eq.~(\ref{genspacei1}) there can be, in principle, other quadratic terms that go like ${\dot \pi}^2$, one has to consider which one is the leading kinetic term and determine the canonically normalized field $\pi_c$ and the energy threshold accordingly. To take the safe route, one might well take the energy threshold, $E_{\rm mix}$, to be the highest one of this set. Since one is concerned with correlators just after horizon crossing, 
one concludes that the decoupling procedure works as long as the decoupling energy is smaller than
the Hubble rate $H$.  More precisely, we can anticipate  that the kinetic terms in $F^{(2)}$ which are going to matter in our discussion 
come with coefficients  $M_2^4$ and $M_{\rm Pl}^2 \epsilon H^2$. The condition   $E_{\rm mix} < H$ is then 
 satisfied if $M_{\rm Pl}^2 \epsilon H^2 > M_2^4$, where $\epsilon=-\dot{H}/H^2$ is a slow-roll parameter; if this is not the case we need to assume $M_2^4< M_{\rm Pl}^2 H^2 $.\\
From now on we will work in the decoupling regime. In considering  the terms of Eq.~(\ref{genspacei1}), we will therefore use only the unperturbed entries of the metric tensor. In order to write the effective Lagrangian up to, say, third order \footnote{We will write the explicit expression up to fourth order when concerned with the trispectrum, here we limit ourselves to third order for simplicity.}, we  start from  Eq.~(\ref{genspacei1}) and follow the algorithm given in \cite{eft08}. Fluctuations are encoded in the $F^{(n)}$ terms. In order to be as general as possible, we also include all possible 
contributions up to third order   coming from extrinsic curvature $K_{\mu\nu}$ terms.  In fact, it is instructive at this stage to step back and consider the action in Eq. (\ref{unper1}). Given a theory which is space diffs-invariant, one can always identify a slicing of spacetime, described by a timelike function $\widetilde t(x)$,  which realizes time diffeomorphism: on surfaces of constant $\widetilde t$ the time symmetry breaking scalar is also constant. Before selecting a gauge, there is still the freedom to make a choice on $\widetilde t(x)$ and working in the unitary gauge amounts to requiring $\widetilde t=t$. In order to describe the geometry of this preferred slicing, one employs the extrinsic curvature tensor. In writing
 down such a  tensor, one needs  two ingredients: the  unit normal vector $n_\mu$, perpendicular to the constant $\widetilde t$ surfaces, and the induced metric $h_{\mu\nu}$. These are defined as
\be
n_\mu=\frac{\partial_\mu \widetilde t}{\sqrt{-g^{\mu\nu}\partial_\mu \widetilde t \partial_\nu \widetilde t}}\label{n-h} \rightarrow \frac{\delta_\mu^0}{\sqrt{-g^{00}}}; \quad h_{\mu\nu}=g_{\mu\nu}+n_\mu n_\nu,
\ee
which allows us to write
\be
\fl K_{\mu \nu} \equiv h^{\sigma}_{\mu} \nabla_{\sigma}n_{\nu}\,\,=\,\, \frac{\delta^{0}_{\nu} \partial_{\mu} g^{00}}{2(-g^{00})^{3/2}} + \frac{\delta^{0}_{\nu} \delta^{0}_{\mu} g^{0\sigma} \partial_{\sigma} g^{00}}{2(-g^{00})^{5/2}}- \frac{g^{0\epsilon}(\partial_{\mu}g_{\epsilon \nu} + \partial_{\nu}g_{\epsilon \mu}-\partial_{\epsilon}g_{\mu \nu})}{2(-g^{00})^{1/2}}.
\ee
The above expressions can be used to  write explicitly the most generic third  order action for the fluctuations around the FRW background:
\begin{eqnarray}
\label{eq:actiontad}
\fl S_{3}& = &\int  d^4 x \; \sqrt{- g} \Big[ \frac12 M_{\rm Pl}^2 R + M_{\rm Pl}^2 \dot H
g^{00} - M_{\rm Pl}^2 (3 H^2 + \dot H) + \frac{1}{2!}M_2(t)^4(g^{00}+1)^2 \nonumber \\
\fl &+&\frac{1}{3!}M_3(t)^4 (g^{00}+1)^3- \frac{\bar M_1(t)^3}{2} (g^{00}+1)\delta K^\mu {}_\mu
-\frac{\bar M_2(t)^2}{2} \delta K^\mu {}_\mu {}^2  \nonumber \\
\fl &-&\frac{\bar M_3(t)^2}{2} \delta K^\mu {}_\nu \delta K^\nu {}_\mu  
 -\frac{\bar M_4(t)^3}{3!} (g^{00}+1)^2 \delta K^\mu {}_\mu -\frac{\bar M_5(t)^2}{3!} (g^{00}+1) \delta K^\mu {}_\mu {}^2  \nonumber \\
\fl &-&\frac{\bar M_6(t)^2}{3!}(g^{00}+1) \delta K^\mu {}_\nu \delta K^\nu {}_\mu  -\frac{\bar M_7(t)}{3!}  \delta K^\mu {}_\mu {}^3-\frac{\bar M_8(t)}{3!}  \delta K^\mu {}_\mu {}  \delta K^\nu {}_\rho \delta K^\rho {}_\nu \nonumber \\
\fl  &-&\frac{\bar M_9(t)}{3!}   \delta K^\mu {}_\nu {}  \delta K^\nu {}_\rho \delta K^\rho {}_\mu
\Big] \; .\label{genv}
\end{eqnarray}
The coefficients $M_{2},M_{3}$ and ${\bar M_1},\cdots,{\bar M_9}$, to which we will often refer to as the $\bf{M}_n$ coefficients, are in principle generic; we will comment on their physical significance as we discuss them more in detail. All the $\bar{M}$ coefficients multiply extrinsic curvature-generated interactions. A given particular set of values (or bounds) for the $\bf{M}_n$'s will specify a given inflationary theory.\\
The action, as written in (\ref{genv}), is not yet  invariant under full diffeormophisms. One needs to follow exactly the steps illustrated in Eqs~ (\ref{ex01}), (\ref{ex02}) and (\ref{ex03}) and promote $\xi_0$ to a field $\pi$ with the proper gauge transformation.

In the decoupling limit we find:

\begin{eqnarray}
\fl  S_3&=&\int d^4 x \sqrt{-g}\left[ M_{\rm Pl}^{2}\dot H (\partial_{\mu} \pi)^2 
+ M_2(t)^4\left(2{\dot\pi}^2 -2\dot\pi \frac{(\partial_i \pi)^2}{a^2}\right) -\frac{4}{3}M_3(t)^4{\dot\pi}^3 
\right.\nonumber\\
\fl &-& \frac{\bar M_1(t)^3}{2}\left(\frac{-2 H (\partial_i \pi)^2 }{a^2} +\frac{(\partial_i \pi)^2 \partial_j^2 \pi}{a^4}  \right) -\frac{2}{3}\bar M_4(t)^3 \frac{1}{a^2}{\dot\pi}^2 \partial_i^2\pi +  \frac{\bar M_5(t)^2}{3}  \frac{\dot\pi}{a^4}(\partial_i^2\pi)^2 
\nonumber\\
\fl &-&\frac{\bar M_2(t)^2}{2} \left( \frac{(\partial_i^2\pi)(\partial_j^2\pi)   +H (\partial_i^2\pi) (\partial_j \pi)^2  +2\dot\pi \partial_{i}^2 \partial_j \pi \partial_j \pi}{a^4} \right) +\frac{\bar M_6(t)^2}{3}  \frac{\dot\pi}{a^4}(\partial_{ij}\pi)^2 
\nonumber\\
\fl &-&\frac{\bar M_3(t)^2}{2} \left( \frac{ (\partial_i^2\pi)(\partial_j^2\pi) +2H(\partial_i \pi)^2\partial_i^2 \pi + 2 \dot\pi \partial_{i  j}^{2}\pi \partial_j \pi  }{a^4} \right)  -\frac{\bar M_7(t)}{3!} \frac{(\partial_i^2 \pi)^3}{a^6}
\nonumber\\
\fl & -& \left.\frac{\bar M_8(t)}{3!}  \frac{\partial_i^2 \pi}{a^6}(\partial_{jk}\pi)^2-\frac{\bar M_9(t)}{3!}\frac{1}{a^6}\partial_{ij} \pi\partial_{jk} \pi\partial_{ki}\pi
\right].\label{action}
\end{eqnarray}
A few clarifying comments are in order:
\begin{itemize}
\item If we consider  terms only up to second-order, for $M_{2}=\bar M_{1,2,3}=0$ one recovers the usual quadratic Lagrangian for the fluctuations, with  sound speed  $c_s^2 =1$ and the standard  solution to the equations of motion. Switching on $M_{2}$ amounts to allowing models with sound speed smaller than unity, $1/c_s^2 =1- 2M_2^4/(M_{\rm Pl}^2 \dot H)$, which are often linked to a high level of primordial non-Gaussianity \cite{chen-bis,eft08}. Furthermore,  turning on $\bar{M}_{2,3}$ in the de Sitter limit,  one recovers  Ghost inflation \cite{ghost}. On the same lines, keeping all the $\bar M$'s vanishing,  but going to third  and higher order with the $M$'s,   one can retrieve the interactions that describe DBI inflation \cite{DBI,chen-bis, chen-tris}. The list of correspondences continues with K-inflation theories and others, thus showing how the effective action approach provides a unifying perspective on inflationary models \cite{eft08}.
\item The action in Eq.~(\ref{action}) has already been  written with  large non-Gaussianities in mind.
This means  that, at every order in fluctuations and for  each $M$ and $ \bar M$ coefficients,  we have  selected  those 
leading terms which will eventually generate large three-point correlators. To clarify this point,  we provide   a simple example. Let us consider the terms up to second order in Eq.~(\ref{action}) and set conveniently   $\bar M_{1,2,3}=0$. The properly normalized solution to the equation of motion will be the usual $\pi_k(\tau) \propto i H e^{-i k c_s \tau}(1+i k c_s \tau)$. It is straightforward to verify that, at the horizon crossing , 
 $\dot \pi \sim H \pi$ and $\nabla \pi \sim H/c_s\,\, \pi$. Therefore, among the $\pi$ terms with the same number of derivatives, the ones with the highest number of space derivatives dominate in the $c_s\ll 1$ limit. Generalizing these estimates for the classical solution (which we will  describe below) 
obtained from the equation of motion of our complete action, one selects  the terms in Eq.~(\ref{action}).
\item 
there is also the  comparison between same perturbative order but different $M$ terms to be made.
In the literature, all non zero coefficients  in front of the various operators are generically assumed to be of the same order (see for example the discussion 
concerning the orthogonal configuration  in Ref. \cite{ssz05} for an interesting perspective). 
We shall not restrict ourselves to this situation. Note that,  were the coefficients to be all of the 
same order, one could already identify the dominant
operators.  For example, consider the third order contributions $\bar M_1^3 (\partial_i \pi)^2 \partial_j^2 \pi/a^4\sim M^3 (H^4/c_s^4) {\pi}^3$; for $c_s\ll 1$ this will be a leading contribution with respect to, say, $\bar M_2(t)^2 H (\partial_i^2\pi) (\partial_j \pi)^2\sim M^2 H (H^4/c_s^4){\pi}^3$. This is due to the fact that in the effective Lagrangian  every additional derivative comes with a $H/M\ll 1$ factor attached: one is basically doing an  $H/M$ expansion where M is roughly the energy range of the underlying theory.  In the last example we have intentionally picked terms with the same power of $c_s$ at the denominator. Let us  now look at the $\bar M_7(t) (\partial_i^2 \pi)^3 \sim M (H^6/c_s^6){\pi}^3 $ term though; comparing this contribution with the $\bar M_1$ term amounts to comparing $M^3$ with $M H^2/c_s^2$. We see that for a very small speed of sound the $\bar M_7$ contribution may still be relevant. 
These examples justify our strategy of   including all the terms in Eq.~(\ref{action}) compatible with  $c_s\ll 1$ (we will make more comments on this point in the next Section).
\end{itemize}
\newpage
\section{Classical solution and power spectrum}
We now proceed to solve the equation of motion for the second-order effective Lagrangian at 
leading order in slow-roll:
\bea
\fl \mathcal{L}_2=a^3\Big(M_{P}^{2}\dot H (\partial_{\mu} \pi)^2 
+ 2 M_2^4{\dot\pi}^2 + \bar M_1^3 H \frac{(\partial_i \pi)^2 }{2 a^2}  
- \frac{\bar M_2^2}{2}\frac{1}{a^4} (\partial_i^2\pi)^2-\frac{\bar M_3^2}{2} \frac{1}{a^4}(\partial_{ij}\pi)^2 \Big) .
\label{l2}
\eea

Let us not a few facts about the above expression.
\begin{itemize}
\item Eq.~(\ref{l2}) is the most general second-order Lagrangian in unitary gauge provided the approximate symmetry of the underlying theory is such that only derivative terms of $\pi$ appear in the action.
\item In order to make contact with more familiar notation, we stress that to first order (which is all we need here) the scalar $\pi$ is lineraly related to the dimensionless gauge invariant quantity $\zeta$, the curvature, by $\zeta=-H \pi$.
\item In full generality the $M$ coefficients above should be time dependent (we will deal with such a case in a specific paragraph). However, if one is only interested in performing leading-order calculations, then, due to a generalized slow-roll approximation, one can safely consider them as constant.
\item The action in Eq.~(\ref{l2}) and, as we have seen, its higher order counterparts, are generally written with large non-Gaussianities in mind. This results in a number of quadratic operators being left out from the formula above. Let us stress already at this stage though that, should one decide to include all these subleading contributions in the action, the \textit{functional} expression of the solution will not change, one merely redefines a couple of approximately time-independent coefficients. This is due to the fact that the types of operator ${\dot \pi}^2, (\partial_i \pi)^2,(\partial_i^2 \pi)^2 $ are already saturated at the level of Eq.~(\ref{l2}). 
\end{itemize}

\noindent We are now ready to tackle the equation of motion. After the usual change of variable,  $\pi(\vec k, t(\tau))=a(\tau) u(\vec k, \tau)$, the equation of motion can be written as:
\be
u'' -\frac{2}{\tau^2} u + \alpha_0 k^2 u + \beta_0 k^4 \tau^2 u=0 \label{eom},
\ee
where  $\alpha_0, \beta_0$ are time independent (again, at leading order) dimensionless coefficients. This equation has been written in the 
context of tilted Ghost Inflation~\cite{tilted} and to our knowledge, it has not been solved analitically before Ref.~\cite{b}, where the analytical solution has been briefly introduced and used for the computation of the three-point function. Here we discuss in much more details the 
properties of this solution. 
\noindent At this stage one can immediately recognize $\alpha_0$ as the more common $c_s^2$ and $\beta_0$ as the constant $\alpha^2 H^2 /M^2$ first introduced in \cite{ghost}. The complete expression for the coefficients is:
\be
\alpha_0 = \frac{-M_{Pl}^{2}\dot H - \bar M_{1}^{3}H}{-M_{Pl}^{2}\dot H +2M_{2}^{4} }; \qquad \beta_{0}=\frac{(\bar M_{2}^{2}+\bar M_{3}^{2})H^2}{2(-M_{Pl}^{2}\dot H +2M_{2}^{4})},
\ee
so that one reobtains the actual $c_s^2$ for $\bar M_{1}=0$. Note that one can simply look up the e.o.m. solution for DBI-like inflation if $\beta_0=0=\bar M_1$ and Ghost Inflation in the de Sitter limit provided $\alpha_0=0$. Let us pause here to comment on the possibility of a negative $\alpha_0$ (see also \cite{ssz05}). Such a scenario would result in a region in the $k$-space, whenever
\be | \alpha_0| k^2 \gg \beta_0 k^4 \tau^2 -\frac{2}{\tau^2},
\ee
for which the solution to the equation of motion will behave exponentially. Such a possibility raises a number of issues we address below. First of all, in order to keep control of the negative $\alpha_0$ region of the parameters space of the theory in the ultraviolet, one requires that the (positive) $\beta_0 k^4 \tau^2$ prevails over the $\alpha_0$ contribution before $k$ reaches the cutoff scale $\Lambda$. Considering that on the IR side, as we will show, the modes will eventually freeze outside the horizon, the case of a negative $\alpha_0$ should not in principle be disregarded. On the other hand, a lot of care should be exerted because an exponential phase of the modes for a sufficiently wide $k$ region could generate values for higher order correlators that directly contradict available observational data. 

\noindent We could now proceed to solve the complete equation of motion but, equipped with just  equation (\ref{eom}), we can already make some educated guesses on the behaviour of the wavefunction. First of all, the typical oscillatory behaviour deep inside the horizon is to be expected in this more general case as well: both $\alpha_0 k^2$ and $\beta_0 k^4 \tau^2$ cause wave-like behaviour (see Fig~1 below) of the wavefunction, while the $(-2/\tau^2)$ contribution is negligible. This is important in that it tells us the main contribution to correlation functions will be coming, as usual, from the horizon-crossing region. Note here that, as far as $\beta_0 \ne 0$, the 'Ghost Inflation' term will eventually lead the oscillation if one goes deep enough inside the horizon.\\
On the other hand, in the $\tau \rightarrow 0$ limit, $(-2/\tau^2)$ will be leading the dynamics and we expect to recover the usual, frozen modes. As is familiar from the DBI-like cases, it is convenient to introduce the notion of an effective horizon, placing it where the oscillatory behaviour stops being dominant. In formulas:
\be
 \alpha_0 k^2 + \beta_0 k^4 \tau_{*}^2 = \frac{2}{\tau_{*}^2} \quad \Rightarrow \quad \tau_{*} = -\frac{2}{k \sqrt{\alpha_0+ \sqrt{\alpha_0^2 + 8 \beta_0}}}\label{h}.
\ee
For $\beta_0=0, \quad \alpha_0\sim 1 $ one recovers $k^2 \tau_{*}^2 \sim 1$ at the horizon.\\
At this stage we can perform a consistency check and show how one can generalize the argument, initially borrowed from DBI-like inflationary models, that in comparing terms at the same order in perturbations and with the same overall number of derivatives, the ones with the most space derivatives are dominating in the $c_s \ll 1 $ limit. The generalization of this argument consists in restricting the parameters space to the $\alpha_0 \ll 1 \quad and \quad \beta_0 \ll 1$ region. Consider Eq.~(\ref{eom}) in Fourier space; in full generality one expects $\nabla \pi \sim k \pi$ and $\dot \pi \sim H \pi$ so what needs to be done is relate $k$ with $H$ at the horizon. Using equation (\ref{h}) and $\tau \sim -1/(a H)$ one obtains 
\be
k= \frac{\sqrt{2}H}{\sqrt{\alpha_0 + \sqrt{\alpha_0^2 + 8 \beta_0}}}. \label{estimates}
\ee
Since the main contributions to correlators comes from the horizon-crossing region, this shows that, for $(\alpha_0, \beta_0) \ll 1$ we can still identify leading terms in the Lagrangian according to the standard procedure. We will strictly follow this procedure when working with the cubic and quartic action for the field $\pi$ in all the following sections. On the other hand, when dealing with the quadratic action, the calculations are simple enough so that we can account for all the terms, not just the leading ones. 
\subsection{Wavefunction}
Let us verify all this quantitatively. The solution to Eq.~(\ref{eom}), being of second order, will come with two $k$-dependent integration constants. We have determined their values by requiring to re-obtain the known DBI and Ghost solutions in the corresponding limits. The general wavefunction reads:

\bea 
u_k(\tau)=\frac{i e^{\frac{1}{2} i \sqrt{\beta_0} k^2 \tau ^2}}{2^{1/4} \tau }\, \mathcal{G}\left[-\frac{1}{4}-\frac{i \alpha_0}{4 \sqrt{\beta_0}},-\frac{1}{2},-i \sqrt{\beta_0} k^2 \tau ^2 \right] C_1(k) 
\nonumber 
\eea
\bea
+ \frac{i e^{\frac{1}{2} i \sqrt{\beta_0} k^2 \tau ^2}}{2^{1/4} \tau } \mathcal{L}\left[\frac{1}{4}+\frac{i \alpha_0}{4 \sqrt{\beta_0}},-\frac{3}{2},-i \sqrt{\beta_0} k^2 \tau ^2\right] C_2(k)\,\, ,\nonumber\\ 
\eea
Where $\mathcal{G}$ stands for the confluent hypergeometric function and $\mathcal{L}$ is the generalized Laguerre polynomial.
We verified that, properly adjusting the integration constants according to
\be
\fl \qquad C_1(k)= \frac{  \left(\alpha_0+\sqrt{\beta_0}\right)^{-3/4}\Gamma\left[\frac{5}{4}-\frac{i \alpha_0}{4 \sqrt{\beta_0}}\right]k^{-3/2}}{ \sqrt{M_{Pl}^2 \epsilon\, H+2M_2^4}\,\, 2^{1/4}\, \Gamma\left[3/2-\frac{ \sqrt{\beta_0}}{4 \left(i \alpha_0+ \sqrt{\beta_0}\right)}\right]}\,\,;\qquad C_2(k)=0,
\ee
\noindent one obtains, in the appropriate limits \cite{b&d}, the wavefunctions of standard inflation and Ghost Inflation \cite{ghost}. We can now write our solution:

\be
\pi_k(\tau)=\frac{H\, e^{\frac{1}{2} i \sqrt{\beta_0} {k}^2 {\tau}^2} k^{-3/2} \Gamma (\frac{5}{4}-\frac{i \alpha_0}{4 \sqrt{\beta_0}}) \mathcal{G}(\alpha_0,\beta_0,k^2,{\tau^2})}{i\sqrt{M_{P}^2\epsilon H^2 +2M_2^4}\,  \sqrt{2}\, \gamma_0^{3/4}  \Gamma(\frac{5}{4}+\frac{\alpha_0}{4 \alpha_0-4 i \sqrt{\beta_0}})} \, , \label{sol}
\ee
where $\gamma =\alpha_0+\sqrt{\beta_0}$ and $\Gamma(x)$ is the Euler gamma function.\\
We note in particular that for $\alpha_0=0$, Eq.~(\ref{sol}) immediately reduces analitically to the Ghost Inflation wavefunction $\pi_k(\tau)=(H (-\tau)^{3/2}/\sqrt{2} M_2^2) \sqrt{\frac{\pi}{8}} \mathcal{H}_{3/4}^{1}(\frac{1}{2}\sqrt{\beta_0}k^2 \tau^2)$ with $\mathcal{H}_{3/4}^{1}$ being the Hankel function of the first kind. On the other hand, one can easily see numerically that the DBI solution is recovered in the $\beta_0 \rightarrow 0$ limit.\\ To give some intuition on the behaviour of the general, interpolating wavefunction, we plot it in several $(\alpha_0, \beta_0)$ configurations. For overall consistency in the comparisons, in all the following pictures we have chosen points in the $(\alpha_0, \beta_0)$-plane so that the horizon crossing always lies at the same point, numerically $\tau_{*}=-\sqrt{2}$, and we have plotted the wavefunction from well inside the horizon ($\tau=-10\,\, \tau_{*}$) up to $\tau=0$.\\
\newpage
\begin{figure}[hp]
	\centering
		\includegraphics[width=0.43\textwidth]{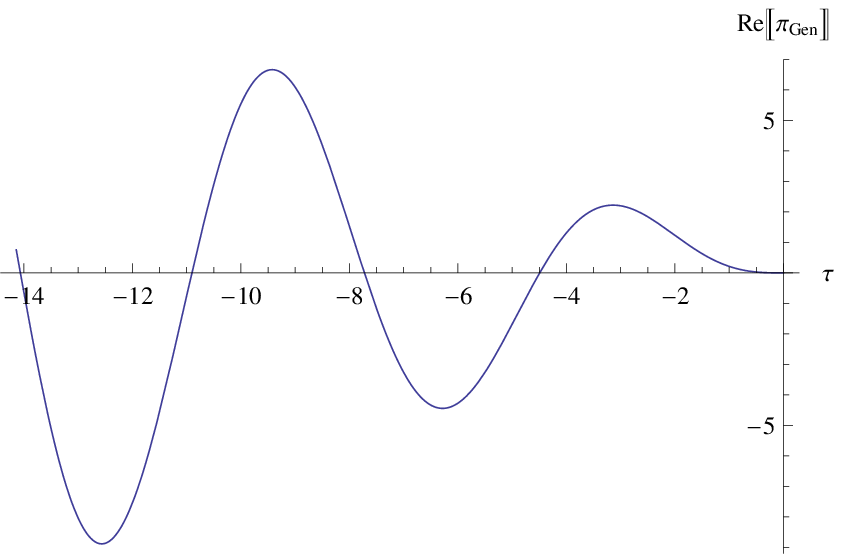}
		\hspace{8mm}
			\includegraphics[width=0.43\textwidth]{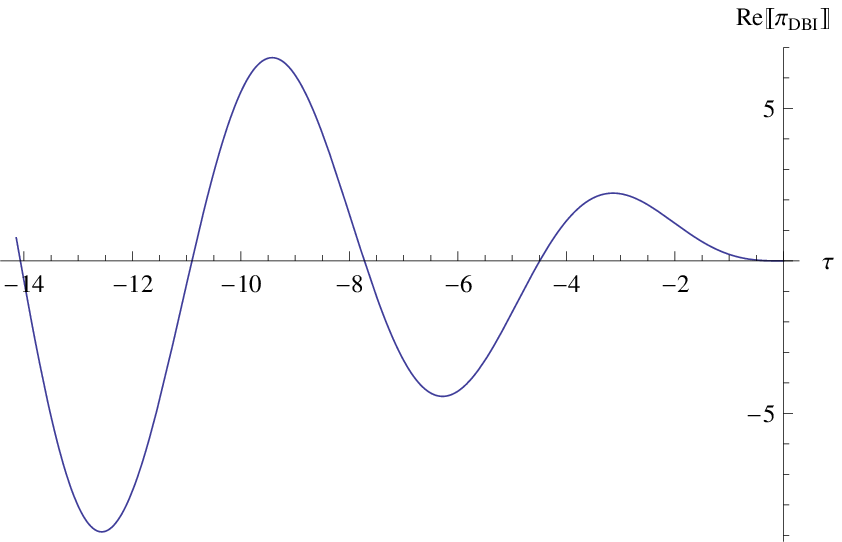}
	\caption{On the left, the general wavefunction in the DBI-like, $\beta_{0}\rightarrow 0$, limit; on the right the DBI-like solution itself. We plot the real part and find perfect agreement, same holds for the imaginary part. To produce the plot the parameters have been set to: $\alpha_0=0.1, \beta_0 \rightarrow 0, k=1, H=1$ and the Planck mass-dependent normalization has been neglected. The corresponding plot can be omitted for the Ghost limit since in that case we recover the Ghost solution analytically.}
	\label{fig:gendbilimit}
\end{figure}

\begin{figure}[hp]
	\centering
		\includegraphics[width=0.43\textwidth]{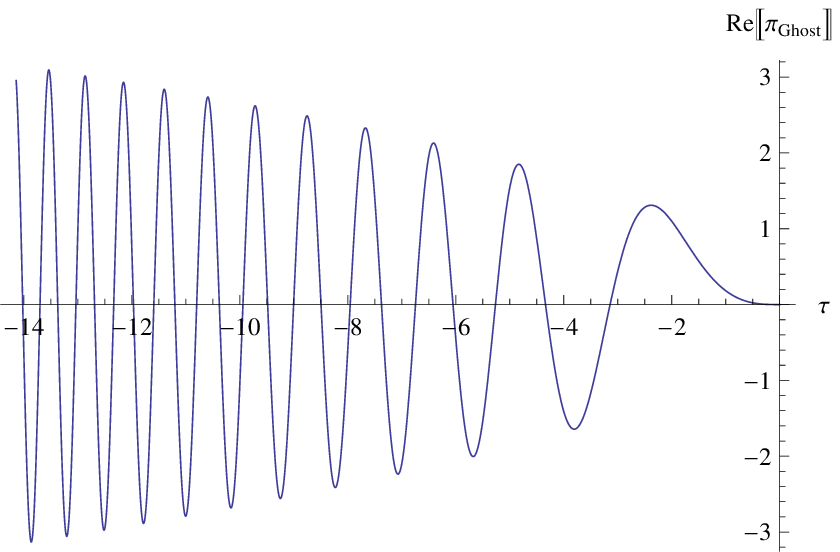}
		\hspace{8mm}
			\includegraphics[width=0.43\textwidth]{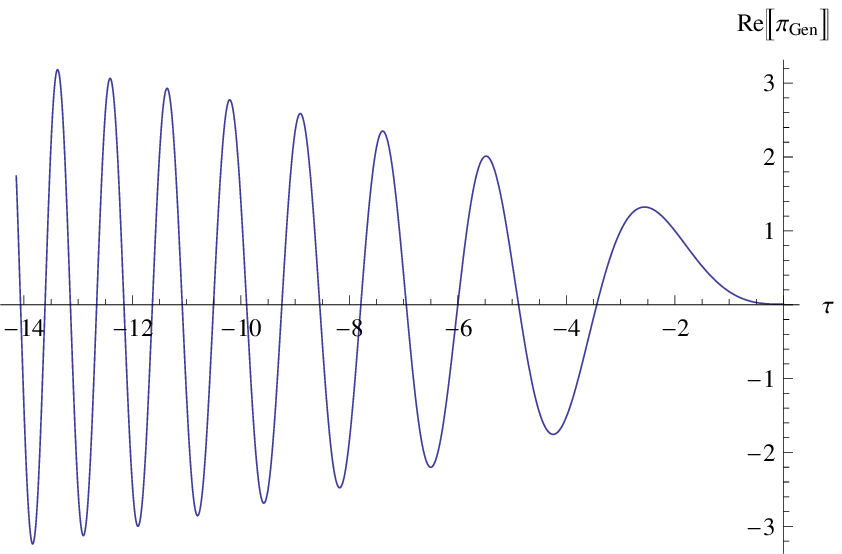}
	\caption{On the left, the Ghost Inflation wavefunction ($\alpha_0=0$). On the right the general interpolating solution calculated for $\alpha_0=1/2\,,\, \beta_0=1/4$.}
	\label{ghostandgen}
\end{figure}

\noindent From these plots we learn several things. First, as argued before, there is a common oscillatory behaviour once inside the horizon. The frequency of these oscillations is more pronounced for the Ghost solution when deeper inside the horizon. In the general solution the frequency varies according to the ``relative weight'' of the Ghost component, $\beta_0$, and the DBI-like one, $\alpha_0$.
\subsection{Power spectrum}
We now turn to the expression of the power spectrum

\bea
\fl P_{\pi}= \frac{k^3}{2 \pi ^2}|\pi(k,\tau \rightarrow 0)|^2=\frac{H^2}{16 \pi  (M_{P}^2 \epsilon H^2 +2M_2^4) (\alpha_0 + \sqrt{\beta_0})^{3/2}\,
|\Gamma(\frac{5}{4}+\frac{\alpha_0}{4 \alpha_0-4 i \sqrt{\beta_0}})|^2} .\nonumber\\
\eea
Clearly there is no time dependence in the above result, the modes freeze outside the horizon. Note here that, to reproduce standard results, we should re-introduce the speed of sound $c_s$ which, in the language we are using, is related to $M_2$ via:
\[
\frac{1}{c_s^2}=1-\frac{2 M_2^4}{M_P^2 \dot H}.
\] 
Upon switching to the gauge invariant quantity related to $\pi$ by $\zeta=-H \pi $ \cite{eft08} (see also the appendix of \cite{3pt} for the relation at second order), and 
reintroducing the proper units with Planck mass we get:
\be
P_\zeta=\frac{ (\alpha_0 + \sqrt{\beta_0})^{-3/2}  H^4}{16 \pi   (M_{Pl}^2 \epsilon H^{2}+2M_2^4) \,
|\Gamma(\frac{5}{4}+\frac{\alpha_0}{4 \alpha_0-4 i \sqrt{\beta_0}})|^2}. \label{p1}
\ee
Again, one could easily check that the above result analytically covers the power spectrum of DBI-like and Ghost Inflation.\\
As we mentioned in the last comments to equation (\ref{l2}), even when including subleading terms in the quadratic action, the functional dependence of our wavefunction does not change, only the definition of $\alpha_0, \beta_0$ does. Since we are now going to set bounds on operators coefficients , we want to be as precise as possible and will therefore extend the definition of the two parameters to cover the subleading terms as well. We now have: 
\bea
\fl \qquad \alpha_0 = \frac{-M_{Pl}^{2}\dot H - \bar M_{1}^{3}H/2 }{-M_{Pl}^{2}\dot H +2M_{2}^{4} - \bar 3 \bar M_{1}^{3}H }\,\, ; \qquad \beta_{0}=\frac{\bar M_{0}^{2}H^2/2}{-M_{Pl}^{2}\dot H +2M_{2}^{4} - 3\bar M_{1}^{3}H}. \label{param}
\eea

\noindent In obtaining Eq.~(\ref{param}), we took into account the fact that the $\bar M_{1},\bar M_{2},\bar M_{3}$-driven terms multiply operators of the type  ${\dot \pi}^2, (\partial_i \pi)^2$ as well. We also choose to replace $\bar M_{2}$ and $\bar M_{3}$ with a linear combination of the two masses: we set $\bar M_3^2 =-3\bar M_2^2$ and $\bar M_0^2 =\bar M_2^2+\bar M_3^2$, see also \cite{b}. This procedure allows one to put to zero all the subleading operators tuned by $\bar M_{2},\bar M_{3}$ and makes the correspondence between inflationary models and the switching of the $M,\bar M$ parameters absolutely sharp \footnote{The reader might worry that one degree of freedom is lost. However, the two coefficients multiply basically the same interaction terms in the action up to
fourth order.}.\\ An immediate simplification is that now, upon requiring $\bar M_0=0 \Leftrightarrow \beta_0 =0$, one goes into DBI inflation, \textit{exactly}. Similarly, now de-Sitter limit and $\bar M_1=0$ give Ghost Inflation with $\alpha_0=0$. The power spectrum looks very similar to the one in Eq.~(\ref{p1})

\be
P_\zeta=\frac{ (\alpha_0 + \sqrt{\beta_0})^{-3/2}  H^4}{16 \pi   (M_{Pl}^2 \epsilon H^{2}+2M_2^4-3 \bar M_1^3 H_{}) \,
|\Gamma(\frac{5}{4}+\frac{\alpha_0}{4 \alpha_0-4 i \sqrt{\beta_0}})|^2}, \label{p2}
\ee
only the definitions of the $\alpha_0,\beta_0$ parameters and the normalization constant have slightly changed. Expressing all the parameters in the spectrum in terms of the $M, \bar M$ coefficients, we count the degrees of freedom as being five, associated to $H, \epsilon, M_2, \bar M_0, \bar M_1$. The first three are the same that appear also in the standard case as $H, \epsilon, c_s$; using $c_s$ instead of $M_2$ is just a matter of dictionary. In the Ghost Inflation case, the quantity $\bar M_0$ replaces the speed of sound and we are again back to three parameters. In the most general case one has to keep both $\bar M_0$ and $\bar M_1$ as well. Bounds can be put on the values of these five parameters by employing the expected value for the power spectrum, $P_{k}^{\zeta}\sim 10^{-10}$ and its tilt \footnote{See also \cite{cora1,cora2}}. Let us also mention here that further mild inequalities must be satisfied by the $\alpha_0, \beta_0$ parameters in order to keep the generalized speed of sound small, \textit{Section 2}. In \textit{Sec. 2.3} below we present a calculation for the tilt and running of the power spectrum. These quantities are essentially obtained from considering the time dependence of the $\bf{M}_n$ coefficients in Eq.~(\ref{p2}). On the other hand, as has been specified above, part of the procedure that led to Eq.~(\ref{p2}) has been to disregard the time dependence of said coefficients \footnote{This is because, at the level of the action, considering the time dependence of these coefficients would automatically translate into going at next-to-leading order in slow-roll; for a leading-order calculation it is therefore sufficient to consider their values at the horizon.}. Restoring it at a later step, as we do below, is standard accepted procedure because generally only at this stage the effect of time-dependence becomes important. A calculation that does without this assumption is presented in \textit{Sec. 3}.
\subsection{Tilt and running}
Below we employ some simplification in order to present our result for the tilt of the power spectrum in a way that resembles as closely as possible the typical expression for  $n_s - 1$. Indeed, the spectrum dependence on the Euler $\Gamma$ function in Eq.~(\ref{p2}) is not to be found in e.g. DBI, Ghost Inflation etc. For simplicity, we choose not to write here the explicit dependence of the Euler function on the $\bf{M}_n$ coefficients and leave it implicit; we report the full explicit dependence in the \textit{Appendix A}.
We apply the following formula,
\be
\fl
n_s-1=  \frac{d}{d \ln{k}}\ln{P_k}=\left(\frac{d \ln{k}}{dt}\Big{|}_{t=t^*}  \right)^{-1}\frac{1}{P_k}\frac{d P_k}{dt}\Big{|}_{t=t^*} \simeq\frac{1}{H P_k}\frac{d P_k}{dt}\Big{|}_{t=t^*} ,
\ee
on the power spectrum, where $t^{*}$ is the time at horizon crossing. The time dependence of the $M, \bar M$ coefficients is taken into account and the time dependence of the Euler function is dealt with as one would do with a generic function  $\Gamma(t)$. One obtains:
\bea
n_s-1= -\frac{\dot \Gamma}{H \Gamma}-\epsilon\times \frac{7 H^2 M_P^2 \epsilon+8 \left(2 M_2{}^4+3 M_1{}^4\right) }{2  \left(H^2 M_P^2 \epsilon+2 M_2{}^4+3 M_1{}^4\right)}+\nonumber\\ 
\fl -\epsilon \times  \frac{8 H^4 M_P^4 \epsilon^2-16 H^2 M_P^2 \epsilon \left(2 M_2{}^4+3 M_1{}^4\right)+H^2 M_P^2 \epsilon  \left(-14 H^2 M_P^2 \epsilon +8 M_2{}^4+15 M_1{}^4\right) }{2  \left(H^2 M_P^2 \epsilon+2 M_2{}^4+3 M_1{}^4\right) \left(2 H^2 M_P^2 \epsilon+M_1{}^4+M_0{}^2 \sqrt{2 H^2 M_P^2 \epsilon +4 M_2{}^4+6 M_1{}^4}\right)}\nonumber\\
\fl- \eta \times \frac{ \left(6 H^2 M_P^2 \epsilon +24 M_2{}^4+33 M_1{}^4\right)H^2 M_P^2 \epsilon }{4 \left(H^2 M_P^2 \epsilon +2 M_2{}^4+3 M_1{}^4\right) \left(2 H^2 M_P^2 \epsilon +M_1{}^4+M_0{}^2 \sqrt{2 H^2 M_P^2 \epsilon +4 M_2{}^4+6 M_1{}^4}\right)}\nonumber\\
\fl - \eta \times \frac{ H^2 M_P^2  \epsilon }{4 \left(H^2 M_P^2 \epsilon +2 M_2{}^4+3 M_1{}^4\right)}-\frac{\dot M_2}{H M_2}\times \frac{ 2 M_2{}^4}{ \left(H^2 M_P^2 \epsilon +2 M_2{}^4+3 M_1{}^4\right)}+\nonumber\\
\fl+ \frac{\dot M_2}{H M_2}\times \frac{ \left(6 H^2 M_P^2 \epsilon +3 M_1{}^4\right)2 M_2{}^4 }{ \left(H^2 M_P^2 \epsilon +2 M_2{}^4+3 M_1{}^4\right) \left(2 H^2 M_P^2 \epsilon +M_1{}^4+M_0{}^2 \sqrt{2 H^2 M_P^2 \epsilon +4 M_2{}^4+6 M_1{}^4}\right)}\nonumber \\
\fl -\frac{\dot M_1}{H M_1}\times \frac{\left(+ 3M_1{}^4-4 H^2 M_P^2 \epsilon +4 M_2{}^4\right)3 M_1{}^4 }{ \left(H^2 M_P^2 \epsilon +2 M_2{}^4+3 M_1{}^4\right) \left(2 H^2 M_P^2 \epsilon +M_1{}^4+M_0{}^2 \sqrt{2 H^2 M_P^2 \epsilon +4 M_2{}^4+6 M_1{}^4}\right)}\nonumber \\ 
\fl  -\frac{\dot M_0}{H M_0}\times \frac{6 M_0^2 }{ \left(2 M_0{}^2+\sqrt{2} \left(2 H^2 M_P^2 \epsilon +M_1{}^4\right) \sqrt{\frac{1}{H^2 M_P^2 \epsilon +2 M_2{}^4+3 M_1{}^4}}\right)} \nonumber \\
\fl -\frac{\dot M_1}{H M_1}\times \frac{3 M_1{}^4 }{ \left(H^2 M_P^2 \epsilon +2 M_2{}^4+3 M_1{}^4\right) } \, ,  \label{ns}
\eea
where all the quantities are to be intended as calculated at horizon crossing.\\
For convenience, we have factored out the usual parameters: $\epsilon$, $\eta$, $s$ (the latter is written in $M_2$ language; we give the dictionary in Eq.~(\ref{e2}) below) and their generalization:
\bea
\epsilon_{0}=\frac{\dot M_0}{H M_0}\,\,; \qquad \epsilon_{1}=\frac{\dot M_1}{H M_1}\,\,; \qquad \epsilon_{\Gamma}=\frac{\dot \Gamma}{H \Gamma}\,\,. \label{genep}
\eea
\bea
 \epsilon_2\equiv \frac{\dot M_2}{H M_2}=\frac{\eta}{2}-\epsilon +\frac{s}{2(c_s^2-1)} \label{e2}\,. 
\eea
\noindent The variable $c_s$ is the one defined in \textit{Sec. 2.1} and $\eta=\dot \epsilon /(H \epsilon)$. For simplicity, we have defined the variables $M_0^4 \equiv \bar M_0^2 H^2$ and $M_1^4 \equiv \bar M_1^3 H$.\\
\noindent Let us briefly comment on the above results. First note that, as expected, the factor that each of these generalized slow-roll parameter multiplies is of order unity or smaller. A quick consistency check consists in specializing the formula above to known inflationary models, for example requiring $\bar M_1=0=\bar M_0$ gives back the usual result \cite{3pt} for DBI-like models:
\bea
n_s-1= -2\epsilon -\eta -s
\eea
Consider now the more general case with $(M_1=0\,,\dot H \not= 0\,,M_0\not= 0\,) ,$ which comprises  DBI-like theories and Ghost Inflation models as limiting cases. In such a scenario there are some mild bounds to be required on $M_0^4, M_{P}^2 \epsilon H^2$. First of all, since we are interested in the $\alpha_0 \ll 1\, , \beta_0 \ll 1$ region of the parameters space this requires:
\bea
M_{P}^2 \epsilon H^2 \ll M_2^4\,\,;  M_0^4 \ll M_2^4 \,.
\eea
On the other hand, the $M_0$-driven slow-roll parameters is $\epsilon_0= \dot M_0 /(H M_0)$ and therefore the above inequalities do not put upper bounds on this slow-roll parameter. Much like we will see in the next section for the running of the bispectrum amplitude, one must instead be careful to account for the fact that too large a value for $\dot M_0$ could give a contribution to the bispectrum amplitude that must be excluded. Indeed, once expanded, the $M_0$-proportional contribution to the quadratic action for the scalar reads:
\bea
\fl \int{ d^4 x \sqrt{-g}\Big[..+ M_0^4(t+\pi) \frac{(\partial_i^2 \pi)^2}{a^4}   \Big]}\sim  \int{ d^4 x \sqrt{-g}\Big[..+ M_0^4(t) \frac{(\partial_i^2 \pi)^2}{a^4}+ 4 M_0^3(t) \dot M_0 \pi\frac{(\partial_i^2 \pi)^2}{a^4}   \Big]},\nonumber\\
\eea
\noindent with the second term on the RHS action clearly contributing to the three-point function of the scalar.  This $\dot M_0$-generated contribution must be weighted against the third order interactions generated by $M_0$ itself, but also by new $\bf{M}_n$ coefficients that first appear at third order in the action (see \cite{b} and the analysis in \textit{Sec. 4}). A similar analysis applies for $\epsilon_1$ if we let $M_1 \not= 0$.\\
We can conclude that the mild bounds on these new, generalized slow-roll parameters come from the obvious fact that we are doing a slow-roll expansion, from the value of the power spectrum itself and from the requirement that they are not so large as to produce too large a value for $f_{NL}$. \\
In order to obtain the running of the power spectrum we proceed as below:
\be
\alpha_{s}= \frac{d n_s}{d\log k} \simeq \frac{1}{H}\frac{d n_s}{d t}.
\ee
We give below a compact results: 
\bea
\alpha_s= &-&\frac{\ddot \Gamma}{H^2 \Gamma} + \epsilon_{\Gamma}^{2}- \frac{\dot \epsilon \Theta_{}}{H }- \frac{ \epsilon \dot \Theta_{}}{H }- \frac{\dot \epsilon_{2} \Theta_{2}}{H }- \frac{ \epsilon_{2} \dot \Theta_{2}}{H }
- \frac{\dot \eta \Theta_{\eta}}{H }
\nonumber \\
&-& \frac{ \eta \dot \Theta_{\eta}}{H }
- \frac{\dot \epsilon_{1} \Theta_{1}}{H }- \frac{ \epsilon_{1} \dot \Theta_{1}}{H }
- \frac{\dot \epsilon_{0} \Theta_{0}}{H }- \frac{ \epsilon_{0} \dot \Theta_{0}}{H },\nonumber\\ \label{run}
\eea
and point the reader to the \textit{Appendix A} for a more explicit expression of the coefficients functions $\Theta,\Theta_{2},\Theta_{\eta},\Theta_{1},\Theta_{0}$.
\subsection{Next-to-leading order}
The discussion presented so far is based on a generalized slow-roll approximation at leading order. In particular, the $M , \bar M$ coefficients driving the various operators in the Lagrangian are assumed to be time independent. This assumption propagates into the equation of motion for the scalar $\pi$, the classical solution itself and the power spectrum. To make up for this approximation when calculating the the tilt of the spectrum, one restores the time dependence of the coefficients at the level of the power spectrum. A more systematic approach consists in accounting for the time dependence of the coefficients already at the Lagrangian level, this is done by taking the generalized slow-roll approximation to next order. Schematically one has:
\bea
\fl S^{\pi}_{2}\propto \int d^4 x \sqrt{-g} \Big[ -M_P^2 (3H^2(t+\pi)+\dot H(t+\pi))+M(t+\pi)\times (\rm{ quadratic})   \Big] \, . \label{2nd}
\eea
\noindent The first term does not appear in the Lagrangian in Eq.~(\ref{l2}) because, at leading order in slow roll, it is not quadratic in fluctuations, it does in fact contribute to the background. We see that accounting for the ``$\pi$'' in the time dependence of $M$ in the last term in Eq.~(\ref{2nd}) would result in a third order operator. This must be considered when studying interactions but it is not what we want to analyze here, the wavefunction comes from the quadratic Lagrangian. On the other hand, the ``$\pi$'' in the first term of the action has to be accounted for; doing so results in just one additional contribution to the action and it turns out  to be proportional to $\epsilon ^2$. This is all consistent with the fact that, at leading order, the action is instead proportional to $\epsilon$. We now have:
\bea
 \mathcal{L}_2=  a^3&\Big[&M_{P}^{2}\dot H(t) (\partial_{\mu} \pi)^2 
+ 2 M_2^4(t){\dot\pi}^2 - \bar M_1^3(t) H(t) (3 {\dot \pi}^2-\frac{(\partial_i \pi)^2 }{2 a^2})  \nonumber\\
 &+& \frac{\bar M_0^2(t)}{2}\frac{1}{a^4} (\partial_i^2\pi)^2 -3 M_P^2 {\dot H(t)}^2 \pi^2  \Big] \, .
\label{l22}
\eea
We proceed to write down the equation of motion as in the leading order case, obtaining:
\bea
\fl \sigma_k^{''}+\alpha_0(1+s_{\alpha})k^2 \sigma_k+\beta_0(1+s_{\beta}) \frac{k^4}{a^2 H^2}\sigma_k = \left(\frac{f^{''}}{f}+3\epsilon a^2 H^2\frac{ M_P^2 {\dot H(t)}}{2M_2^4-M_P^2 \dot H -\bar M_1^3 H}   \right)\sigma_k\, ,\nonumber\\ \label{eoma}
\eea
where the following definitions have been employed:
\bea
\fl s_{\alpha}=\frac{\dot \alpha_0}{ H \alpha_0}\ll 1\,; \qquad s_{\beta}=\frac{\dot \beta_0}{ H \beta_0}\ll 1\,;\quad f^2=a^2( -M_P^2 \dot H +2 M_2^4 -3\bar M_1^3 H)\,; \quad \pi=\frac{\sigma}{f}. \nonumber \\
\eea
In order to solve the equation of motion one needs to calculate $f^{''}/f$ explicitly; the result is given in the Appendix A.
A compact expression for the equation of motion is given by:
\bea
 \sigma_k^{''}+\tilde \alpha_0 k^2 \sigma_k+\tilde \beta_0 {k^4 \tau^2}\sigma_k =\frac{2}{\tau^2}(1+x_0)\sigma_k\, \label{eomb} \, ,
\eea
where $x_0$ is a linear combination of slow roll parameters and $\tilde \alpha_0,\tilde \beta_0$ represent a slight redefinition of the initial parameters. For explicit expression we refer once again the reader to the Appendix A. Equipped with Eq.~(\ref{eomb}), one uses the Bunch-Davies vacuum condition to write down the solution:
\bea
\fl \qquad \sigma(x,k)=C_1 (k) \frac{ \mathcal{G}\left[\frac{-i \tilde\alpha_0 \sqrt{\tilde \beta_0}+2 \tilde \beta_0 + \tilde \beta_0 \sqrt{9+8 x_0}}{4 \tilde\beta_0},\frac{1}{2} \left(2+\sqrt{9+8 x_0}\right),-i \sqrt{\tilde \beta_0} k^2 x^2\right]}{2^{-\frac{1}{4} \left(2+\sqrt{9+8 x_0}\right)} e^{-\frac{1}{2} i \sqrt{\tilde \beta_0} k^2 x^2} \left(x^2\right)^{\frac{1}{4} \left(2+\sqrt{9+8 x_0}\right)}\sqrt{x}} \, ,
\eea
where $\mathcal{G}$ is the usual hypergeometric function, $x = -\tau$, and the wavefunction must be expanded to first order in $x_0 = 0$. In the first section, we gave the exact expression for the solution above at leading order and went on to calculate the resulting power spectrum, its tilt and running. The calculation at next-to-leading order has already been performed for DBI-like theories of inflation, using the same formalism employed here, in \cite{3pt}. Here instead, we choose to calculate the next-to-leading order Ghost Inflation solution obtaining also the tilt of the power spectrum and the running. The procedure is a standard one, so we briefly sketch it. The two $k$-dependent constant of the Ghost equation of motion
\bea
 \sigma_k^{''}+\tilde \beta\, {k^4 \tau^2}\sigma_k =\frac{2}{\tau^2}(1+x_0^{G})\sigma_k\,, \label{eomghost}
\eea
are reduced to one by imposing the correct leading order limit on the wavefunction. The remaining constant is obtained by requiring the proper normalization, that is by imposing the following commutation relations to hold:
\bea
[\pi(\vec x),P(\vec y)]=i\, \delta^3(\vec x -\vec y)\,; \qquad [a_{\vec k},a^{\dagger}_{\vec p}]= (2\pi)^3 \delta^3(\vec k -\vec p) \, ,
\eea
where $P$ is the momentum conjugate of the scalar $\pi$ and the creation and annihilation operators $a,a^{\dagger}$ are the usual operators in the free field expansion for the quantized field $\pi$. Proceeding as prescribed above, one obtains:
\bea
\fl \pi(k,\tau)= \frac{H \sqrt{2 \pi } (1+s_{\beta})^{1/8} (-\tau )^{3/2} \, H^{(1)}\left[\frac{1}{4} \sqrt{9+8 x_0},\frac{1}{2} k^2 \sqrt{\beta_0 (1+s_{\beta})} \tau^2\right]}{3 \left(1+\frac{s_{\beta}}{8}-x_0 (-\frac{5}{9}+\gamma/3-\frac{\pi }{6}+\log{2})\right)\left( \Gamma \left[1+\frac{1}{4} \sqrt{9+8 x_0}\right]\right)^{-1} \Gamma \left[\frac{3}{4}\right]}.\label{ghostsol}
\eea
The wavefunction above is the correct Ghost Inflation wavefunction up to next-to-leading order in generalized slow-roll parameters; it gives back the leading order solution and respects the proper Bunch-Davies vacuum requirement. From Eq.~(\ref{ghostsol}) one can readily calculate the tilt of the spectrum: one simply considers the leading behaviour of the wavefunction as $k$ goes to zero. In fact, upon expanding for small $k$ one finds that our solution goes like:
\bea
 H^{(1)}\left[\frac{3}{4}+\frac{x_0}{3},\frac{1}{2} \sqrt{\beta_0}\left(1+\frac{s_{\beta}}{2}\right) k^2 \tau^2\right]\sim k^{-\frac{3}{2}-\frac{2 x_0}{3}} \, ,
\eea
from which we obtain that 
\bea n_s-1 = -4/3 x_0.
\eea 
\noindent When specialized to Ghost Inflation, the value of the $x_0$ parameter (which always constists of a linear combination of the generalized slow roll parameters) is given by:
\bea
x^{G}_0=3 \frac{\dot M_2}{H M_2},
\eea
and the running amounts to simply 
\bea \frac{d n_s}{d\ln{k}}= -4\, \dot{x_0}/3H .
\eea

\noindent Let us briefly recall what has been done so far at second order in perturbations: we obtained a solution to the equation of motion of the $\pi$ effective action. This, properly normalized, has been used to obtain the power spectrum of curvature perturbations, its tilt and running. For the specific case of Ghost inflation we have obtained all of the above also at next-to-leading order in generalized slow-roll approximation.\\
We now move on to the study of cosmological observables relevant to non-Gaussianity: the bispectrum, its running, and the trispectrum. For the bispectrum calculations one starts from the cubic effective action as written in Eq.~(\ref{action}). We note here that, for the bispectrum calculations, we employed both the exact, general wavefunction as given in  Eq.(\ref{sol}) and the simplified solution which is found in models such as DBI inflation or K-inflation. By comparing the shape-functions obtained in these two different ways, we verified that the simplified solution is indeed a good approximation to the exact wavefunction and used only the former in the calculations for the trispectrum.
\newpage
\section*{*}
\newpage
\section{Amplitude of the primordial non-Gaussianity: Bispectrum}
\label{ampl}
In this section we wish to perform a general analysis of the   amplitude of the bispectra stemming from the 
general  third-order interaction terms. The shape analysis will be done in the following section. In the calculations that follow we employ the so-called {\rm in-in} 
formalism \cite{in-in1, in-in3, in-in2, w-qccc}. To compute the amplitude of the non-Gaussianity, indicated by   $f_{\rm NL}$, we proceed as traditionally done in the literature and 
evaluate  the three-point correlator in the so-called  equilateral configuration where all momenta are taken to be equal: $k_1 = k_2 = k_3$. In other words one can write the bispectra of the gauge-invariant 
curvature perturbation $\zeta$ generated by each interaction term (I) as  
\begin{eqnarray}
\langle \zeta(\textbf{k}_1)\zeta(\textbf{k}_2)\zeta(\textbf{k}_3)\rangle_{I}
=(2\pi)^3\delta^3(\textbf{k}_1+\textbf{k}_2+\textbf{k}_3) B_I(k_1,k_2,k_3)\, ,
\end{eqnarray} 
with an amplitude $f^I_{\rm NL}$ defined so that with all three-momenta equal $f^I_{\rm NL}=(6/5) B_I(k,k,k)/P_{\zeta}(k)^2$, where $P_{\zeta}(k)$ is the power spectrum of the curvature perturbation. 
For large non-Gaussianities the linear relation  $\zeta=-H \pi$ suffices  for the bispectrum calculations since quadratic corrections give a negligible contributions. 
Notice also that the values of the integrals appearing in the in-in computations have been specified at horizon crossing 
and the contribution of the integral function at $\tau=-\infty$  has been put to zero mimicking  the effect of the slight rotation of the  $\tau$-axis into the imaginary plane.

Given the broad number of possibilities,  we choose to compute numerically the amplitude of the various bispectra  identifying  six  benchmark points in the $(\alpha_0, \beta_0)$  plane and numerically integrating the exact wavefunctions. Since one is interested in probing models with large non-Gaussianity, the values of $(\alpha_0, \beta_0)$ are taken much smaller than unity as described in the following Table:

\begin{center}
\begin{tabular}{| l || l | l | l | l | l| l | }
\hline			
       Benchmarks   & 1 & 2 &3 &4 &5 &6 \\ \hline 
  $\alpha_0$ & $10^{-2}$  & 0 & $0.5\cdot 10^{-2}$ & $2 \cdot 10^{-7}$ &$10^{-4}$  &$10^{-6}$ \\ \hline
  $\beta_0$  &  0 & $0.5\cdot 10^{-4} $ & $0.25\cdot 10^{-4} $ & $5\cdot 10^{-5}$ & 0&0  \\
\hline  
\end{tabular}
\\
\vspace{5mm}
\end{center}
 Upon using Eq.~(\ref{h}) one can check that the first four benchmark points   correspond to the same choice of the 
 effective horizon. 
 This choice has been made to suitably perform a comparison between the various cases and in particular against the values of  $f_{\rm NL}$ of purely $P(X,\phi)$ and  Ghost models which correspond to benchmarks 
1 and 2. 
 The last two configurations (which are $P(X,\phi)$-type) probe the space of extremely small $\alpha_0$ and, in interactions with at least two space derivatives, are expected to give a larger amplitude than the first four points, at least if the  interaction terms are regulated by unconstrained masses. To get the feeling of the figures involved, if we restrict ourselves  to the case of theories for which $\alpha_0$ reduces to  the usual sound speed $c_s^2$, typical values of $c_s$ are between  $10^{-3}$ and $10^{-2}$. Note also that the definition of $c_s$ varies according to which operators are switched on in the action (one can well be 
in the de Sitter limit where the only spatial quadratic term has four derivatives; this leads 
to a different $c_s$, see also \cite{ssz05}). Therefore we choose to specify all the amplitudes as a function of the various $M$ and $\bar M$ masses.

It is convenient at this stage to clearly point out which $M$ coefficients in Eq.~(\ref{action}) are free and underline the relations among the constrained ones. From Eq.~(\ref{param}) one can see that, despite fixing $\alpha_0$, as we did in the Table, there is still  the freedom to pick any reasonable value for either $ M_2^4$ or $\bar M_1^3$. Similarly, fixing $\beta_0$ does not completely specify $\bar M_0$. In other words, both $\bar M_1^{3}$ and  $\bar M_0^{2}$ are constrained by our choice of the $(\alpha_0, \beta_0)$ parameters; all the other coefficients are unconstrained. As the  $M$'s are expected to set the   energy  scale of the various underlying theories, they should be  larger than the Hubble rate $H$, and can go up to  $M_{\rm Pl}$.
As elucidated in Ref. \cite{w-e} though , for the action (\ref{action}) to be as general as possible,  one might want to require the $M$'s to be smaller than the Planck mass. That said, some useful inequalities that the mass coefficients must respect can now be reminded. Due to the fact we are working in the decoupling regime, we must require $M_2^4 < M_{\rm Pl}^2 H^2$. Also, the fact that we are probing the 
$(\alpha_0, \beta_0) \ll 1 $ space, imposes  bounds on some masses.  Consider the parameter $\alpha_0$ in Eq.~(\ref{param}). There are two ways this coefficient can be much smaller than unity. The first and perhaps most natural way, is to ask $M_2^4 \gg {\rm Max}\,(-M_{\rm Pl}^2 \dot H, -\bar M_1^3 H)$ which, due to decoupling inequalities on $M_2$ puts a bound on $\bar M_1$,  $\bar M_1^3 \ll M_{\rm Pl}^2 H$. The other possibility requires a partial cancellation in the numerator of $\alpha_0$, $-M_{\rm Pl}^2 \dot H \sim \bar M_1^3 H/2$ which is certainly possible but it implies  we are neither in the DBI, nor in the ghost regime, both of which have $\bar M_1=0$. Looking at $\beta_0$ we see it is  enough to require $M_0^2 H^2 \ll M_2^4$ or $M_0^2 H^2 \ll -\bar M_1^3 H$ and again, the first condition seems more natural. Let us stress here that, upon requiring the masses to be all of the same order, $M$,  and using that $H/M\ll 1$ in the effective theory,  one can easily obtain small $\alpha_0, \beta_0$ coefficient. 
However, when employing a single mass scale $M$  in the whole Lagrangian, working with tiny values for $\alpha_0$ and $\beta_0$ would put a bound on $M$ and necessarily influence the magnitude of all the interaction terms.
In our analysis we let the $M,\bar M$'s coefficients  be not all of the same order (with some important caveats upon which we expand at the end of this section).

 Below we present the results for each interaction term. All the amplitudes can be written as a dimensionless coefficient, $\gamma_n$, times an ($\alpha_0,\beta_0$)-dependent numerical coefficient.
The terms described in the first subsection  are interactions that have already been discussed in the literature. The novelty here is represented by the fact we are able to study also  interpolating configurations through the third and fourth benchmark points. In the second subsection we report on the amplitudes of the contributions from some curvature-generated terms that have never been discussed in the literature.
 
\subsection{Amplitudes from $P(X,\phi)$-type interactions and first two curvature-generated terms}
The amplitudes from ``DBI-like'' (an expression which we use as a synonym of $P(X,\phi)$ models) interactions and first two curvature-generated terms are the following: 
\vskip 0.5cm

 $\bullet$ $ {\cal O}_1=-2 M_2^4 \dot\pi (\partial_i \pi)^2/a^2$

\begin{center}
\begin{tabular}{ |l || l | l | l | l | l| l | }
\hline			
benchmarks   & 1 & 2 &3 &4 &5 &6 \\ \hline 
$f^{M_2}_{\rm NL}$   &     $\,\,\, 10^{2} \gamma_1 \,\,\, $     &    $\,\,\, 8 \cdot  10^{1}\gamma_1\,\,\, $       &     $\,\,\, 6\cdot 10^{1}\gamma_1\,\,\,$     &      $\,\,\, 4\cdot  10^{1}\gamma_1\,\,\,$      &      $\,\,\,  10^{4}\gamma_1\,\,\,$    &    $\,\,\,  10^{6}\gamma_1\,\,\,$  \\ 
  
\hline  
\end{tabular}
\end{center}
 where 
\be 
\gamma_1= \frac{M_2^4}{2M_2^4 +M_{\rm Pl}^2 \epsilon H^2-3 \bar M_{1}^{3}H}.
\ee
We see the parameter $\gamma_1$ can in principle be of order unity. Indeed, if one assumes $M_2^4$ is the largest  term in the denominator (in DBI this would correspond to a very small speed of sound), $\gamma_1$ is roughly $1/2$.\\

$\bullet$ ${\cal O}_2=-4/3 \,\,  M_3^4 {\dot\pi}^3 $\\

\begin{center}
\begin{tabular}{| l || l | l | l | l | l| l | }
\hline			
benchmarks   & 1 & 2 &3 &4 &5 &6 \\ \hline 
$f^{M_3}_{\rm NL}$   &     $\,\,\, 1/2  \gamma_2 \,\,\, $     &    $\,\,\,  10^{-2}\gamma_2\,\,\, $       &     $\,\,\,5 \cdot  10^{-2}\gamma_2\,\,\,$     &      $\,\,\, 10^{-2}\gamma_2\,\,\,$      &      $\,\,\,  1/2 \gamma_2\,\,\,$    &    $\,\,\,  1/2\gamma_2\,\,\,$  \\ 
  
\hline  
\end{tabular}
\end{center}

\noindent where 
\be \gamma_2= \frac{M_3^4}{2M_2^4 +M_{\rm Pl}^2 \epsilon H^2-3 \bar M_{1}^{3}H}.
\ee
The parameter $\gamma_2$ can be even  larger than unity if, for instance,   $M_3$ is larger than  $M_2$. We can see, already at this stage, the effect of small values of $\alpha_0$ and  $\beta_0$ at work: the numerical factor of a spatial derivative-free interaction is much smaller than that of a third order term like the $M_2$ one calculated above which has two spatial derivatives. \\

$\bullet$  \textbf{$ {\cal O}_3=-1/2\,\,  \bar M_1^3 (\partial_i \pi)^2 \partial_j^2 \pi/a^4  $}\\

\begin{center}
\begin{tabular}{| l || l | l | l | l | l| l | }
\hline			
benchmarks   & 1 & 2 &3 &4 &5 &6 \\ \hline 
$f^{\bar M_1}_{\rm NL}$   &     $\,\,\,  10^{5} \gamma_3 \,\,\, $     &    $\,\,\,  10^{3}\gamma_3\,\,\, $       &     $\,\,\,  4 \cdot 10^{4}\gamma_3\,\,\,$     &      $\,\,\, 1.5\cdot  10^{3}\gamma_3\,\,\,$      &      $\,\,\, 10^9 \gamma_3\,\,\,$    &    $\,\,\,  10^{13}\gamma_3\,\,\,$  \\ 
  
\hline  
\end{tabular}
\end{center}
\vspace{5mm}
Due to the fact that $\bar M_1^3$ is a constrained parameter we find that the amplitude equals an $(\alpha_0, \beta_0)$-dependent number times the paramter
\be
\gamma_3 = \frac{-M_{\rm Pl}^2 \epsilon H^2 (1-\alpha_0)+ 2 \alpha_0 M_2^4}{M_{\rm Pl}^2 \epsilon H^2 + 2  M_2^4-3\bar M_1^3 H}.
\ee
One should substitute the various values of $\alpha_0$ in the Table in the $\gamma_3$ expression given above. Barring cancellation between different mass terms, $\gamma_3$ is generally smaller than one and can be as small as $\alpha_0$ itself.\\

$\bullet$  \textbf{$ {\cal O}_4=-1/2 \,\, \bar M_0^2/4\,\,  \left(5H (\partial_i^2\pi) (\partial_j \pi)^2  +4\dot\pi \partial_{i}^2 \partial_j \pi \partial_j \pi \right)/ a^4 $}\\

\begin{center}
\begin{tabular}{| l || l | l | l | l | l| l | }
\hline			
benchmarks   & 1 & 2 &3 &4 &5 &6 \\ \hline 
$f^{\bar M_0}_{\rm NL}$   &     $\,\,\, 3\cdot  10^{5}\,  \gamma_4 \,\,\,$     &    $\,\,\,  10^{5}\, \gamma_4\,\,\, $       &     $\,\,\, 1.3 \cdot 10^{4}\, \gamma_4\,\,\,$     &      $\,\,\,  10^{5}\, \gamma_4\,\,\,$      &      $\,\,\, 3\cdot 10^9 \, \gamma_4\,\,\,$    &    $\,\,\, 3\cdot  10^{13}\, \gamma_4\,\,\,$  \\ 
\hline  
\end{tabular}
\end{center}
\vspace{5mm}\
with 
\be
 \gamma_4 = (\bar M_0^2 H^2 )/(M_{\rm Pl}^2 \epsilon H^2 + 2 M_2^4-3\bar M_1^3 H).
\ee
We see that large numerical coefficients appear in such a case, nevertheless    $\gamma_4=\beta_0\ll 1$.




\subsection{ Amplitudes from curvature-generated novel interaction terms}
We come to the curvature-generated  interaction terms that generate novel bispectra. Their amplitudes are given by
\vskip 0.5cm

$\bullet$  {\bf$ {\cal O}_5=-2/3\,\,  \bar M_4^3 {\dot\pi}^2 \partial_i^2\pi \,\,/ a^2 $}\\

\begin{center}
\begin{tabular}{| l || l | l | l | l | l| l | }
\hline			
benchmarks   & 1 & 2 &3 &4 &5 &6 \\ \hline 
$f^{\bar M_4}_{\rm NL}$   &     $\,\,\, 7 \cdot 10^{2}\, \gamma_5 \,\,\, $     &    $\,\,\,  10^{2} \,\gamma_5\,\,\, $       &     $\,\,\,2 \cdot 10^{2}\, \gamma_5\,\,\,$     &      $\,\,\,  10^{2}\, \gamma_5\,\,\,$      &      $\,\,\, 7\cdot 10^4\, \gamma_5\,\,\,$    &    $\,\,\,   7\cdot 10^6 \,\gamma_5\,\,\,$  \\ 
\hline  
\end{tabular}
\end{center}
\vspace{5mm}

where 
\be 
\gamma_5 = (\bar M_4^3 H )/(M_{\rm Pl}^2 \epsilon H^2 + 2 M_2^4-3\bar M_1^3 H).
\ee 
The coefficient $\gamma_5$ can be larger than unity. Barring cancellations in the denominator, a $\gamma_5\gg1 $ larger than unity  imposes  $\bar M_4 \gg M_2, \bar M_1$. \\

$\bullet$ {\bf $ {\cal O}_6=1/3\,\,  \bar M_5^2 \dot\pi (\partial_i^2\pi)^2 \,\,/ a^4 $}\\

\begin{center}
\begin{tabular}{ |l || l | l | l | l | l| l | }
\hline			
benchmarks   & 1 & 2 &3 &4 &5 &6 \\ \hline 
$f^{\bar M_5}_{\rm NL}$   &     $\,\,\, 5\cdot  10^{4}\, \gamma_6 \,\,\, $     &    $\,\,\, 1.6 \cdot  10^{4}\,\gamma_6\,\,\, $       &     $\,\,\,2 \cdot 10^{4}\, \gamma_6\,\,\,$     &      $\,\,\, 1.6 \cdot  10^{4}\,\gamma_6\,\,\,$      &      $\,\,\, 4\cdot 10^8\, \gamma_6\,\,\,$    &    $\,\,\, 4 \cdot 10^{12}\gamma_6\,\,\,$  \\ 
\hline  
\end{tabular}
\end{center}
\vspace{5mm}
 where 
\be
\gamma_6 = (\bar M_5^2 H^2 )/(M_{\rm Pl}^2 \epsilon H^2 + 2 M_2^4 -3\bar M_1^3 H).
\ee
To get $\gamma_6\gg 1$, one needs to impose a less natural condition $ \bar M_5^2 H^2\gg M_2^4, \bar M_1^3 H $. \\

$\bullet$  {\bf $ {\cal O}_7=1/3\,\,  \bar M_6^2 \dot\pi (\partial_{ij}\pi)^2 \,\,/ a^4 $}\\

\begin{center}
\begin{tabular}{| l || l | l | l | l | l| l | }
\hline			
benchmarks   & 1 & 2 &3 &4 &5 &6 \\ \hline 
$f^{\bar M_6}_{\rm NL}$   &     $\,\,\,  10^{4}\, \gamma_7 \,\,\, $     &    $\,\,\, 4\cdot  10^{3}\,\gamma_7\,\,\, $       &     $\,\,\,5\cdot 10^{3}\,\gamma_7\,\,\,$     &      $\,\,\, 4\cdot 10^{3}\,\gamma_7\,\,\,$      &      $\,\,\, 10^8\,\gamma_7\,\,\,$    &    $\,\,\, 10^{12}\,\gamma_7\,\,\,$  \\ 
\hline  
\end{tabular}
\end{center}
\vspace{5mm}

where
\be
\gamma_7 = (\bar M_6^2 H^2 )/(M_{\rm Pl}^2 \epsilon H^2 + 2 M_2^4 -3\bar M_1^3 H).
\ee 
The same consideration as for the case of $\bar M_5$ apply here. Note again that the numerical values, especially in the fifth and sixth benchmark points tend to be much larger for the $M$ coefficients with the most spatial derivatives, thus conferming our expectations.\\

$\bullet$  {\bf ${\cal O}_8 = -1/6\,\,  \bar M_7 (\partial_i^2 \pi)^3 \,\,/ a^6 $}\\

\begin{center}
\begin{tabular}{| l || l | l | l | l | l| l | }
\hline			
benchmarks   & 1 & 2 &3 &4 &5 &6 \\ \hline 
$f^{\bar{M}_7}_{\rm NL}$   &     $\,\, 8 \cdot 10^6\,\gamma_8 \,\, $     &    $\,\, 2.6 \cdot 10^{6}\,\gamma_8\,\,$       &     $\,\,3.5 \cdot 10^{6}\,\gamma_8\,\,$     &      $\,\,  2.6 \cdot 10^{6}\,\gamma_8\,\,$      &      $\,\, 8\cdot 10^12 \gamma_8\,\,$    &    $\,\, 8\cdot 10^{18}\gamma_8\,\,$  \\ 
\hline  
\end{tabular}
\end{center}
\vspace{5mm}

with 
\be
\gamma_8 = (\bar M_7 H^3 )/(M_{\rm Pl}^2 \epsilon H^2 + 2 M_2^4  -3\bar M_1^3 H).
\ee 
Notice that in this case , and in the following ones, the $\gamma$ coefficients are naturally expected to be smaller than unity. 
For this interaction, as well as the two following ones, the numerical factor coming from the integration can be quite large, especially in the fifth and sixth benchmark points. 
This is clearly due to the six space derivatives that characterize these interactions. \\

$\bullet$  {\bf $ {\cal O}_9=-1/6\,\,  \bar M_8 \, \partial_i^2 \pi (\partial_{jk} \pi)^2 \,\,/ a^6 $}\\

\begin{center}
\begin{tabular}{| l || l | l | l | l | l| l | }
\hline			
benchmarks   & 1 & 2 &3 &4 &5 &6 \\ \hline 
$f^{\bar{M}_8}_{\rm NL}$   &     $\,\, 2\cdot 10^6 \gamma_9 \,\, $     &    $\,\,  6 \cdot 10^5 \gamma_9\,\, $       &     $\,\, 8\cdot  10^{5}\gamma_9\,\,$     &      $\,\, 6\cdot  10^{5}\gamma_9\,\,$      &      $\,\, 2\cdot 10^12 \gamma_9\,\,$    &    $\,\,  2 \cdot  10^{18}\gamma_9\,\,$  \\ 
\hline  
\end{tabular}
\end{center}
\vspace{5mm}

with 
\be 
\gamma_9 = (\bar{M}_8 H^3 )/(M_{\rm Pl}^2 \epsilon H^2 + 2 M_2^4 -3\bar M_1^3 H).
\ee
Let us stress that one could guess the amplitude of $\bar M_{8,9}$ terms by simply looking at the results obtained for the interaction term tuned by $\bar M_7$ because, althought these terms might produce a different shape for non-Gaussianities, they have essentially the same structure as far as the integration is concerned.\\

$\bullet$  {\bf $ {\cal O}_{10}=-1/6\,\,  \bar M_9 \, \partial_{ij} \pi \partial_{jk} \pi  \partial_{ki} \pi \,\,/ a^6 $}\\

\begin{center}
\begin{tabular}{ |l || l | l | l | l | l| l | }
\hline			
benchmarks   & 1 & 2 &3 &4 &5 &6 \\ \hline 
$f^{\bar M_9}_{\rm NL}$   &     $\,\,10^6 \gamma_{10} \,\, $     &    $\,\,  3\cdot 10^{5}\gamma_{10}\,\, $       &     $\,\,  4\cdot  10^{5} \gamma_{10}\,\,$     &      $\,\, 3\cdot  10^{5}\gamma_{10}\,\,$      &      $\,\, 10^{12} \gamma_{10}\,\,$    &    $\,\, 10^{18}\gamma_{10}\,\,$  \\ 
\hline  
\end{tabular}
\end{center}
\vspace{5mm}

where
\be
\gamma_{10} = (\bar M_9 H^3 )/(M_{\rm Pl}^2 \epsilon H^2 + 2 M_2^4 -3\bar M_1^3 H).
\ee 
One could again just read off the maximum value from $f^{\bar M_7}_{\rm NL}$, they differ by just a factor $1/8$. 
\subsection{Some general considerations on the amplitudes of non-Gaussianities}
It might be worth now to pause and   comment on our findings in relation to earlier results in the literature. As we have stressed, the non-Gaussianity generated from operators proportional 
to the  $\bar M$'s masses arises from curvature terms that are often neglected.  From general considerations, one expects that  the curvature terms, 
especially the ones coming from ${\delta K}^3$-type of contributions,  can often be neglected. This is not generically true though. The general structure of the amplitudes we have found for the bispectra can be schematically written as
\be
{\cal A}_n \sim \frac{1}{\left(\sqrt{\alpha_0 +\sqrt{\alpha_0^2 +8 \beta_0}}\right)^q} \frac{{\bf M}_n^{4-q} H^q}{2M_2^4 + M_{\rm Pl}^2 \epsilon H^2  -3\bar M_1^3 H}\, ,
\ee
where $\bf{M}_n$ stands for a generic mass coefficient of the type $M, \bar M$.\\ If $\bf{M}_n \gg H/(\alpha_0 +(\alpha_0^2+8 \beta_0)^{1/2})^{1/2}$, and the masses are all of the same order, then it is natural to expect the $\bf{M}_n$ with the largest exponent to dominate. In this case the dominant terms are those associated 
with $M_2$ and $M_3$, corresponding to DBI-like inflation.  
Furthemore, we may also recall that large masses appear also in the definition of $\alpha_0, \beta_0$ and, at least in $P(X,\phi)$ models, a large $M_2$ is needed to have a small speed of sound. 

On the other hand, one can employ the freedom for all the $M$'s not to be of the same order in magnitude. The $\bf{M}_n$'s have a natural upper bound that must be smaller than the Planck mass and in the general theory with both DBI and ghost (and more in general curvature terms) operators switched on, one can allow for a very small speed of sound without assuming much on the unconstrained masses. In such cases, the generalized sound speed can be so small that $H/(\alpha_0 +(\alpha_0^2 +8\beta_0)^{1/2})^{1/2}$ is actually larger than $\bf{M}_n$. Consequently, 
  the amplitudes of the masses with a smaller exponent, typically the barred $\bar M$'s, are not negligible any longer.  We conclude
  that large non-Gaussianities may be induced by theories parametrized by  suitable values of the $\bar M$ masses.\\

It is important at this stage to offer some additional comments on the general structure of the Lagrangian and of the resulting amplitudes.
As clear from our discussion in \textit{Section 2}, each space derivative acting on the scalar $\pi$ can be schematically written in Fourier space as a term $H \pi$ divided by a number which is much smaller than unity in the cases of interest. Now, it is clear from dimensional analysis that other terms in the action, specifically the ones with less derivatives, will have their own $ H/(\alpha_0+(\alpha_0^2 +8\beta_0)^{1/2})^{1/2}$ factors coming from derivatives but, most importantly, will also  need some generic $M$ factor to make the action properly dimensionless. This is exactly what happens in our theory (see Eq.~(\ref{action})) as well: the exponent of $\bar M_1$ is bigger than the one of $\bar M_2$ and so on. We have employed in the analysis of the amplitudes some freedom on these $\bar M_n$ parameters to show that even terms with higher spatial derivatives can give non negligible contributions to the amplitude. Of course, if one wants to have a reliable effective theory the $\bar M_n$ coefficients must eventually prevail over $H/(\alpha_0+(\alpha_0^2 +8\beta_0)^{1/2})^{1/2}$ (in a Lorentz invariant theory it would suffice to ask for $\bar M_n\gg H$, here we need more) so that it makes sense to consider higher derivatives up to some given finite order, but not further.
\newpage
\section*{*}
\newpage

\section{The shapes of non-Gaussianities: Bispectrum}
In this section we wish to analyze the shapes of the bispectra generated by the   various operators analyzed above.
In calculating the amplitudes of the bispectra in the previous section we have chosen to perform, albeit numerically, calculations with the exact wavefunctions. We decided to counterbalance the loss of information in not having $\alpha_0, \beta_0$ explicit in the result by running the same procedure for six different  benchmark values of the parameters.  In analaysing the shapes we do not enjoy the possibility to use the exact wavefunction any longer as we  must approximate the wavefunctions inside the conformal time $\tau$ integral(s) calculated as prescribed by the in-in formalism. Of course, exact results are always available when the classical solution reduces to the usual, Hankel function $H_{(3/2)}(\tau)$ .
In computing terms like

\be
\int^{\tau}_{-\infty} d \widetilde\tau \frac{1}{H^4 \widetilde\tau^{4}} \left[ \dot\pi_{k1}(t(\widetilde\tau)) \dot\pi_{k2}(t(\widetilde\tau)) \dot\pi_{k3}(t(\widetilde\tau))         \right]\, ,
\ee
we have  choosen  to expand in series each $k$-mode inside the integrand within a region that starts from slightly inside its effective horizon (setting to zero the function in the rest of the interval). This enables us to keep both, parameters and external momenta, arbitrary. This approximation  is justified by the fact that, due to the oscillatory behaviour of the wavefunctions\footnote{Of course, an oscillatory behaviour is not, by itself, enough to provide a cancellation all over the -inside the horizon- region. Indeed in some cases for $\tau \rightarrow - \infty$ the amplitude of the wavefunctions increases and one certainly does not expect this to give zero contribution. On the other hand, much like in the simplest single field slow-roll calculations, one is expected to slightly rotate $\tau$ to the imaginary plane to match the vacuum thus basically putting to zero the contribution at $- \infty$.}, the main contribution to the integral comes only  from the region where  all the wavefunctions are not oscillating anymore (to be safe, 
we actually choose to include the region where the $k$-mode with the latest effective horizon is still within its horizon). Furthermore, it is reasonable to assume that, even if some non-negligible contribution is being left over because of this approximation procedure, it might have a systematic effect on all $k$-modes and would therefore not change the  shapes of  the bispectrum of perturbations. This procedure has been calibrated with the calculations of Ref. \cite{chen-bis} which were performed exactly: we are able to reproduce  the very same shapes for the bispectrum.

Following the scheme used for the amplitudes, we now report below the shape of non-Gaussianity for each interaction term. Within a single interaction term, we consider four configurations in the $(\alpha_0, \beta_0)$-plane. We preliminary found that the shape are not particularly sensible to their absolute values. This is expected as these coefficients,  being the same for each $k$-mode, enter mainly  in the amplitudes, not in  the shapes. Therefore,  we limited our attention to the $\alpha_0/\beta_0$ ratios. We probed the following cases: $\alpha_0\not= 0, \beta_0 \simeq 0$  (DBI-like);  $\alpha_0\simeq 0, \beta_0\not= 0$ (ghost); $\alpha_0\not= 0 \not= \beta_0$ (with the solution interpolating between DBI-like models and ghost inflation and $\alpha_0$ playing the dominating role in determining where is the effective horizon, we call this configuration $A$) and  $\alpha_0\not= 0 \not= \beta_0$ (with the effect of $\beta_0$ leading in the expression for the horizon, we call this configuration $B$). 

Let us briefly add some general preliminary comments.  We need to point out that when we use the name DBI might generate  some confusion. Indeed,  we use it also to  describe shapes due to interaction terms coming from curvature perturbations, such as $\bar M_{1,2,3....}$. What we mean here  is that we can employ in the integrations the usual wavefunction $\pi_k \propto e^{-i \sqrt{\alpha_0}k \tau }(1+i \sqrt{\alpha_0}k \tau)$ without resorting to any approximation. This wavefunction can be used as far as it is the solution to the equation of motion. This is a condition concerning only second order perturbations:  we need only require $\beta_0=0\Leftrightarrow 0= M_0$ to employ it. We can, at the same time, have extrinsic curvature-driven interaction terms, the $\bar M$'s, and yet find exact results.
 The shapes corresponding to what we denominate DBI configuration have been thoroughly investigated in a number of papers \cite{bispectrum}, but what has been generally left out is the contribution coming from the extrinsic curvature terms: it is indeed possible to have the usual Hankel, $H_{3/2}(\tau)$, wavefunction as a solution to the classical equations and, at the same time, switch on curvature operators like $\bar M_{4,5,6,7,8,9}$. All the shapes obtained in the DBI configuration are generated through exact analytical methods. We will call them as ``exact-DBI configurations''. Should we find, as we will, a shape which is not equilateral in the first configuration, that shape suffers none of any possible limitations the approximated method might introduce. The ghost configuration has been analyzed in depth in many articles, among which \cite{tilted,ssz05}, and again, not all the terms coming from extrinsic curvature have been taken into account. It is important to note though that in  \cite{eft08,tilted,ssz05} the $M$ coefficients multiplying curvature terms have all been chosen of the same order thus resulting in only a couple of leading curvature terms (they correspond to the $\bar M_1, \bar M_0$  contributions).
 Finally, the shapes in configuration \textit{$A$, $B$} have never been analyzed before.

The shapes reported below have been  obtained by employing the  shape function  $B(1,x_2,x_3)$,  where $x_i=k_i/k_1$,  which posseses  the same $k_{1,2,3}$-dependence as the three point function. What is plotted exactly is $x_2^2 x_3^2 B(1,x_2,x_3)/B(1,1,1)$ in the region satisfying $x_2 \geq x_3 \geq 1-x_2$. 
Let us remind the reader that a shape is called local when it peaks for a small value of, say,  $x_2$, with $x_1=x_3 \sim 1$; 
equilateral when it  peaks in the equilateral configuration $x_1=1, x_2=1=x_3$ and is called flat for squashed triangles with $x_1=1=x_2+x_3$. In particular, we find, as detailed below,  that some novel curvature-generated terms produce a flat bispectrum which specifically peaks for $x_2=x_3=1/2=1/2\, x_1$.   
In presenting the shapes we follow the same order and organization we employed for the amplitudes:

\subsection{Shapes from $P(X,\phi)$-type interactions and first two curvature-generated terms.} 
\vskip 0.5cm

$\bullet$  ${\cal O}_1=-2 M_2^4 \dot\pi (\partial_i \pi)^2/a^2$

\begin{figure}[h]
\includegraphics[scale=0.50]{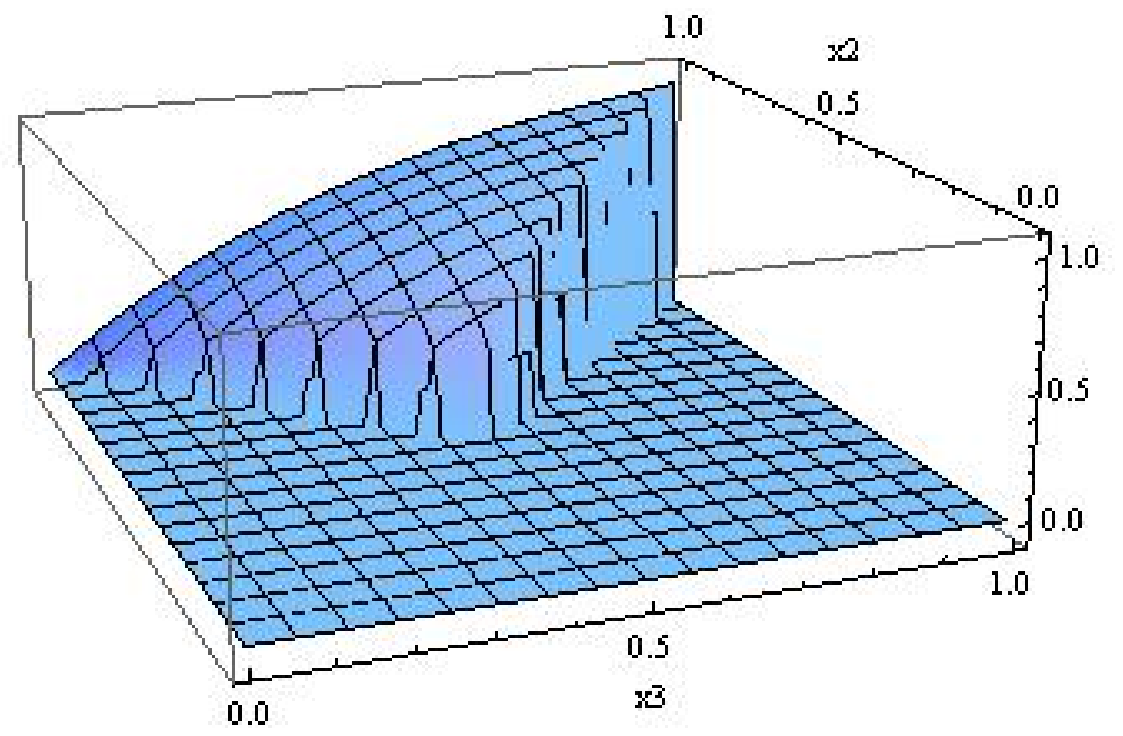}
\hspace{10mm}
\includegraphics[scale=0.43]{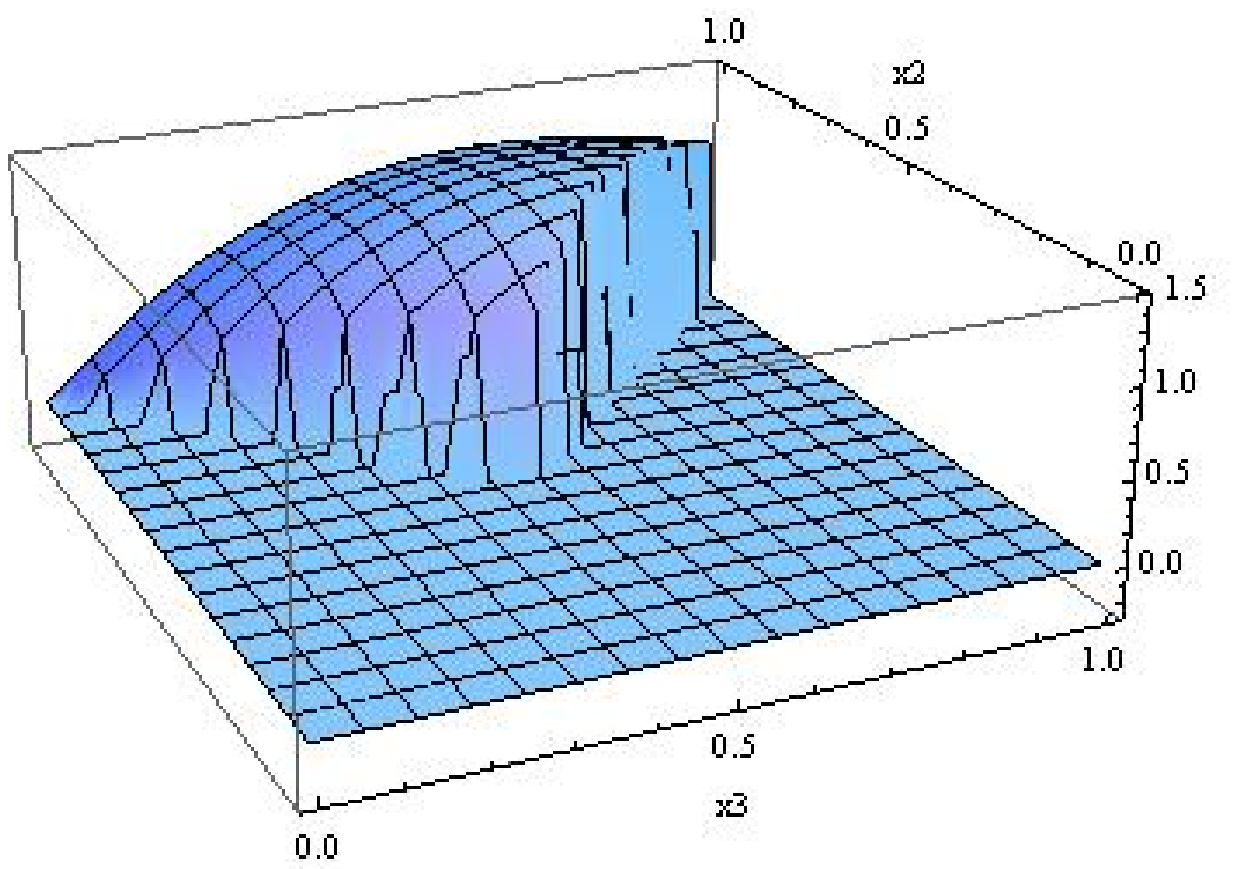}
	\label{fig:M2-1} 
	\caption{DBI configuration on the left, obtained using exact methods; the approximated ghost shape on the right.}
\end{figure}

\begin{figure}[h]
	\includegraphics[scale=0.50]{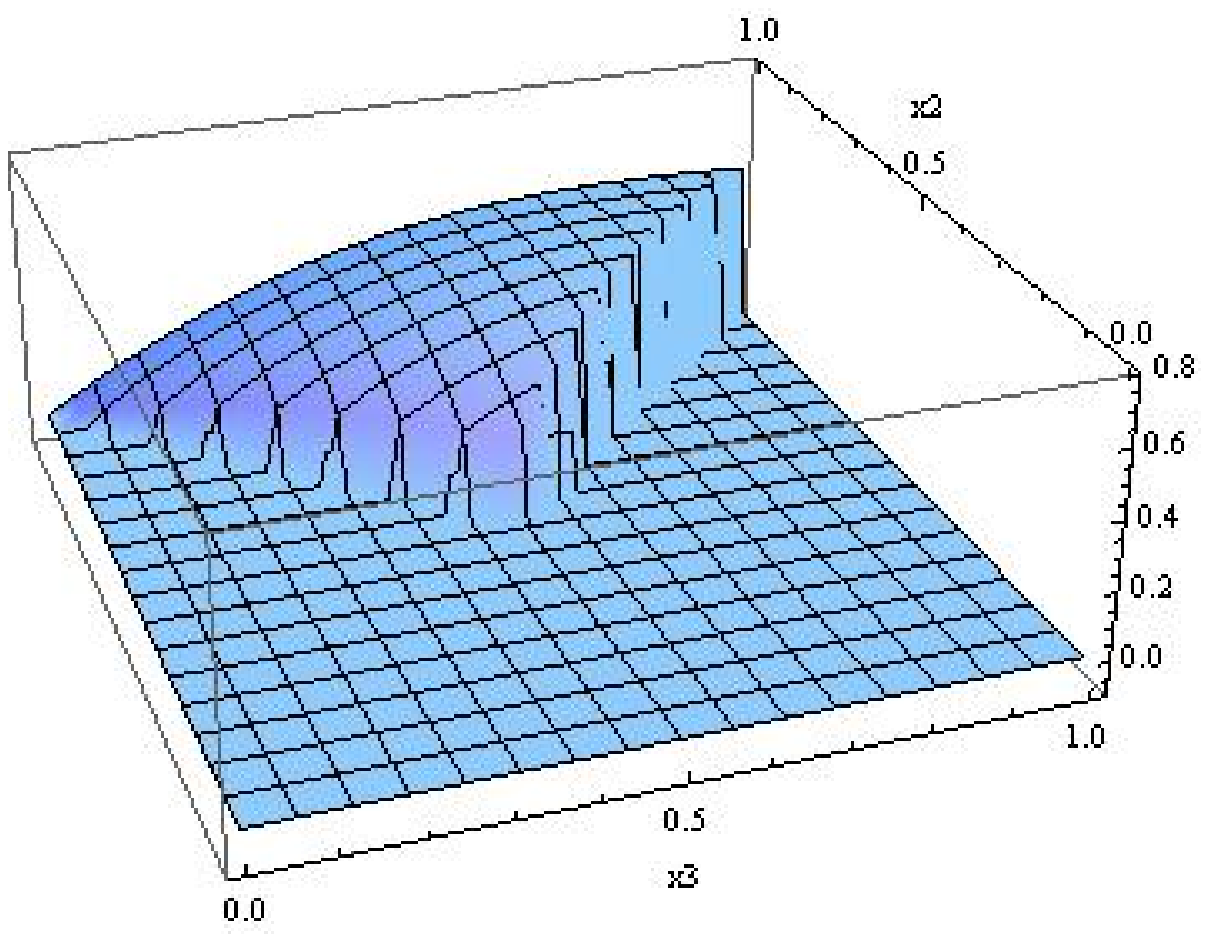}
\hspace{15mm}
	\includegraphics[scale=0.55]{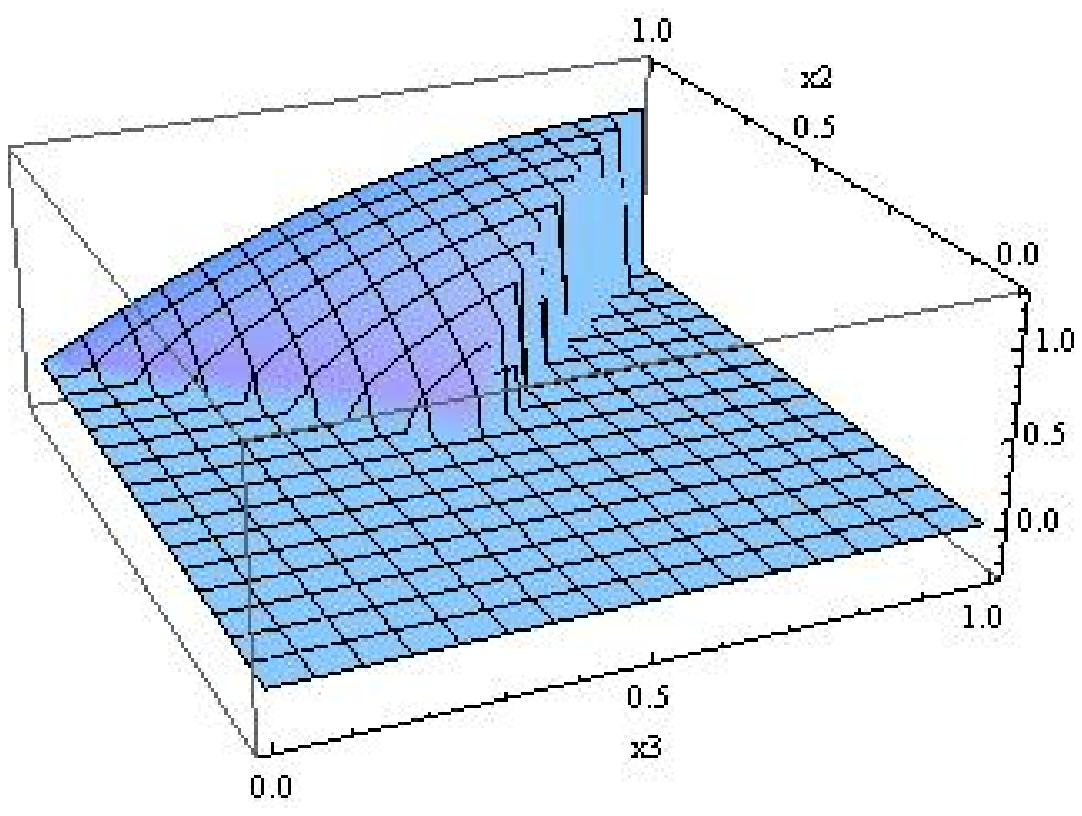}
	\label{fig:M2-2} 
	\caption{$A$ configuration on the left, $B$ configuration on the right.}
\end{figure}
Note that, although more or less sharply, all the four plots are peaked in the equilateral configuration.\\
\newpage 
$\bullet$  ${\cal O}_2=4/3 \,\,  M_3^4 {\dot\pi}^3 $\\

\begin{figure}[h]
	\includegraphics[scale=0.50]{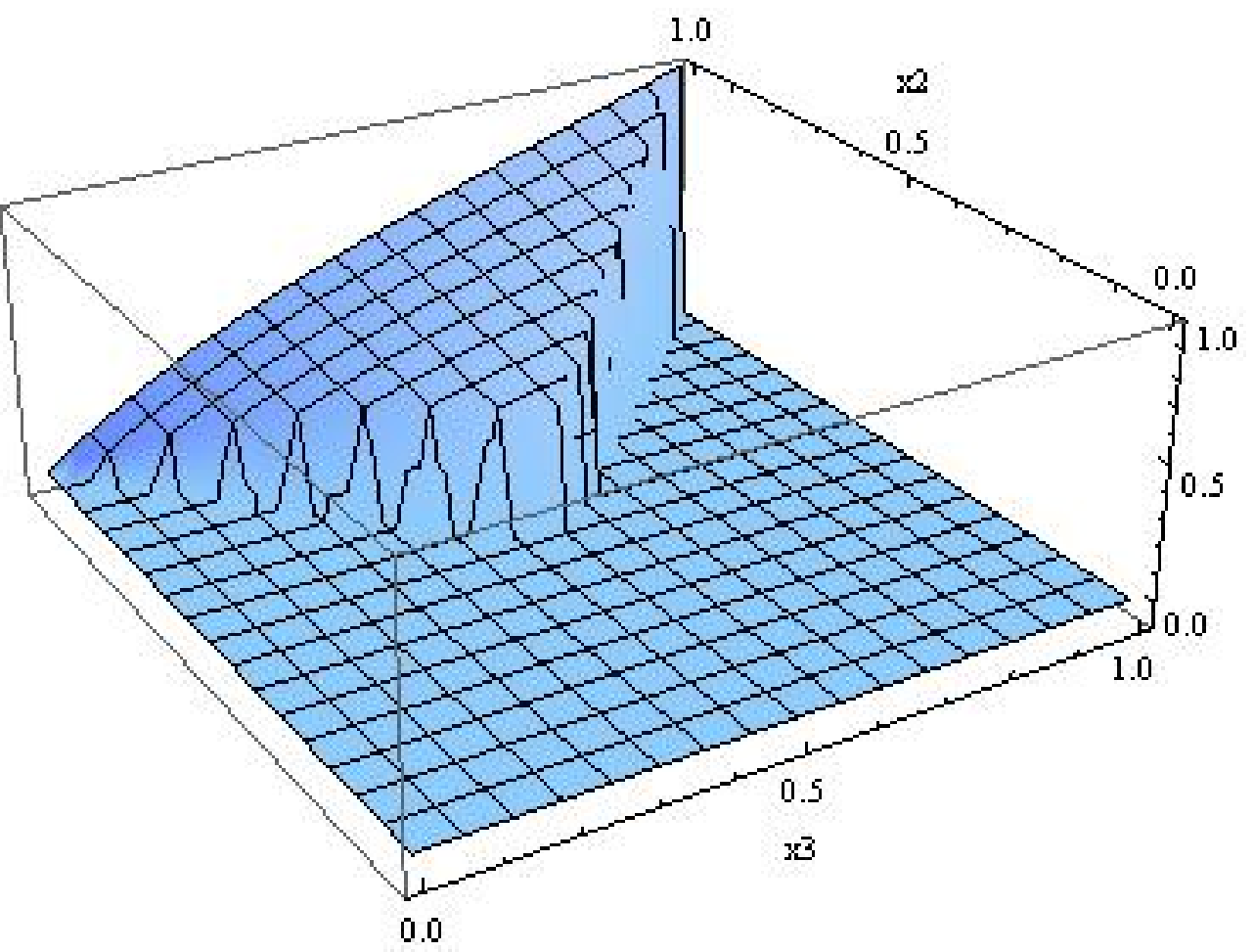}
\hspace{10mm}
	\includegraphics[scale=0.60]{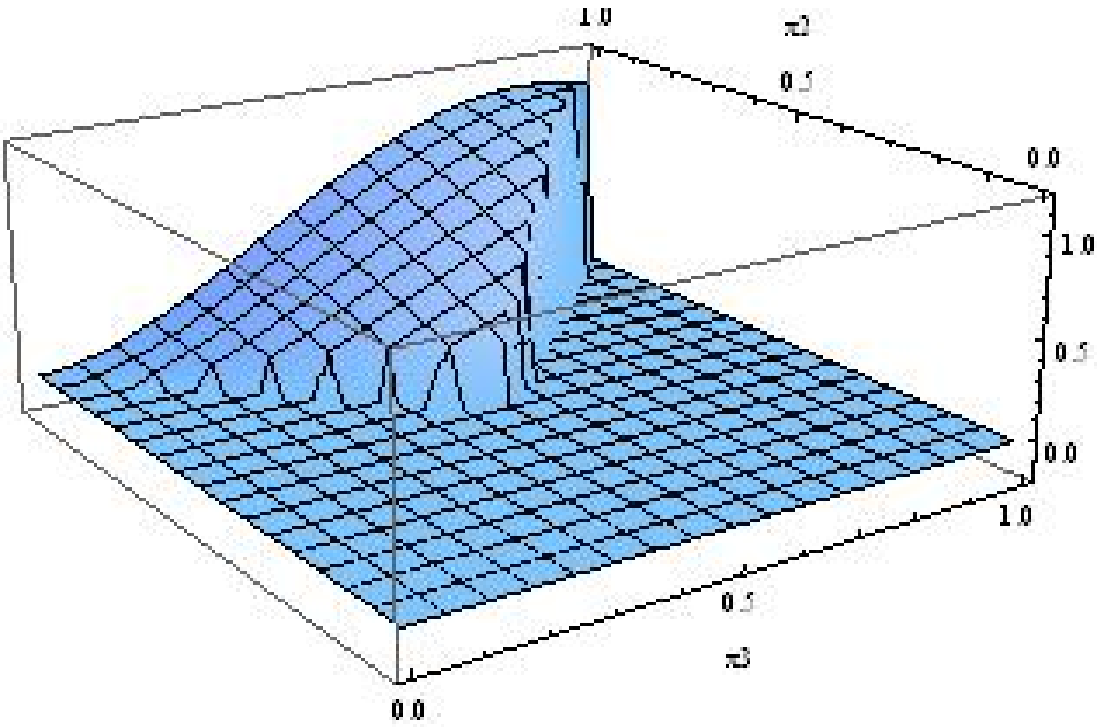}
	\label{fig:M3-1} 
	\caption{exact DBI configuration on the left; approximated ghost shape on the right.}
\end{figure}
\begin{figure}[h]
	\includegraphics[scale=0.65]{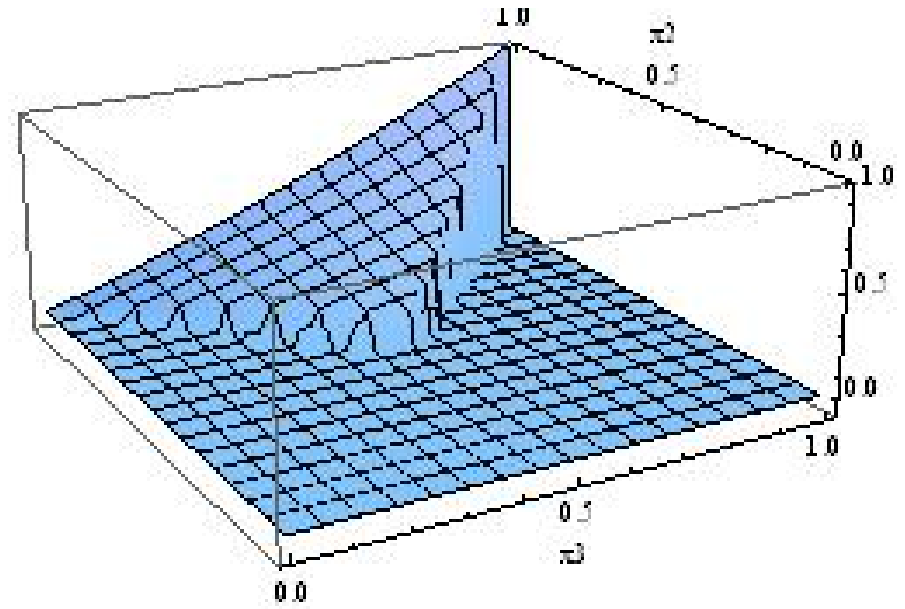}
\hspace{15mm}
	\includegraphics[scale=0.55]{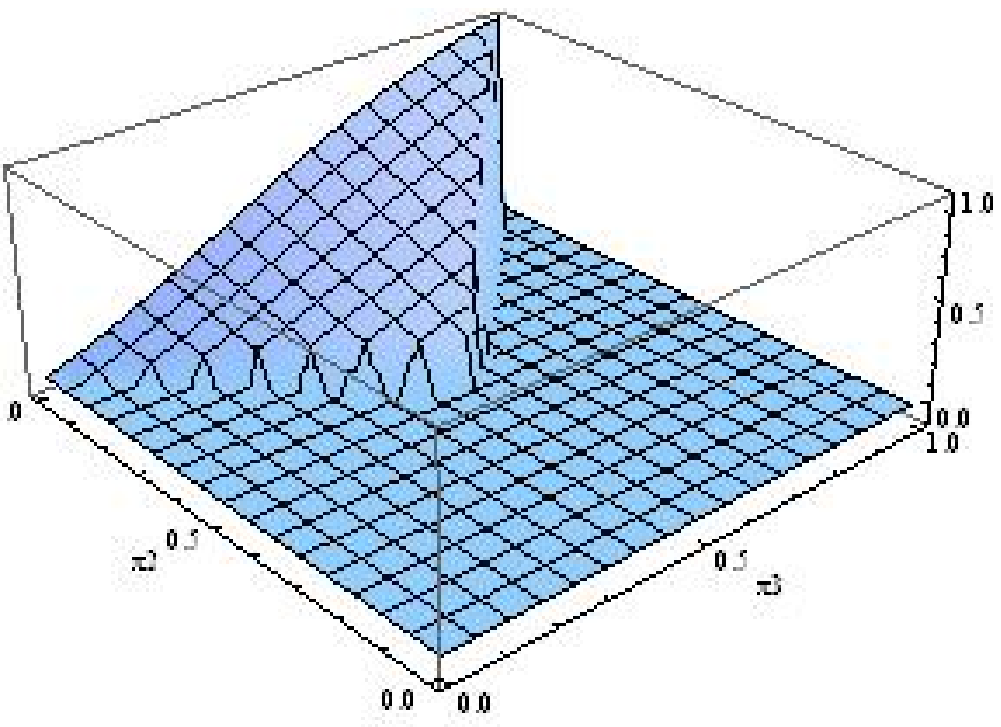}
	\label{fig:M3-2} 
	\caption{$A$ on the left, $B$ configuration on the right for the $M_3$-driven interaction term.}
\end{figure}
We obtain equilateral shapes in all four cases. It is somewhat expected that the general, interpolating 
solution employed in configurations \textit{$A$} and \textit{$B$}, will give qualitatively the same plot, we have verified it in these first two rounds of shapes.\\
\newpage 
$\bullet$  ${\cal O}_3= -1/2\,\,  \bar M_1^3 (\partial_i \pi)^2 \partial_j^2 \pi/a^4  $\\

\begin{figure}[h]
	\includegraphics[scale=0.55]{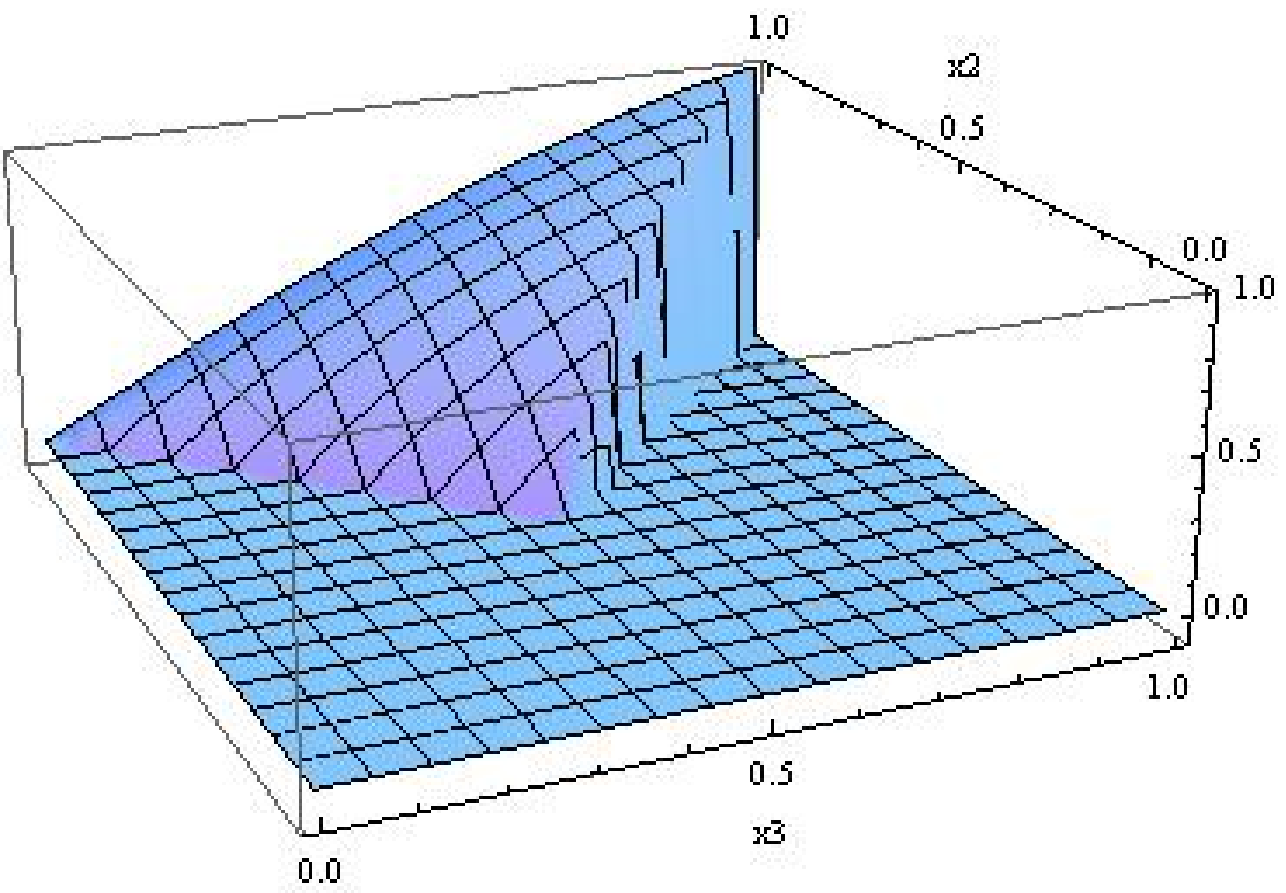}
\hspace{10mm}
	\includegraphics[scale=0.55]{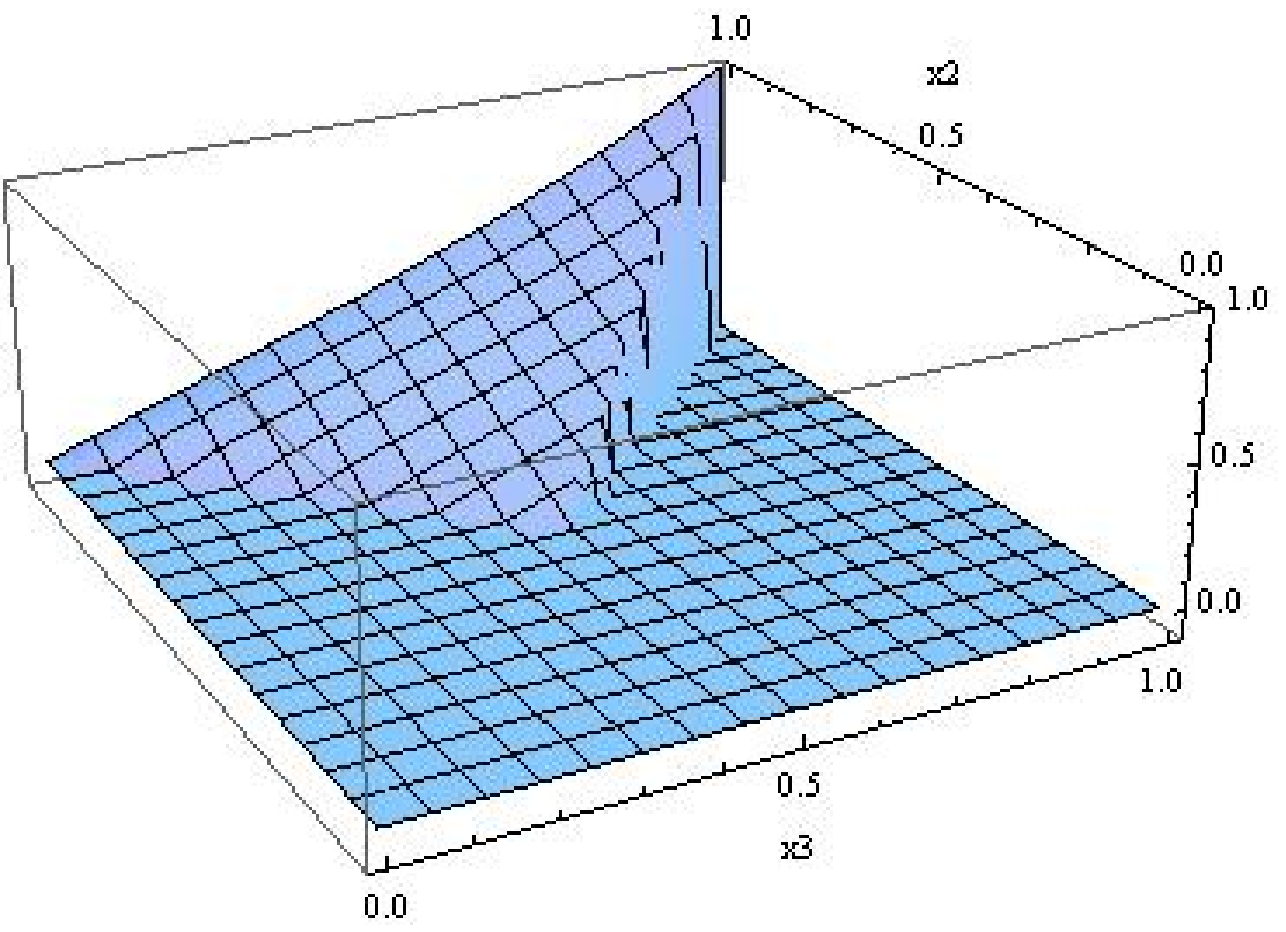}
	\label{Mb1-1} 
	\caption{exact DBI configuration on the left; approximated ghost shape on the right.}
\end{figure}

\begin{figure}[h]
	\includegraphics[scale=0.55]{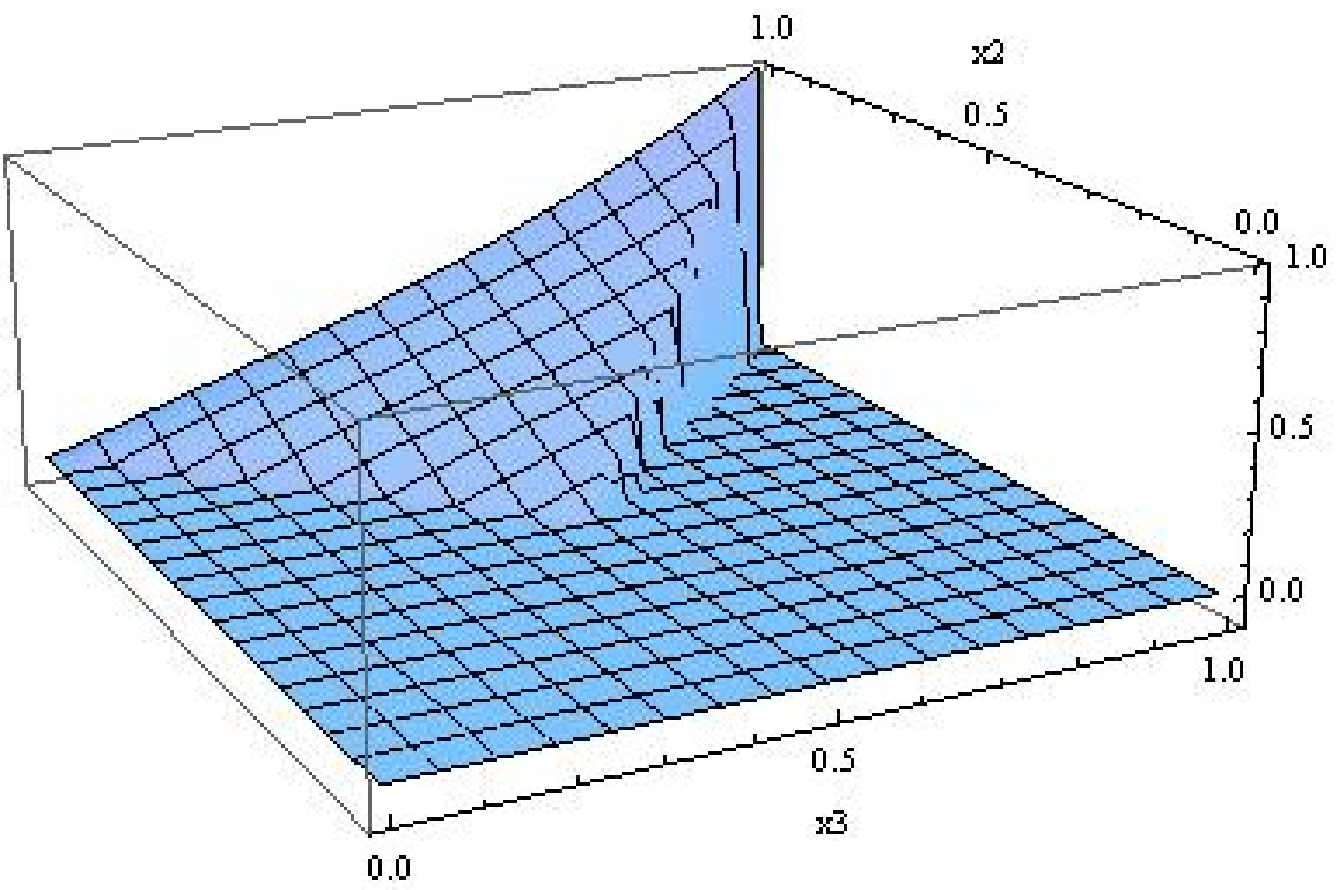}
\hspace{15mm}
	\includegraphics[scale=0.52]{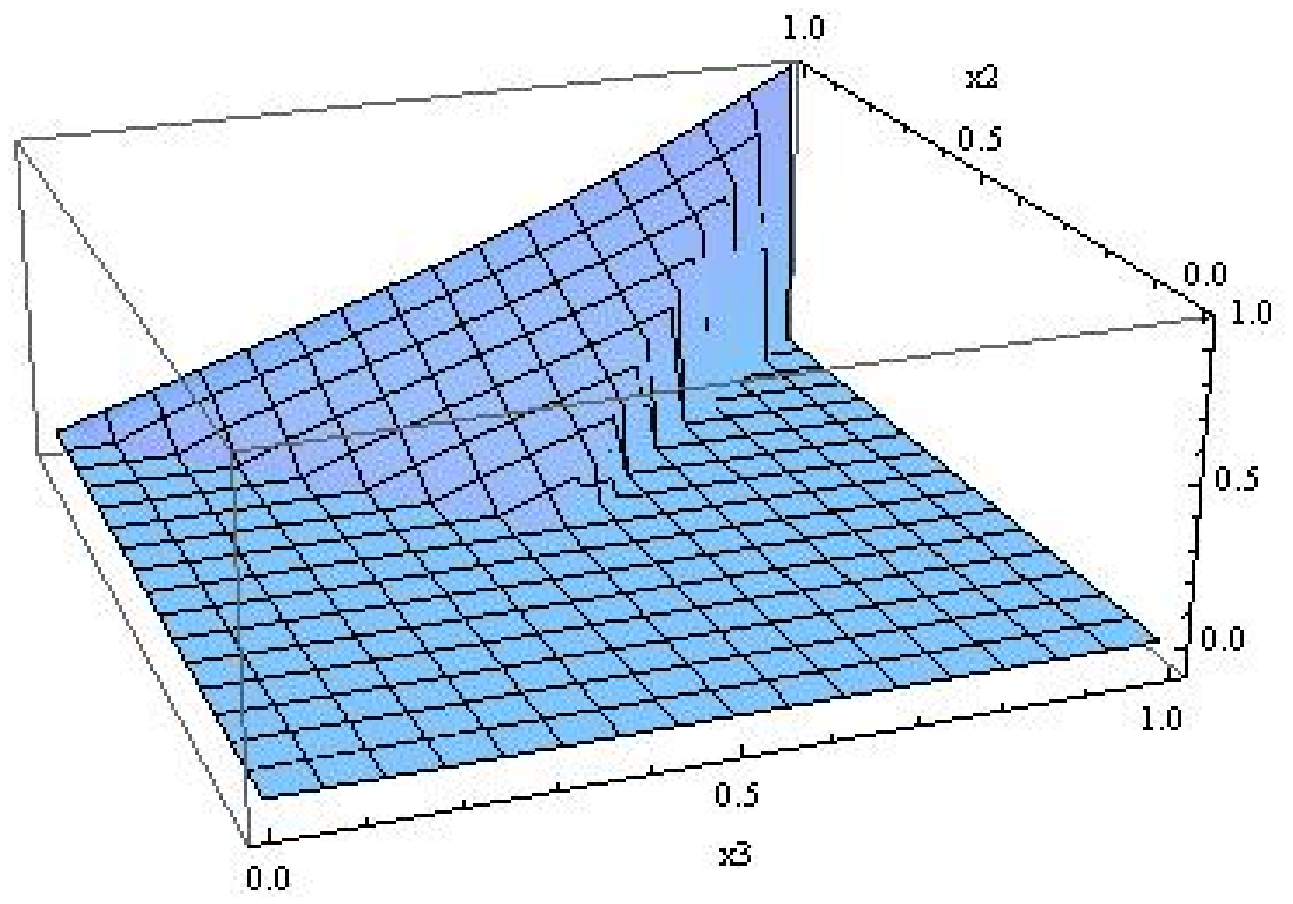}
	\label{Mb1-2} 
	\caption{$A$ on the left, $B$ configuration on the right for the $\bar M_1$-driven interaction term.}
\end{figure}
Also the bispectrum generated by this interaction term has an equilateral shape; the last two plots show, employing the general wavefunction, that also the interpolating models produce an equilateral shape.\\

$\bullet$  ${\cal O}_4= -1/2 \,\, \bar M_0^2/4\,\,  \left(5H (\partial_i^2\pi) (\partial_j \pi)^2  +4\dot\pi \partial_{i}^2 \partial_j \pi \partial_j \pi \right)/ a^4 $\\

\begin{figure}[h]
	\includegraphics[scale=0.55]{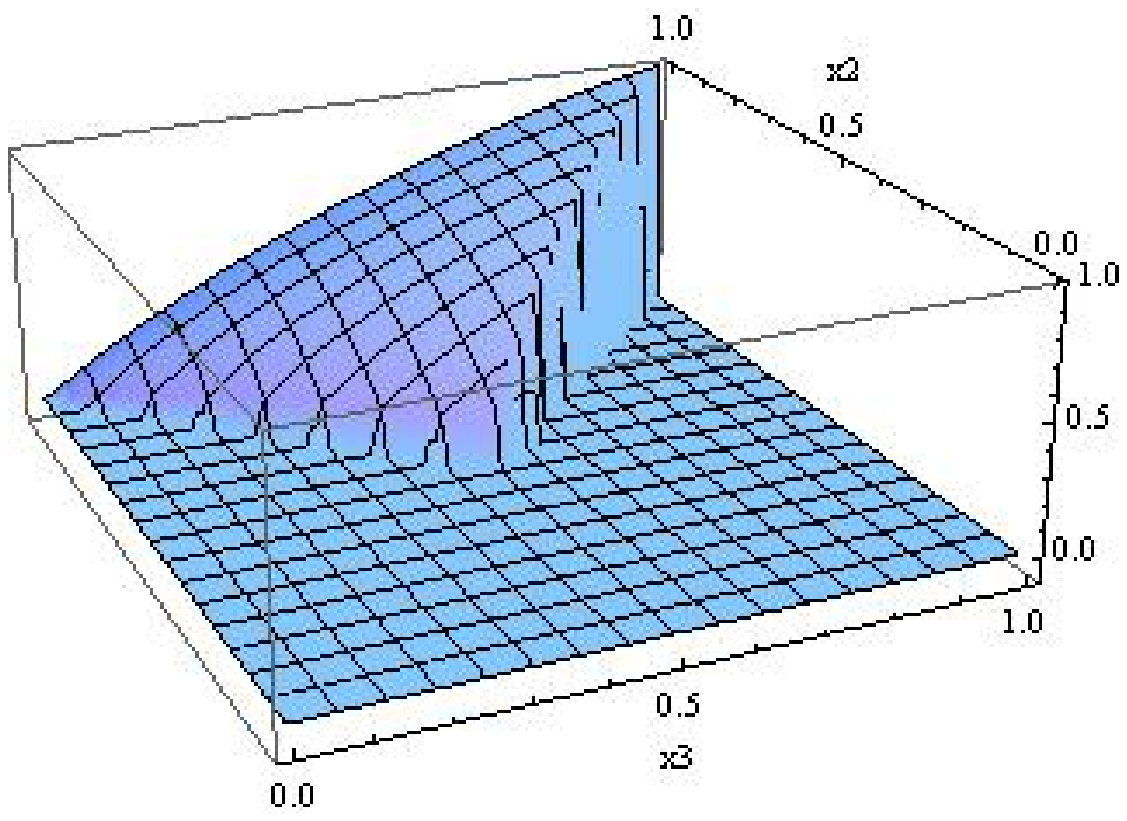}
\hspace{10mm}
	\includegraphics[scale=0.55]{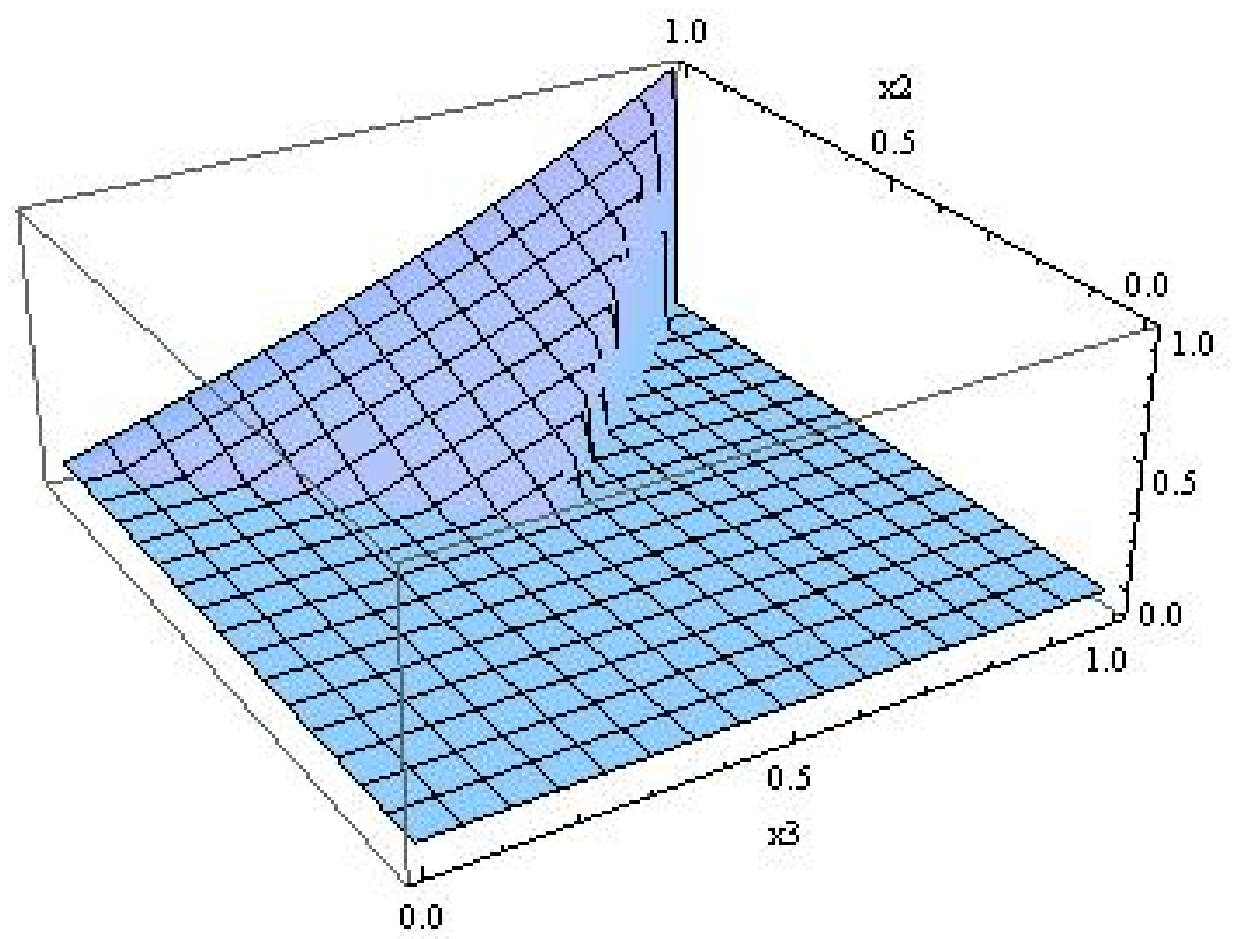}
	\label{fig:Mb2-1} 
	\caption{exact DBI configuration on the left; approx. ghost shape on the right.}
\end{figure}

\begin{figure}[h]
	\includegraphics[scale=0.55]{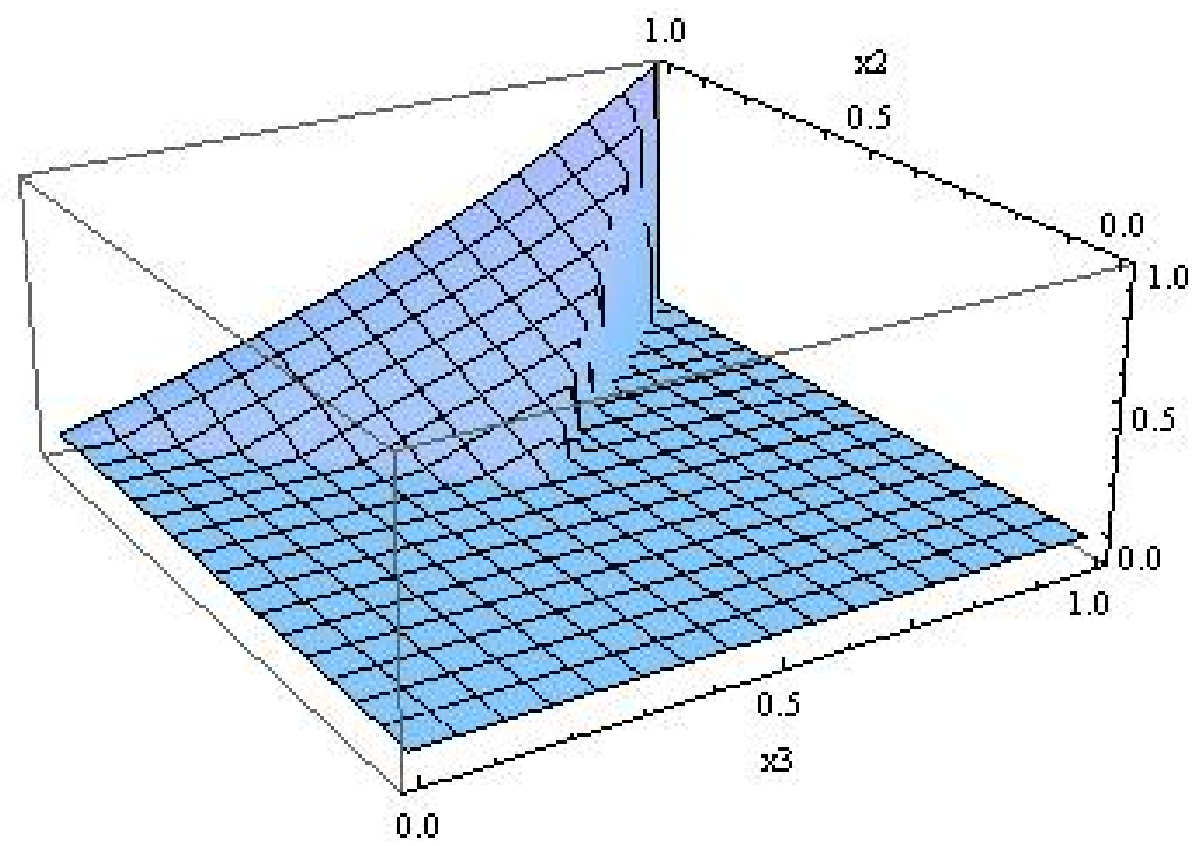}
\hspace{15mm}
	\includegraphics[scale=0.55]{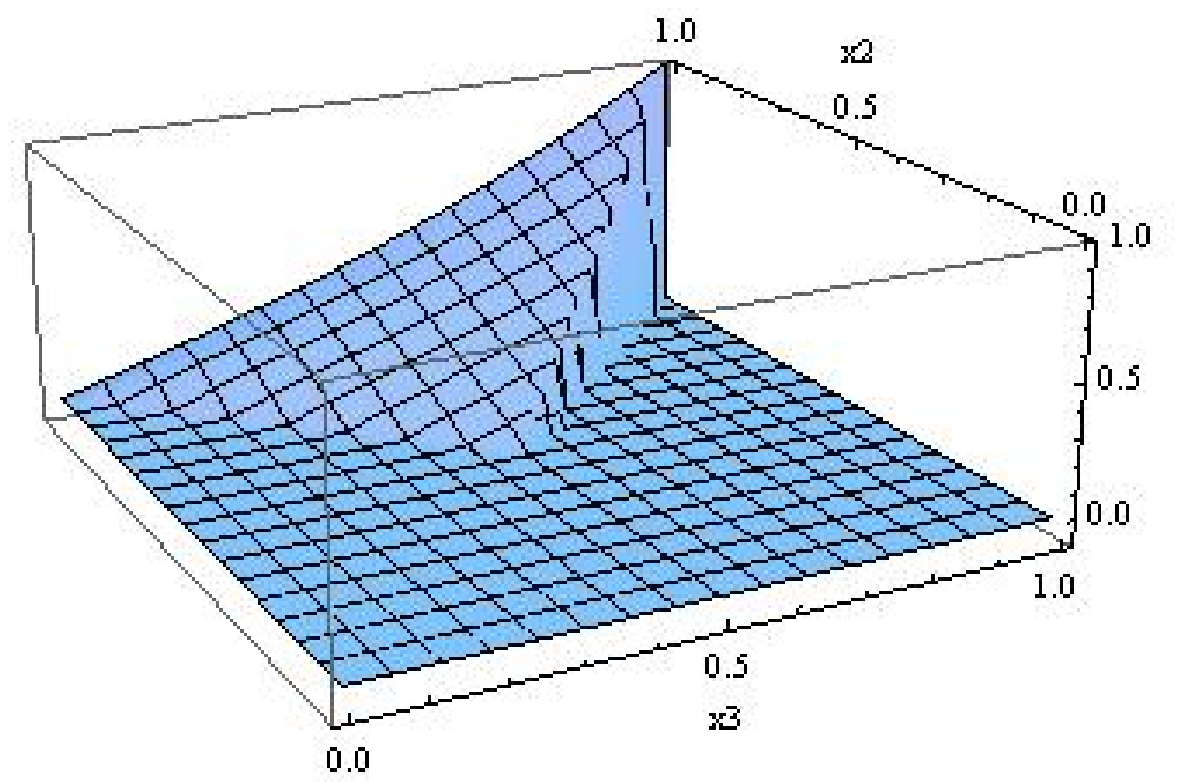}
	\label{Mb2-2} 
	\caption{$A$ on the left, $B$ configuration on the right for the $\bar M_0$ interaction term.}
\end{figure}
The various bispectra peak in the equilateral configuration. 
\newpage
\subsection{ Shapes from curvature-generated novel interaction terms.}
\vskip 0.5cm

$\bullet$  ${\cal O}_5= -2/3\,\,  \bar M_4^3 {\dot\pi}^2 \partial_i^2\pi \,\,/ a^2 $\\

With the term tuned by $\bar M_4$ we start including in our description the contributions that have so far been neglected in the literature.

\begin{figure}[h]
	\includegraphics[scale=0.45]{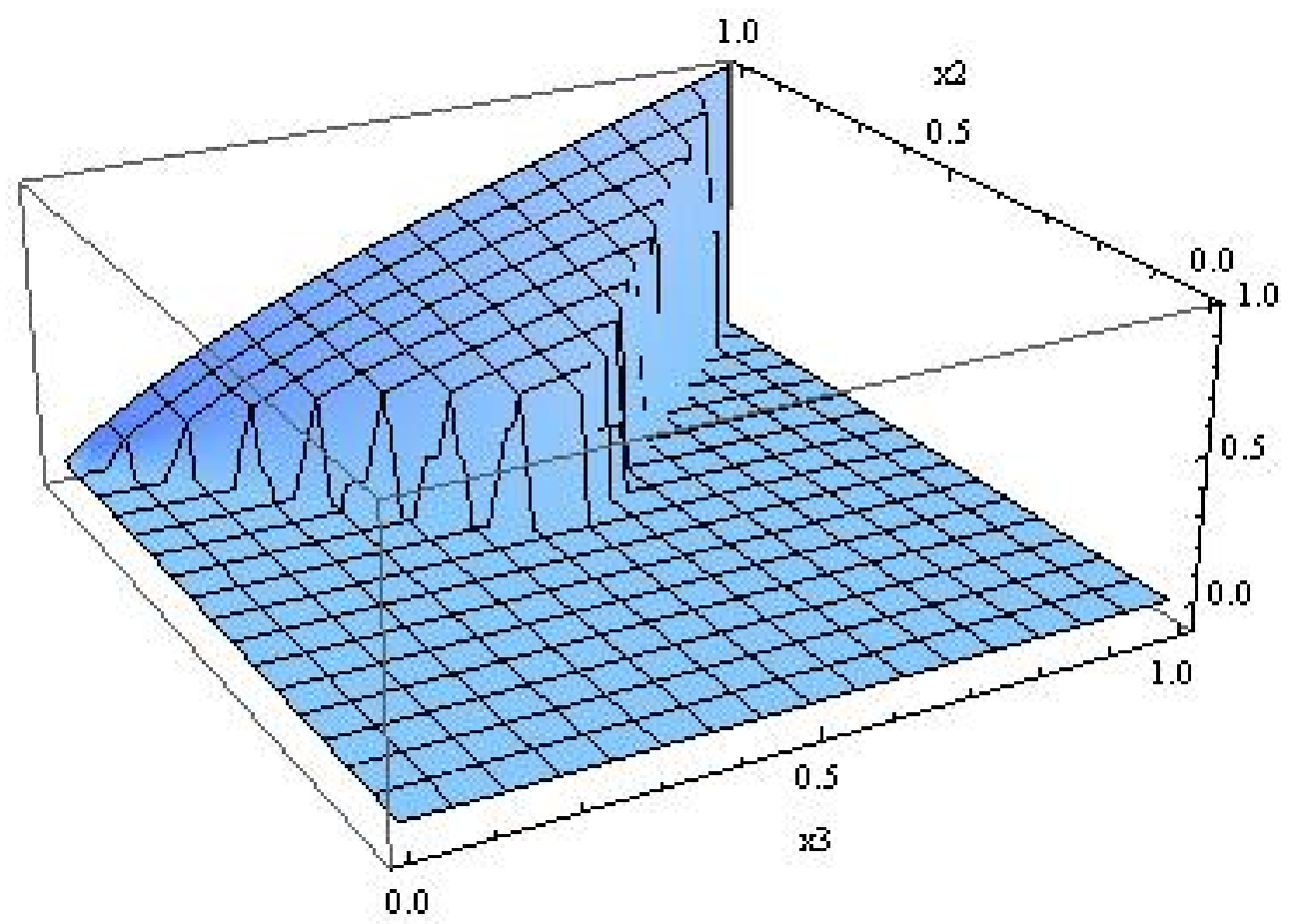}
\hspace{10mm}
	\includegraphics[scale=0.50]{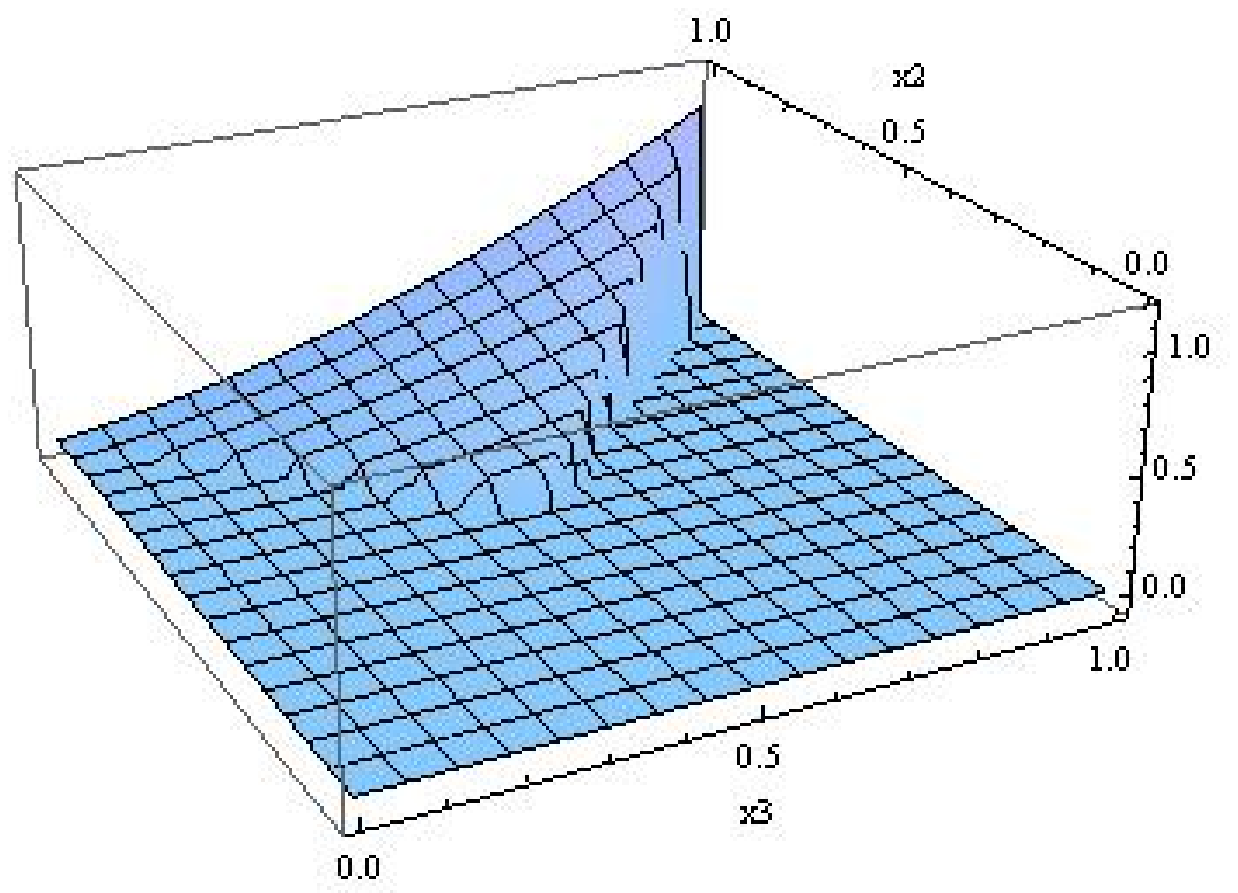}
	\label{fig:Mb4-1} 
	\caption{exact DBI configuration on the left; approximated ghost shape on the right.}
\end{figure}

\begin{figure}[h]
	\includegraphics[scale=0.53]{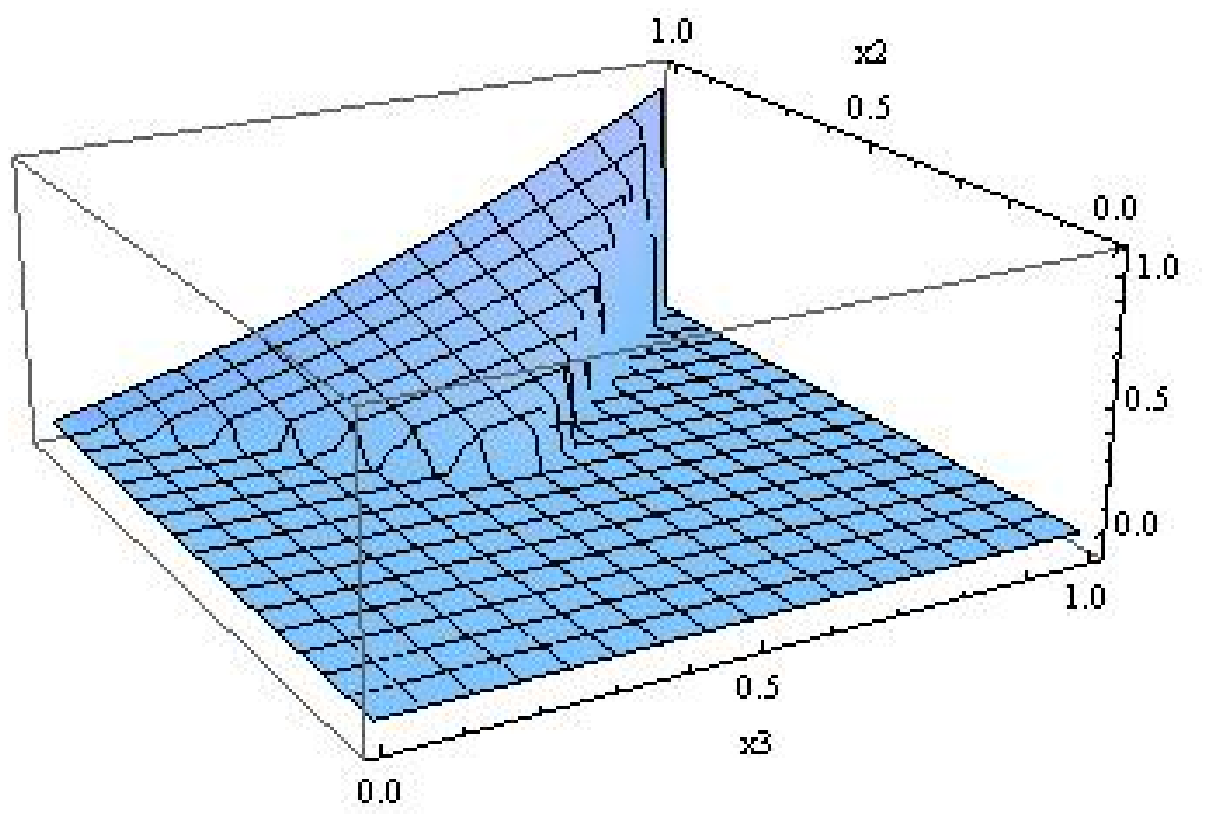}
\hspace{15mm}
	\includegraphics[scale=0.53]{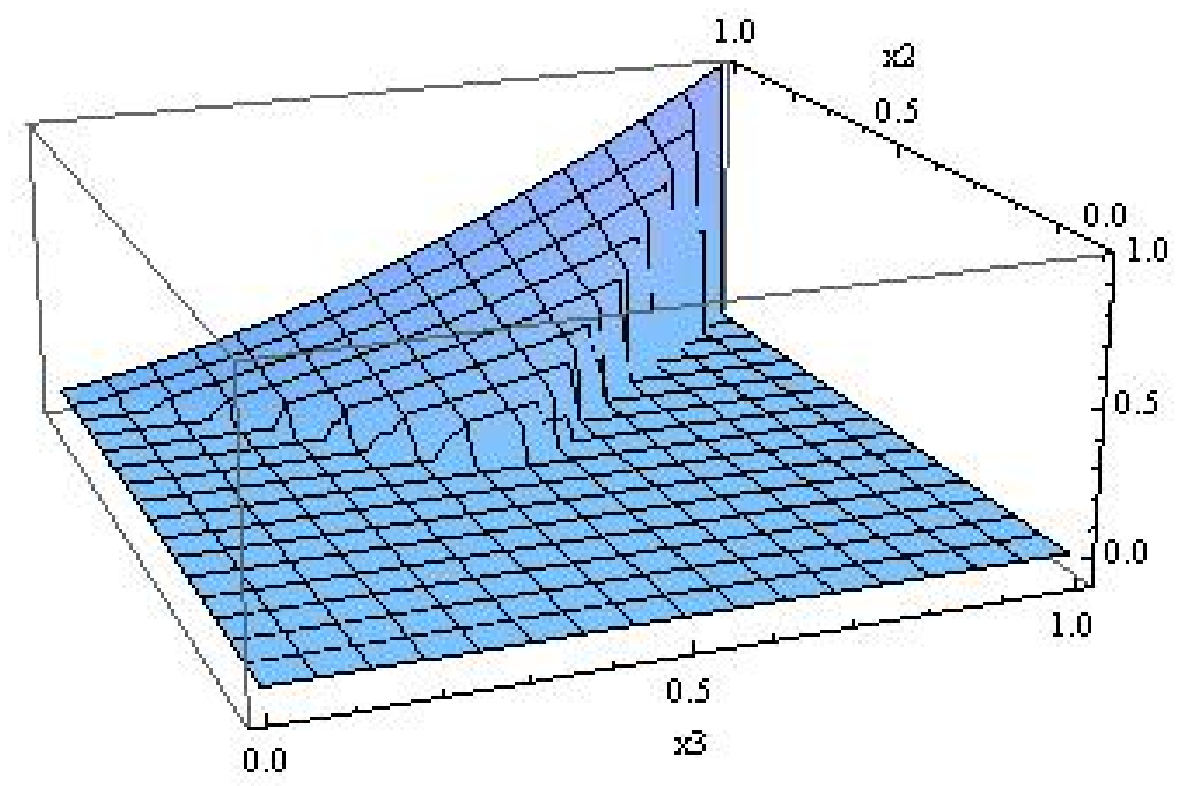}
	\label{fig:Mb4-2} 
	\caption{$A$ on the left, $B$ configuration on the right for the $\bar M_4$-driven interaction term.}
\end{figure}
Equilateral shapes for all configurations are obtained.\\

$\bullet$  ${\cal O}_6= 1/3\,\,  \bar M_5^2 \dot\pi (\partial_i^2\pi)^2 \,\,/ a^4 $\label{mb5}

\begin{figure}[h]
	\includegraphics[scale=0.45]{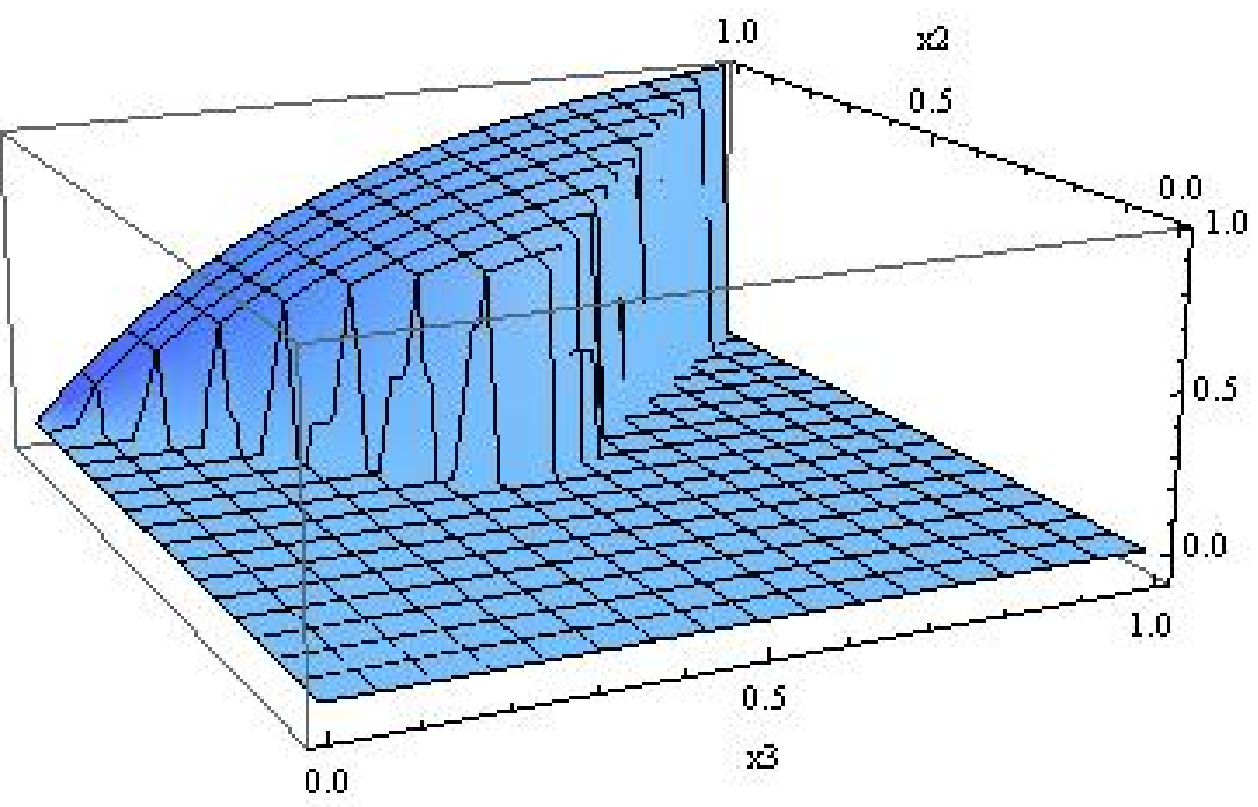}
\hspace{10mm}
	\includegraphics[scale=0.50]{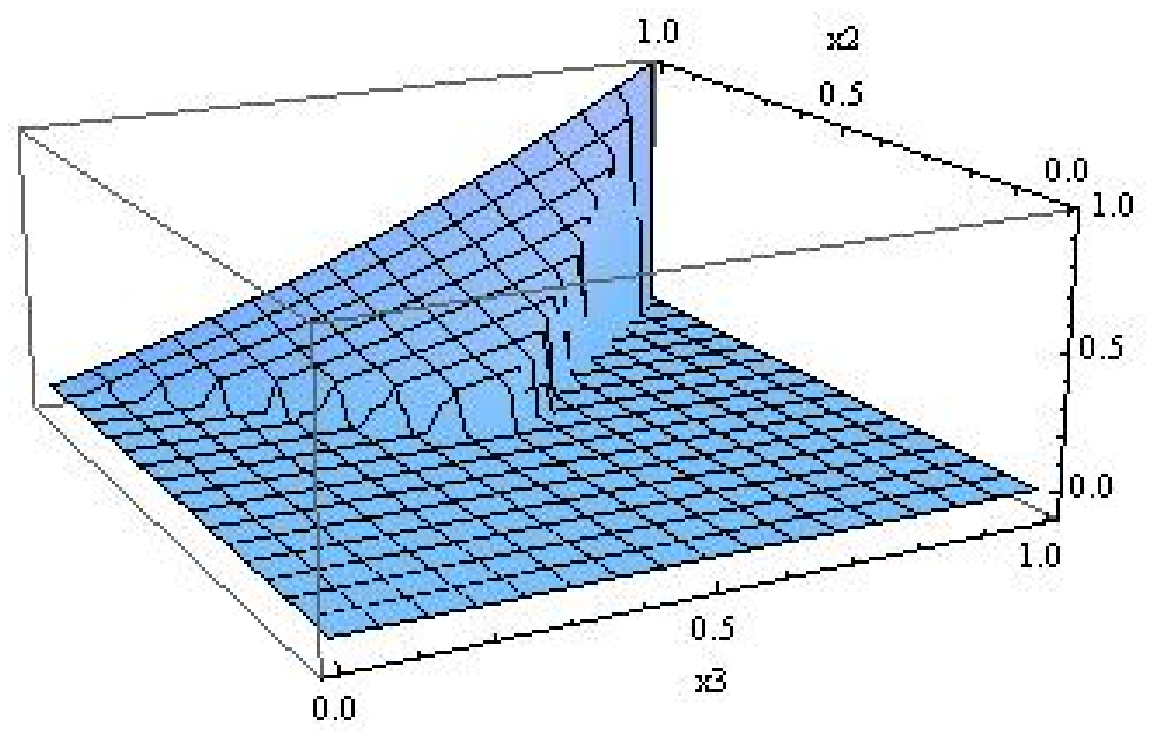}
	\label{fig:Mb5-1} 
	\caption{exact DBI configuration on the left; approximated ghost shape on the right.}
\end{figure}

\begin{figure}[h]
	\includegraphics[scale=0.60]{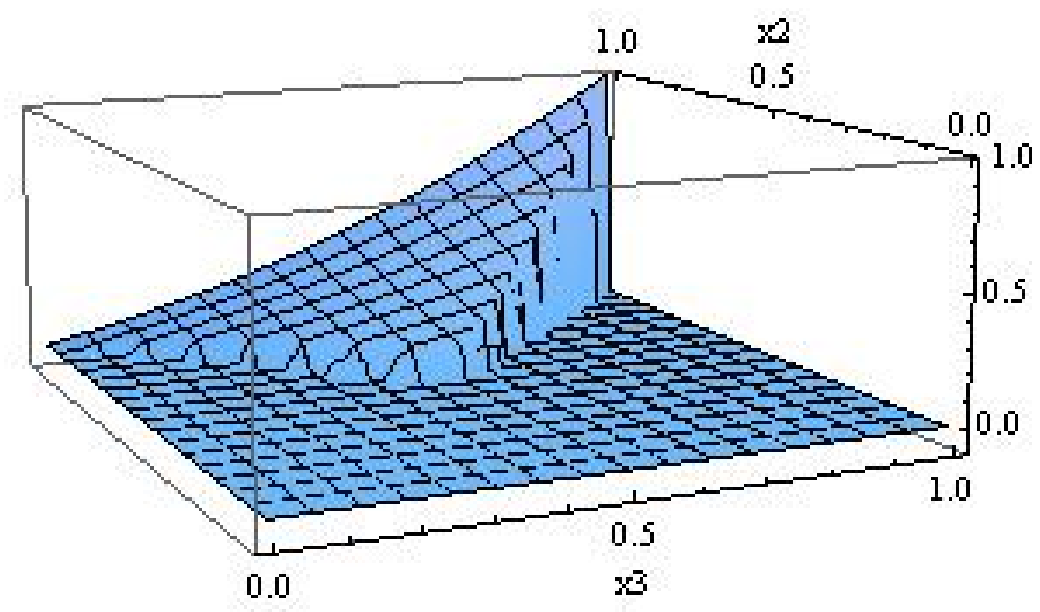}
\hspace{15mm}
	\includegraphics[scale=0.47]{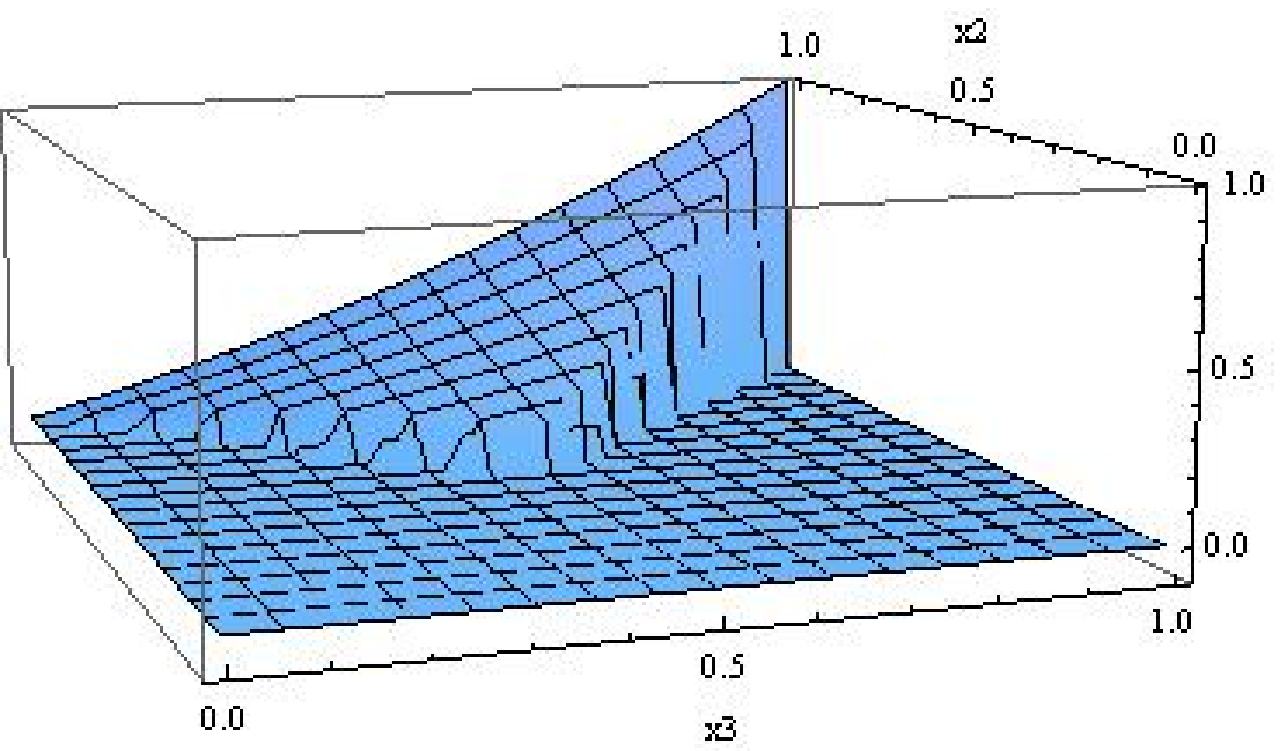}
	\label{fig:Mb5-2} 
	\caption{$A$ on the left, $B$ configuration on the right for the $\bar M_5$-driven interaction term.}
\end{figure}

Again, all the four bispectra peak in the equilateral configuration.\\
\newpage

$\bullet$  ${\cal O}_7= $ $1/3\,\,  \bar M_6^2 \dot\pi (\partial_{ij}\pi)^2 \,\,/ a^4 \label{mb6}$\\

\begin{figure}[h]
	\includegraphics[scale=0.53]{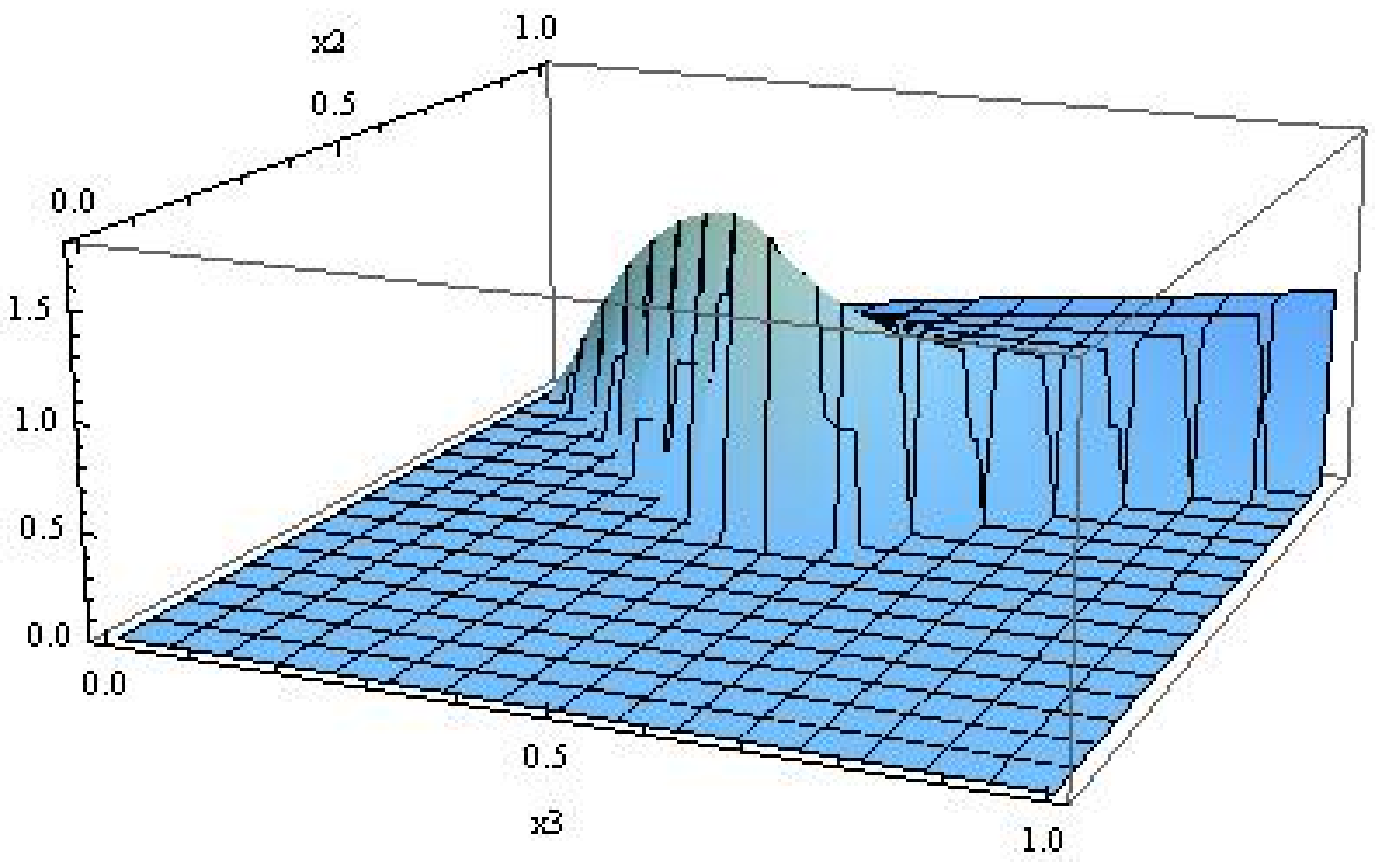}
\hspace{10mm}
	\includegraphics[scale=0.53]{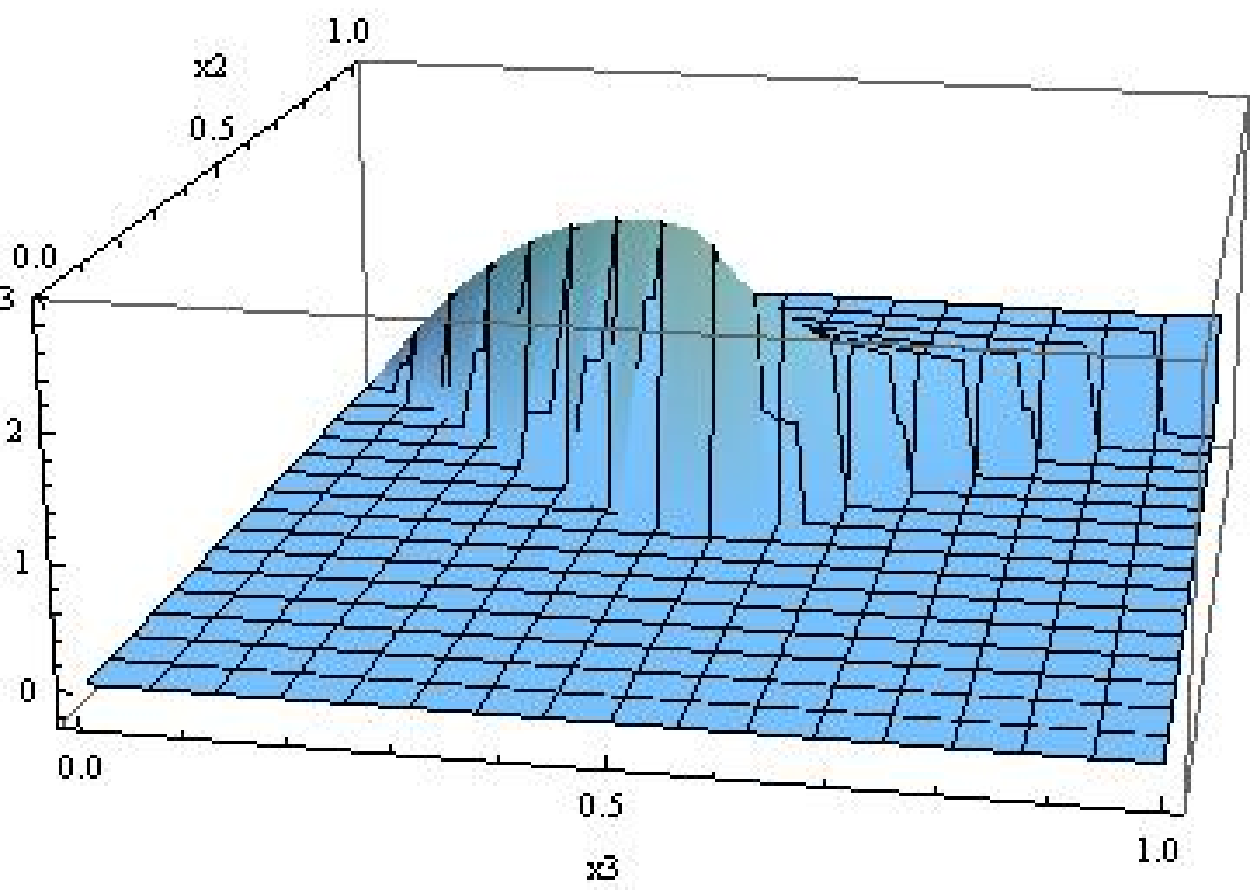}
	\label{fig:M_6_DBI} 
	\caption{exact DBI configuration on the left; approximated ghost shape on the right.}
\end{figure}

\begin{figure}[h]
	\includegraphics[scale=0.58]{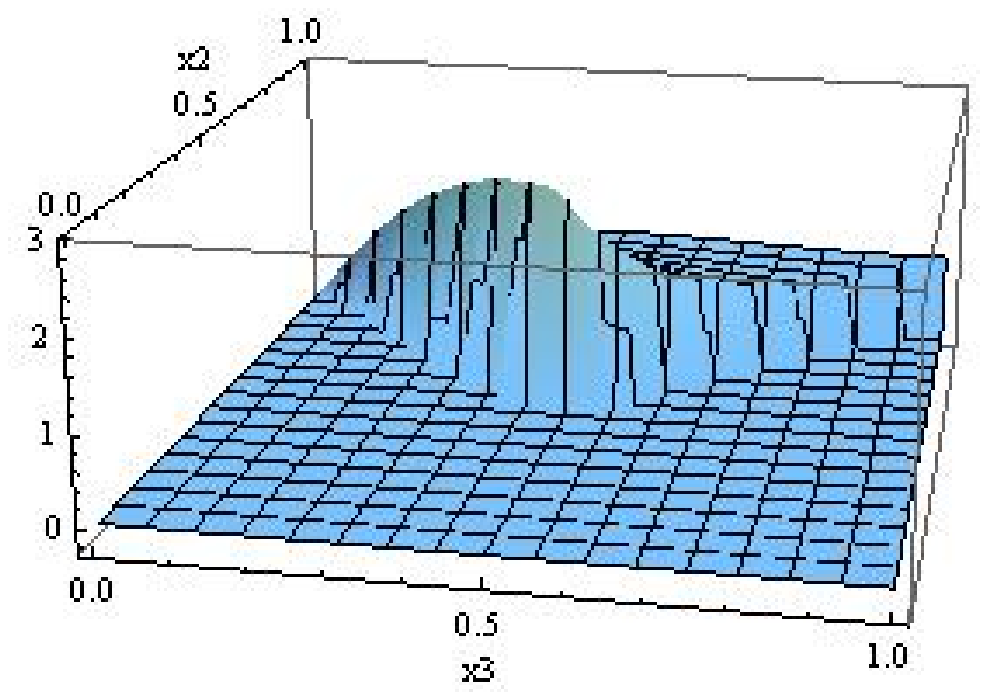}
\hspace{15mm}
	\includegraphics[scale=0.52]{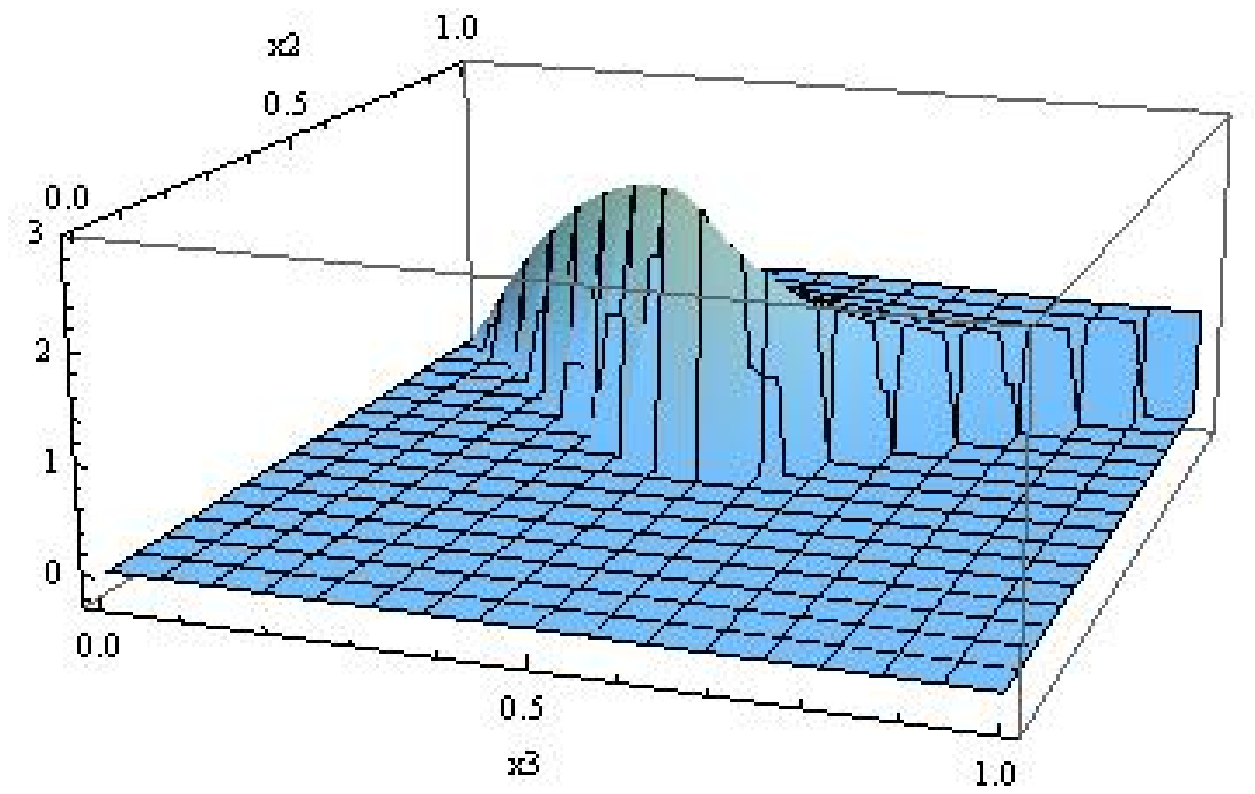}
	\label{fig:M_6b_DBI} 
	\caption{$A$ on the left, $B$ configuration on the right for the $\bar M_6$-driven interaction term.}
\end{figure}
This is one of the interesting novel curvature-generated terms that give rise to a flat shape (more precisely, the plot peaks at $k_1=1, k_2 \sim 1/2 \sim k_3$). 
Note that for a very similar interaction, namely the one generated by ${\cal O}_6$, we saw an equilateral plot. Here, derivatives combine to provide a different $k$-dependent factor outside the integral. Writing in Fourier space the interaction term ${\cal O}_6$  we obtain something proportional to ${k_2}^2 {k_3}^2$, while here we obtain a contribution proportional to $(\vec k_2 \cdot \vec k_3)^2$.\\
\newpage
$\bullet$  ${\cal O}_8= -1/6\,\,  \bar M_7 (\partial_i^2 \pi)^3 \,\,/ a^6 $\label{mb7}\\

\begin{figure}[h]
	\includegraphics[scale=0.50]{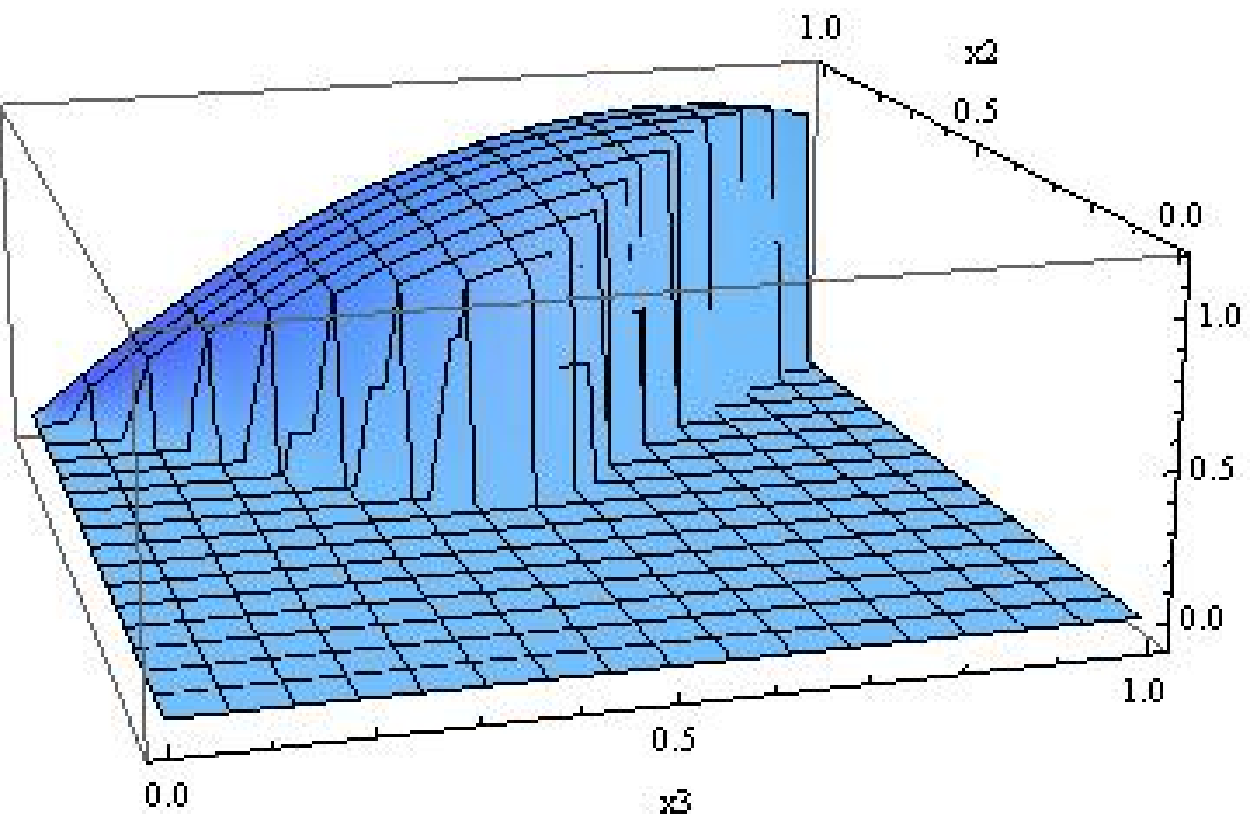}
\hspace{10mm}
	\includegraphics[scale=0.50]{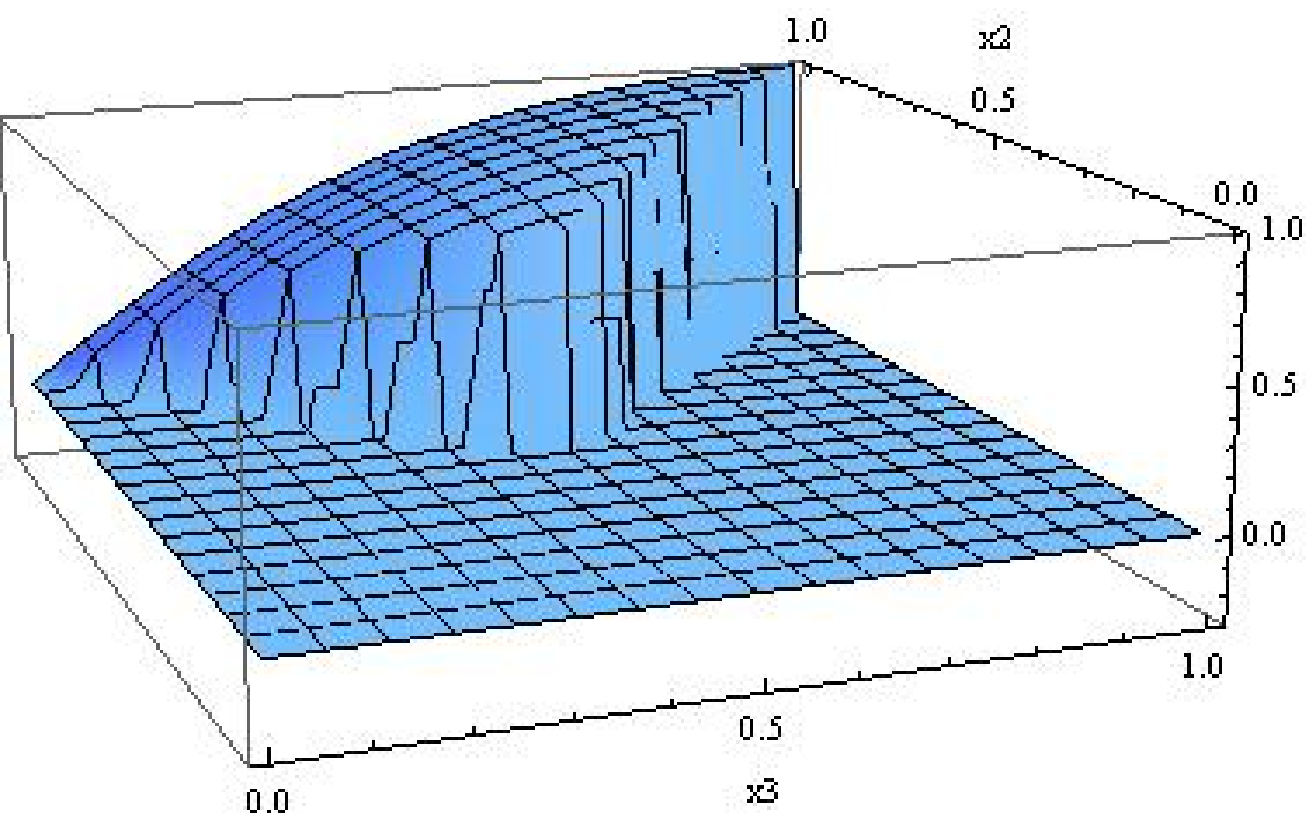}
	\label{fig:M_7_DBI} 
	\caption{exact DBI configuration on the left; approximated ghost shape on the right.}
\end{figure}

\begin{figure}[h]
	\includegraphics[scale=0.50]{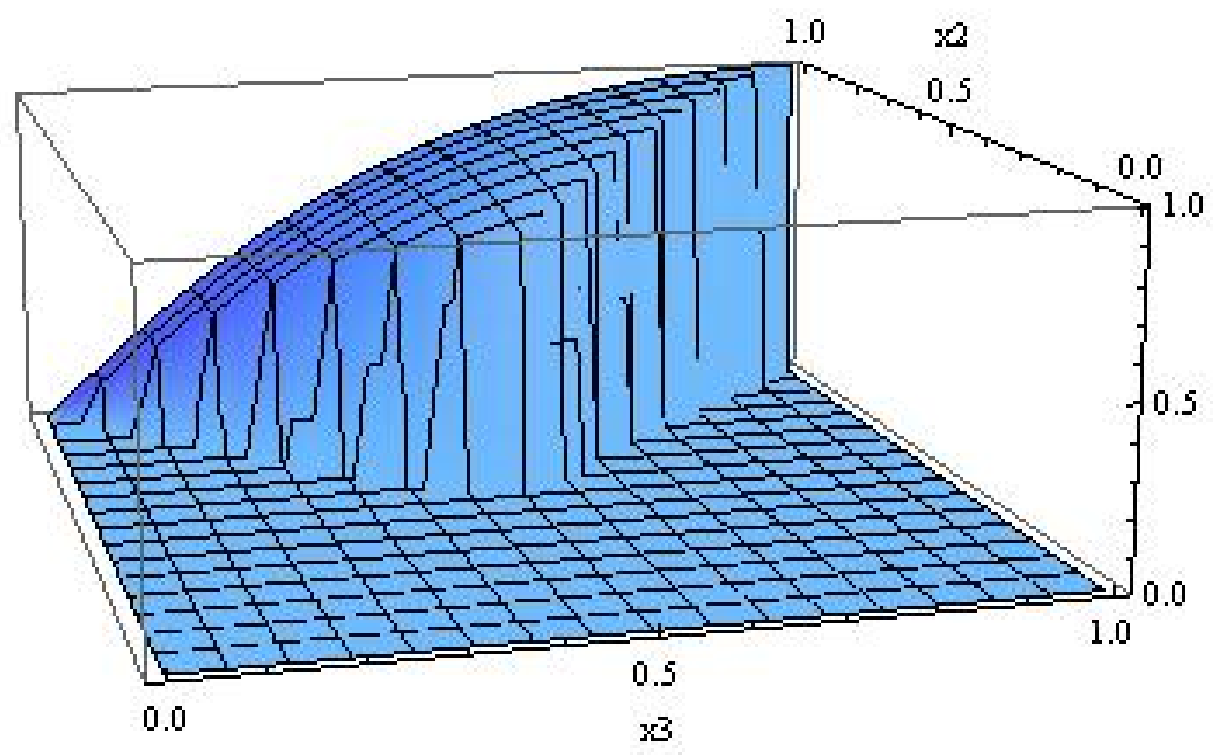}
\hspace{15mm}
	\includegraphics[scale=0.52]{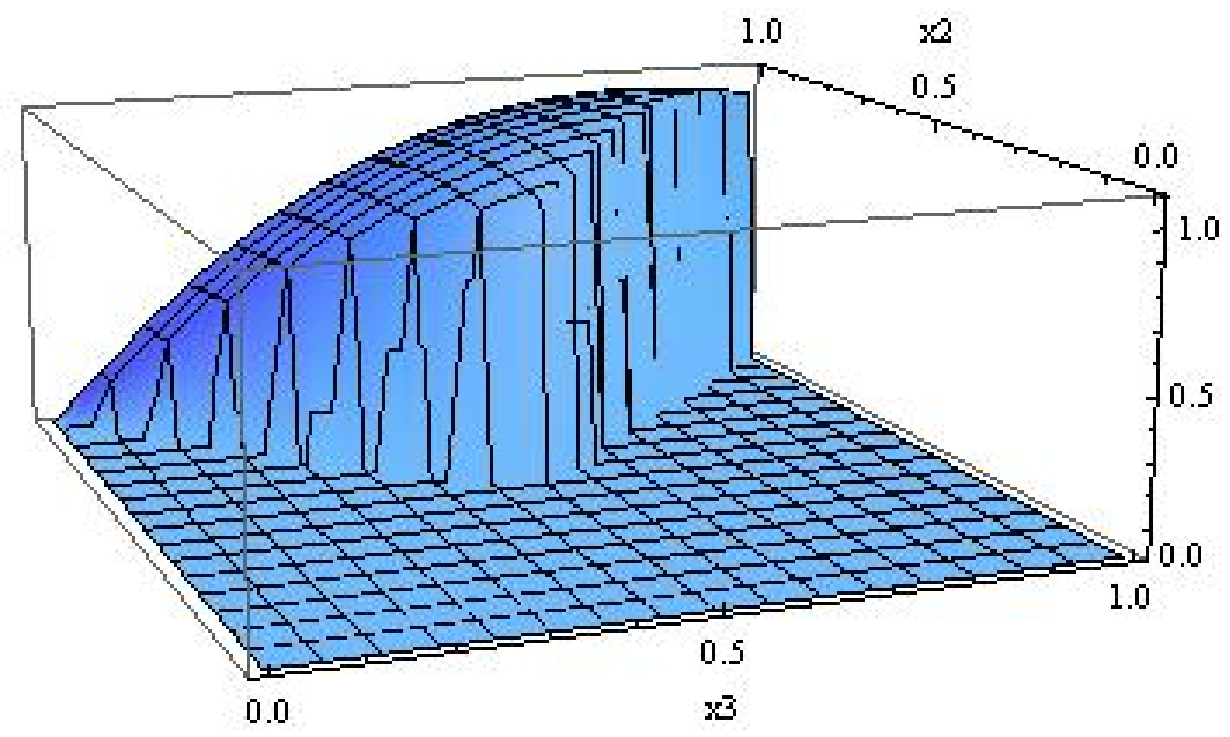}
	\label{fig:M_7b_DBI} 
	\caption{$A$ on the left, $B$ configuration on the right for the $\bar M_7$-driven interaction term.}
\end{figure}
All shapes peak in the equilateral configuration.\\
\newpage

$\bullet$  ${\cal O}_9= -1/6\,\,  \bar M_8 \, \partial_i^2 \pi (\partial_{jk} \pi)^2 \,\,/ a^6 $\label{mb8}\\
\begin{figure}[h]
	\includegraphics[scale=0.45]{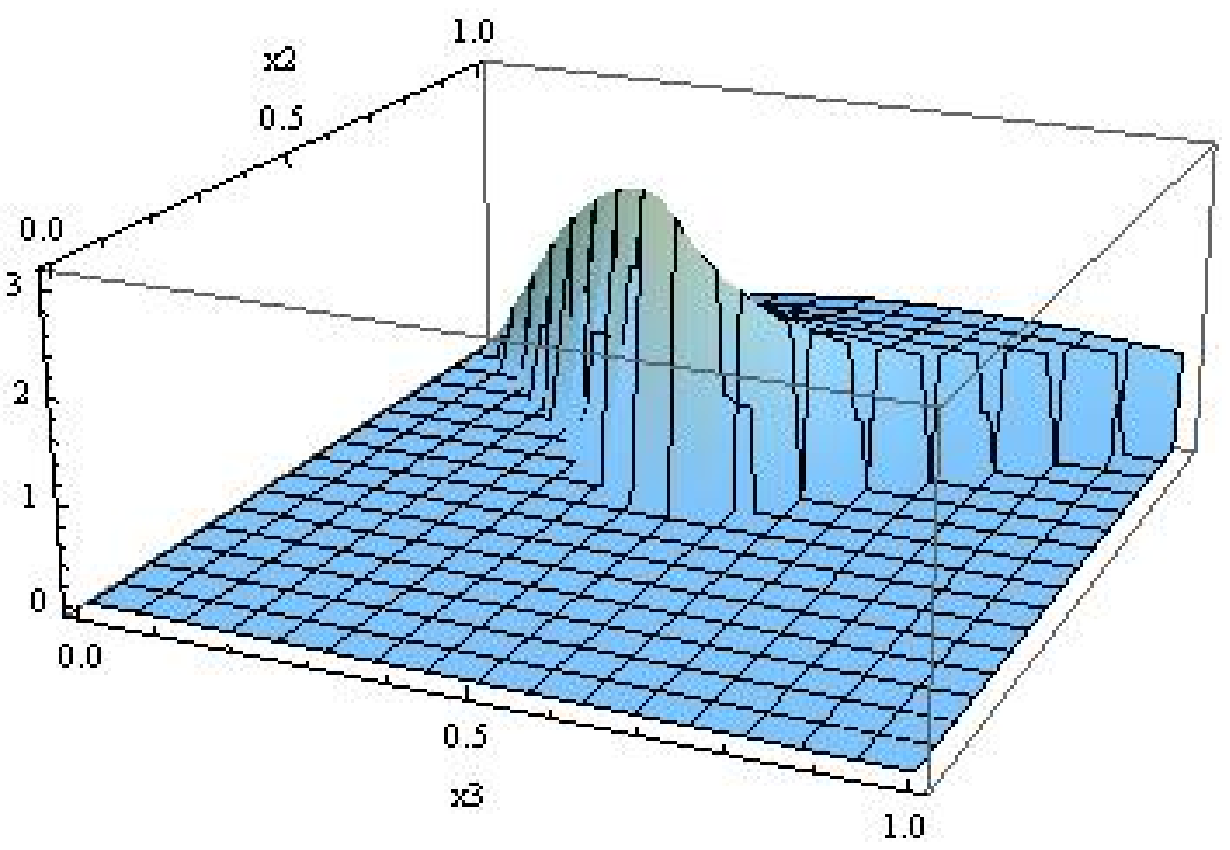}
\hspace{10mm}
	\includegraphics[scale=0.50]{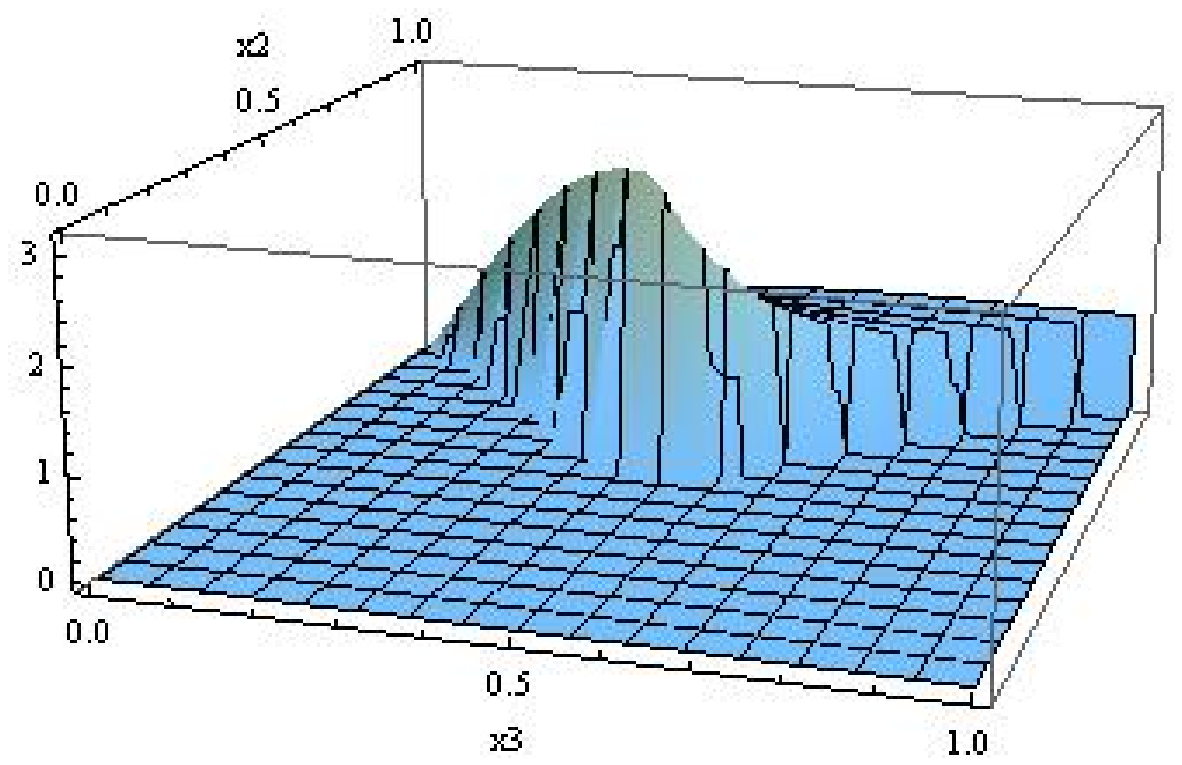}
	\label{fig:M_8_DBI} 
	\caption{exact DBI configuration on the left; approximated ghost shape on the right.}
\end{figure}

\begin{figure}[h]
	\includegraphics[scale=0.53]{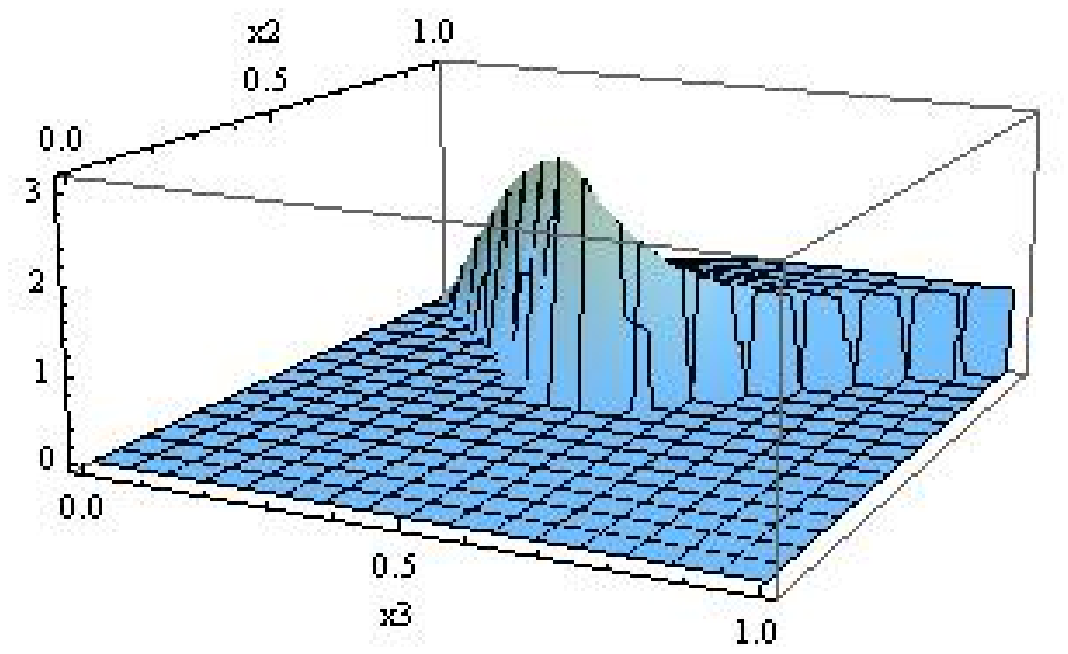}
\hspace{15mm}
	\includegraphics[scale=0.50]{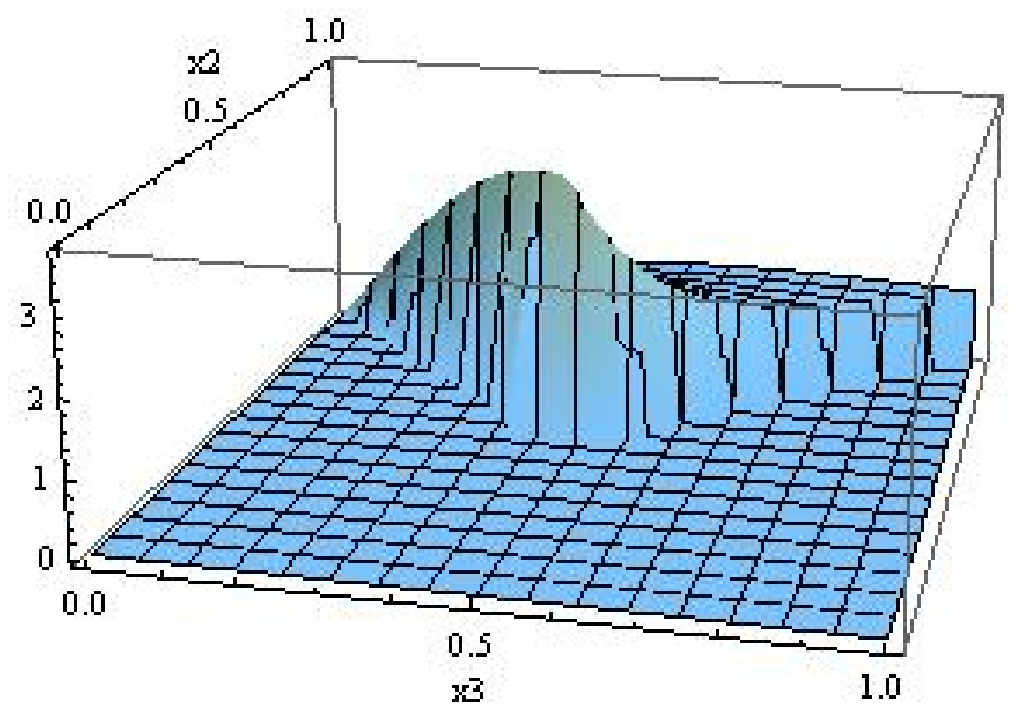}
	\label{fig:M_8b_DBI} 
	\caption{$A$ on the left, $B$ configuration on the right for the $\bar M_8$-driven interaction term}
\end{figure}
This is a second interaction term that produces, just as for  ${\cal O}_7$, a flat shape for the bispectra. Comparing it in Fourier space with our findings for ${\cal O}_8$, one can see that it is due to the way the spatial derivatives are combined. As shown in Figs. 21 and 22, also ${\cal O}_{10}$ gives rise to flat-shape bispectrum. We can see that 
the interactions ${\cal O}_8$, ${\cal O}_9$ and ${\cal O}_{10}$ have the same structure as far as the integral is concerned; on the other hand 
their $k$-dependence goes like $k_1^2 \, k_2^2 \, k_3^2 $, 
$k_1^2 (\vec{k_2}\cdot \vec{k_3})^2+ {\rm perm}$ and $(\vec{k_1}\cdot \vec{k_2})(\vec{k_2}\cdot \vec{k_3})(\vec{k_3}\cdot \vec{k_1})$, respectively. The last two produce a flat shape.\\
\newpage
$\bullet$ ${\cal O}_{10}=-1/6\,\,  \bar M_9 \, \partial_{ij} \pi \partial_{jk} \pi  \partial_{ki} \pi \,\,/ a^6 \label{mb9}$

\begin{figure}[h]
	\includegraphics[scale=0.50]{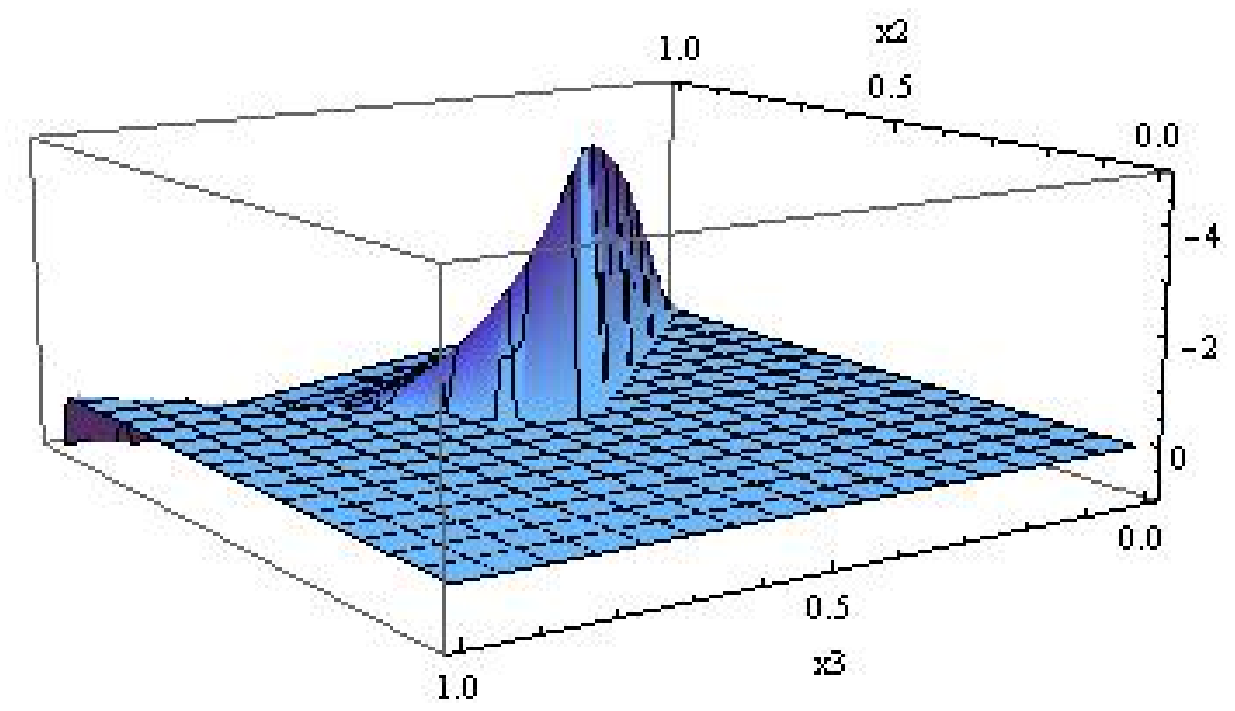}
\hspace{10mm}
	\includegraphics[scale=0.48]{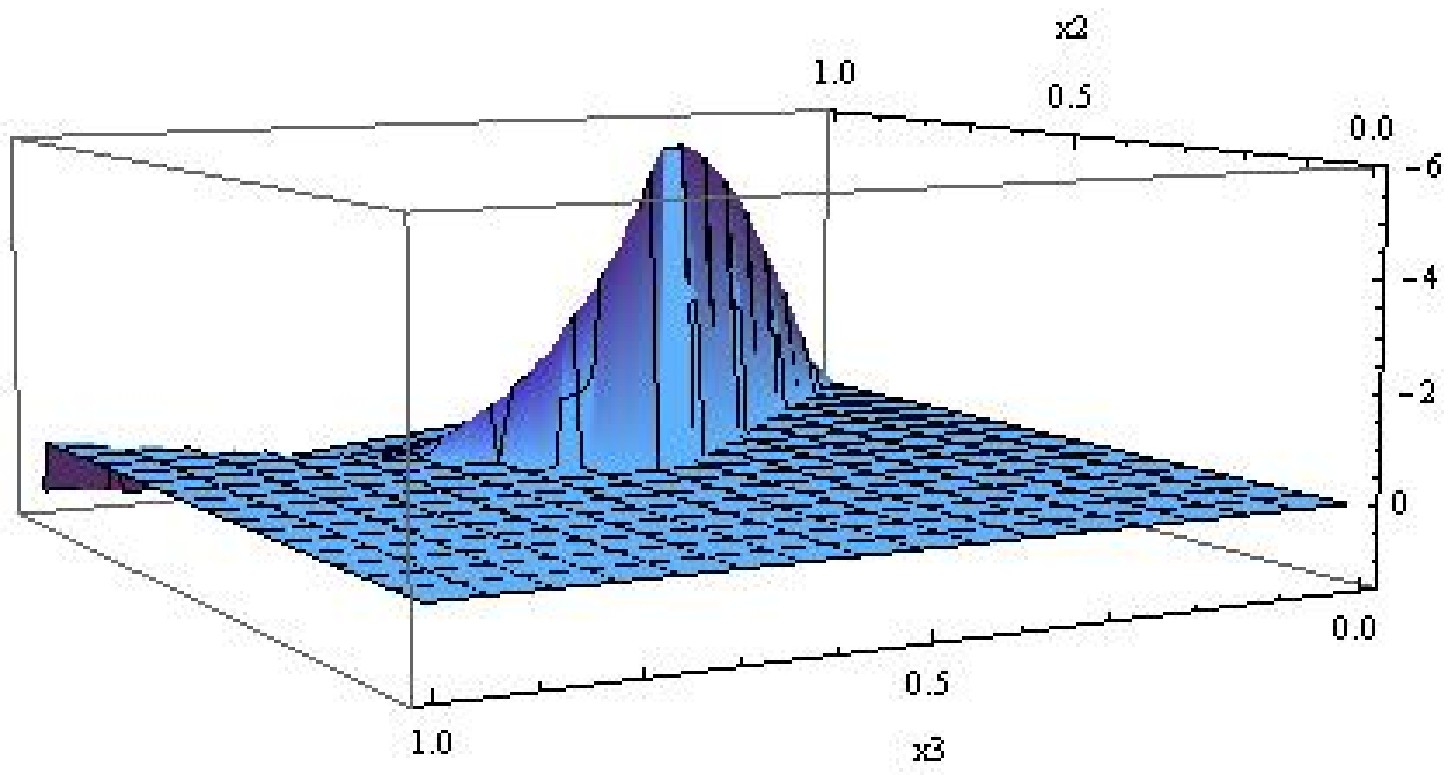}
	\label{fig:M_9_DBI} 
	\caption{exact DBI configuration on the left; approximated ghost shape on the right.}
\end{figure}

\begin{figure}[h]
	\includegraphics[scale=0.53]{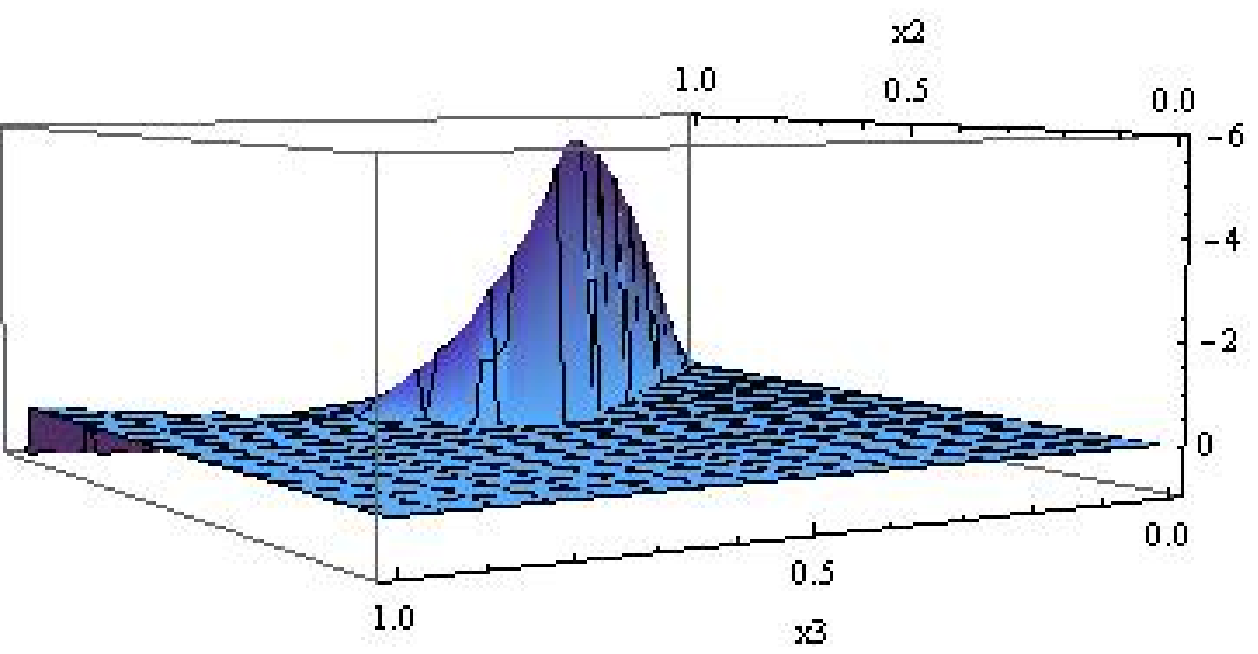}
\hspace{15mm}
	\includegraphics[scale=0.53]{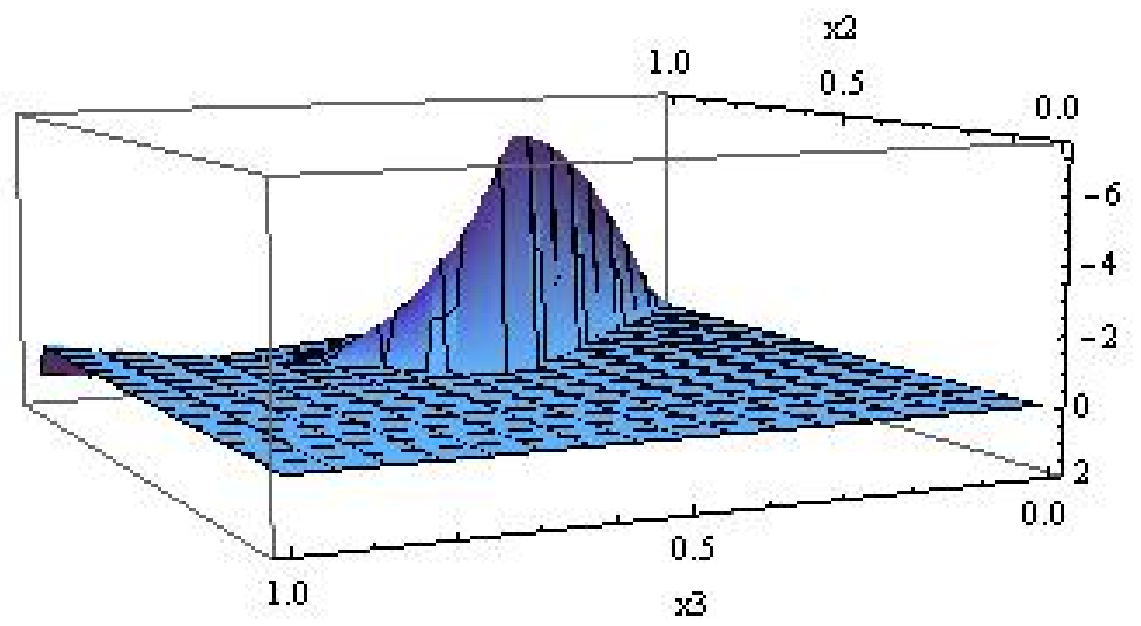}
	\label{fig:M_9b_DBI} 
	\caption{$A$ on the left, $B$ configuration on the right for the $\bar M_9$-driven interaction term.}
\end{figure}
The shapes are peaked in the flat configuration.

Obviously, having presented all the shapes due to each individual interaction, one might proceed with the study of the shape of linear combinations of them, much in the spirit of the orhtogonal shape recently introduced \cite{ssz05}. On the other hand, we are using approximated methods for three of the four configurations under scrutiny and it is therefore not a cautious step to infer new shapes from linear combinations of approximated ones, especially when delicate substractions are involved. One could proceed to study linear combinations in the case of the first configuration. We leave this to future work.

\subsection{Some general considerations on the shapes of non-Gaussianities}
From the above discussion we can read off some general qualitative features characterizing  the shapes of the bispectrum of a generic single-field model of inflation:
\begin{itemize}
\item Two qualitatively very different kinds of shapes appear: equilateral and flat.  As pointed out before, the next natural step would be to consider the shapes resulting from linear combination of the various interaction terms in the spirit of Ref. \cite{ssz05}.

\item In single field-models of inflation usually large non-Gaussianities are associated to equilateral shapes. 
In order to obtain a flat shape one needs to consider linear combinations of interaction operators such as what is done in \cite{ssz05} or models with an initial vacuum different from Bunch-Davies.
 Interestingly, in our case a flat shape emerges simply from individual operators generated by
curvature-related terms. When the flat shape appears, it does so  
 in all four configurations considered. We stress this point because it implies that this result does not depend on the type of wavefunction one employs in the calculation, be it the DBI inflation solution, the ghost inflationary one and the exact wavefunction that covers also more general models. The results of the DBI-like configuration are exact and easily reproducible with analytical methods. In fact, the DBI wavefunction is the usual solution of the standard single-field slow-roll inflation.
 Adopting such a wavefunction we may  provide  analytic results for the flat bispectra
\[
\fl  \langle\zeta_{k1} \zeta_{k2}\zeta_{k3}\rangle_{\bar M_6}= \frac{\left(60 a^6 b^2+11 a^3 b^5+b^8+\left(48 a^6-4 a^3 b^3-3 b^6\right) c^2-4 b^4 c^4\right) H^4 \bar M_6^2}{4 a^9\, \alpha_0^2\, b^5\, \epsilon ^3\, M_{\rm Pl}^6},
\]
\[
\fl \langle\zeta_{k1} \zeta_{k2}\zeta_{k3}\rangle_{\bar M_8}=   \frac{\left(15 a^3+b^3+3 b c^2\right) \left(12 a^6+b^6-6 b^4 c^2+8 b^2 c^4+8 a^3 \left(b^3-2 b c^2\right)\right) H^5 \bar M_8}{ a^9\, \alpha_0^3\, b^6 \,\epsilon ^3\, M_{\rm Pl}^6},
\]
\bea
\fl \langle\zeta_{k1} \zeta_{k2}\zeta_{k3}\rangle_{\bar M_9}=   \frac{3 \left(15 a^3+b^3+3 b c^2\right) \left(8 a^6+b^6-6 b^4 c^2+8 b^2 c^4+8 a^3 \left(b^3-2 b c^2\right)\right) H^5 \bar M_9}{2 a^9 \,\alpha_0^3\, b^6\, \epsilon ^3 \,M_{\rm Pl}^6}\, ,\nonumber\\ \label{bisflat}
\eea
where the overall momentum conservation delta has been omitted, 
\[
a=(k_1 k_2 k_3)^{1/3};\quad b={k_1+k_2+k_3};\quad c=(k_1 k_2+ k_1 k_3+k_2 k_3)^{1/2}, 
\]
and $\alpha_0$ is now the usual speed of sound:
\[
 \alpha_0= -M_{\rm Pl}^2 \dot H/(-M_{\rm Pl}^2 \dot H +2M_2^4)=c_s^2\, .
\]
Note that all the three expression given above have a maximum precisely in the flat configuration ($k_1=1,k_2=1/2=k_3 $). What immediately stands out in Eq.~(\ref{bisflat}) is the presence in the numerator of factors consisting  of subtractions between generally positive $k$-symmetrized terms: it is this characteristic that selects a flat, rather than a  equilteral shape, as one can readily verify by checking the bispectrum of the ``equilateral'' interaction terms.\\ The expression in Eq.~(\ref{bisflat}) is exact for all $P(X,\phi)$ models and can be employed to get a shape qualitatively similar in the other three configurations.  For practical purposes, we give below a very simple expression that very closely mimics the behaviour  of the typical bispectrum contribution that generates a flat shape:
\bea
\fl \langle\zeta_{k1} \zeta_{k2}\zeta_{k3}\rangle_{\bar M}\sim   \frac{\left(-k_1^2+k_2{}^2-k_3{}^2\right) \left(k_1^2+k_2{}^2-k_3{}^2\right) \left(-k_1^2+k_2{}^2+k_3{}^2\right)}{k_1^3 k_2{}^3 k_3{}^3 \left(k_1+k_2+k_3\right){}^6}  . \label{bisflats}
\eea

\item As a general rule, the terms which are going to generate a flat shape can be read off already at  the Lagrangian level: indeed the flatness originates from the way the external momenta combine with each other and are summed over. Whenever mixed space derivatives act on a single $\pi$ term and the mixing is repeated on at least another $\pi$ field, the shape turns out to be flat (note that this criterium puts $\bar M_{6,8,9}$ contributions in the same, ``flat'' class, but correctly excludes apparently very similar ones such as $\bar M_{2,3,7}$). 
\end{itemize}

\newpage

\section{Running of $f_{NL}$}
In the quest for properties that help in removing degeneracies among the many inflationary models one generally considers another observable beyond the power spectrum and its running,  i.e. the analysis of non-Gaussianities. Starting with the bispectrum, one can study its amplitude, shape and running. In the same spirit of the analysis we performed for the power spectrum, we now want to estimate the value for the running of the bispectrum amplitude, $f_{NL}$. In performing the calculation for $P_{\zeta}$, $n_s-1$ and then $\alpha_s$, we used the fact that the coefficients driving quadratic operators in the Lagrangian are nearly constant, up to slow-roll corrections. At first approximation, one writes down the power spectrum as a function of these parameters calculated at the horizon. Only when calculating the tilt of the spectrum and its running one does consider the time dependence on the $\bf{M}_n$'s, thus obtaining Eq.~(\ref{ns}),(\ref{run}). Similarly here, we will employ the results on the bispectrum amplitude contributions generated by independent interaction terms as given in \cite{b}. More precisely, we are going to focus on the running of $f_{NL}$ generated by a third-order interaction term whose bispectrum shape-function peaks is an uncommon flat configuration (see also \cite{ssz05}). In fact, as we have seen in the previous section and as detailed in \cite{b}, there are several independent terms that generate at least two qualitatively different flat shape-functions; we choose here to concentrate on the analysis of just one term as the same considerations and conclusions can be straighforwardly adapted to all of them.\\
The analysis for the running of $f_{ NL}$ has been done for several inflationary mechanisms such as DBI inflation, and others. We are going to extend this type of study to a third order interaction term driven by a nearly constant coefficient, $\bar M_6(t)$, and which is generated by an extrinsic curvature contribution which we reproduce here for convenience:
\be
\int d^3x dt \sqrt{-g}\Big[...+\frac{\bar M_6(t+\pi)^2}{3} \frac{\dot \pi}{a^4} \sum_{i,j}(\partial_{ij}\pi)^2+...\Big].\label{m6}
\ee
We have already calculated the corresponding contribution to $f_{NL}$ in six different points of the $(\alpha_0,\beta_0)$ plane, we report them in the \textit{Table} below.
\begin{center}
\begin{tabular}{| l || l | l | l | l | l| l | }
\hline			
       Benchmarks   & 1 & 2 &3 &4 &5 &6 \\ \hline 
  $\alpha_0$ & $10^{-2}$  & 0 & $0.5\cdot 10^{-2}$ & $2 \cdot 10^{-7}$ &$10^{-4}$  &$10^{-6}$ \\ \hline
  $\beta_0$  &  0 & $0.5\cdot 10^{-4} $ & $0.25\cdot 10^{-4} $ & $5\cdot 10^{-5}$ & 0&0  \\
\hline  
\end{tabular}
\end{center}
\noindent{\scriptsize Configuration \textit{1} describes pure DBI-like theories, pure ghost corresponds to configuration \textit{2}. In \textit{3,4} a more general model is considered while in the last two configurations one aims at considering the cases characterized by very small generalized speed of sound, $\sqrt{\alpha_0}$.
}\\

\vspace{2mm}
\noindent Corresponding to each one of the configurations of the table above, we report below the values for $f_{NL}$.
\begin{center}
\begin{tabular}{| l || l | l | l | l | l| l | }
\hline			
benchmarks   & 1 & 2 &3 &4 &5 &6 \\ \hline 
$f^{\bar M_6}_{ NL}$   &     $\,\,\,  10^{4}\, \gamma_6 \,\,\, $     &    $\,\,\, 4\cdot  10^{3}\,\gamma_6\,\,\, $       &     $\,\,\,5\cdot 10^{3}\,\gamma_6\,\,\,$     &      $\,\,\, 4\cdot 10^{3}\,\gamma_6\,\,\,$      &      $\,\,\, 10^8\,\gamma_6\,\,\,$    &    $\,\,\, 10^{12}\,\gamma_6\,\,\,$  \\ 
\hline  
\end{tabular}
\end{center}
\noindent{\scriptsize \textbf{ The value of the bispectrum amplitude $f^{\bar M_6}_{ NL}$ in the six different configurations.}
}
\vspace{5mm}

\noindent The dimensionless quantity $\gamma_6$, which was introduced in \textit{Section} 4, is given by
\be
\gamma_6 = (\bar M_6^2 H^2 )/(M_{ Pl}^2 \epsilon H^2 + 2 M_2^4 -3\bar M_1^3 H). \label{gam6}
\ee 
Computing the running of $f_{NL}$ amounts then to calculating
\bea
n_{NG}\equiv \frac{d\, \ln{|f_{NL}(k)|}}{d \ln{k}}\simeq \frac{1}{H f_{ NL}}\frac{d\,f_{ NL}}{dt}.
\eea
For the specific case at hand, we have:
\bea
\fl \frac{1}{H f_{NL}^{\bar M_6}}\frac{d}{dt} f_{NL}^{\bar M_6}=\frac{1}{H f_{NL}^{\bar M_6}}\frac{d}{dt}\Big[  N(t) \gamma_6(t)\Big]\Big{|}_{t=t^{*}}=
   -\epsilon \frac{4 M_2{}^4 +2 M_1{}^4}{\left(H^2 M_P^2 \epsilon +2 M_2{}^4+3 M_1{}^4\right)}+2 \epsilon_6+\epsilon_{N}+\nonumber\\
\fl -\eta \frac{H^2 M_P^2\epsilon  }{H^2 M_P^2 \epsilon +2 M_2{}^4+3 M_1{}^4}-\epsilon_2\frac{8 M_2{}^4 }{ \left(H^2 M_P^2 \epsilon +2 M_2{}^4+3 M_1{}^4\right)}-\epsilon_1\frac{4 M_1{}^4 }{ \left(H^2 M_P^2 \epsilon +2 M_2{}^4+3 M_1{}^4\right)}.   \nonumber\\ \label{f6}
\eea
In the above expression $N$ stands for a numerical factor which is dependent on $\alpha_0$ and $\beta_0$, specifically it goes like $(\alpha_0+\beta_0)^{-n}$ with $n$ small positive integer. The generalized slow-roll parameters are defined as usual, with the only new ones being $\epsilon_6$,$\epsilon_{N}$:
\bea
\epsilon_6 = \frac{\dot{\bar M_6}}{H \bar M_6}\,\,; \qquad \epsilon_{N}={\cal O}(1)\times (\epsilon,\eta,\epsilon_{2},\epsilon_{1},\epsilon_{0})\label{e6}.
\eea
The last expression in Eq.~(\ref{e6}) means that $\epsilon_{N}$ can be expressed as a linear combination of the generalized slow roll parameters introduced before multiplied at most by an order unity constant. Overall, we can then conclude that the running of $f_{NL}^{\bar M_6}$ is of the order of the generalized slow roll parameters (or a linear combination thereof). Now, all these parameters, except for $\epsilon_6$, can be expressed as functions of $M,\bar M$ coefficients that first appear in the quadratic Lagrangian of the theory. As such, four of these parameters could in principle be expressed as a function of the observables one usually uses, namely the scalar and tensor spectral indices and their running. Not so for $\epsilon_6$, on which, in principle, we enjoy more freedom as it drives terms in the action that are at least cubic. It could indeed be that the running of $f_{NL}^{\bar M_6}$, and of $f_{NL}$ itself, is dominated by this contribution and could therefore be larger than one finds in some general single-field slow-roll models where the extrinsic curvature-generated interaction terms are not accounted for. The same considerations apply to other interaction terms such as the $\bar M_9$-driven one. On the other hand, special care must be exerted so as to make sure that requiring $\epsilon_6$ to be the leading generalized slow-roll parameter in Eq.~(\ref{f6})
does not spoil the possibility to have the contribution in Eq.~(\ref{m6}) dominate the overall bispectrum amplitude, which is what made this type of contribution interesting in the first place. Indeed, there are two ways of making $\epsilon_6$ the leading parameter, a large $\dot{\bar M_6}$ and a small $\bar M_6$. Pushing the latter option too far the interesting and possibly leading bispectrum and trispectrum amplitudes~\cite{b} would become subdominant and this would make the corresponding flat shape-functions a mere curiosity. But also the former option has to be discussed and this is clear from the expansion of Eq.~(\ref{m6}) resulting from considering the $\bar M_6$ time dependence:

\bea
\fl \int d^3x dt \sqrt{-g}\Big[...+\frac{\bar M_6(t)^2}{3} \frac{\dot \pi}{a^4} \sum_{i,j}(\partial_{ij}\pi)^2+\frac{ 2\bar M_6(t)\dot{\bar M_6}(t)}{3} \frac{\dot \pi}{a^4} \sum_{i,j}(\partial_{ij}\pi)^2 \times \pi...\Big].
\eea
Indeed the quartic interaction term that appears is proportional to $\dot{\bar M_6}$ and if the latter is too big it could give rise to too large a contribution to the power spectrum at one loop and would have to be ruled out. Below we give a number of inequalities that the quantities $\dot{\bar M_6}$, $\bar M_6$ need to satisfy in order not to spoil the appealing bispectrum and trispectrum features outlined above. It turns out they are not too restrictive and that the running of $f_{NL}^{\bar M_6}$ can be safely ruled by $\epsilon_6$. We first write down the inequalities that stem from requiring that the $\dot{\bar M_6}$-proportional quartic interaction term is not the leading interaction in the fourth order Lagrangian (this would have consequences on the loop corrections to the power spectrum as well) as compared to the usual single-field interactions (for the complete action at fourth-order see Ref.~\cite{t} or the detailed derivation in the next Section):
\bea
\fl \frac{M_2^4 }{\bar M_6^2 H^2}> \epsilon_6 \gg \epsilon\,; \qquad \frac{(\alpha_0+\sqrt{\beta_0}) M_3^4 }{\bar M_6^2 H^2}> \epsilon_6 \gg \epsilon\,;\qquad \frac{(\alpha_0+\sqrt{\beta_0})^2 M_4^4 }{\bar M_6^2 H^2} > \epsilon_6 \gg \epsilon\,.\label{ineq1}
\eea
Notice that only one of these inequalities need be satisfied. The $\epsilon_6 \gg \epsilon$ part of the inequalities above ensures that indeed $\epsilon_6$ is the dominating generalized slow-roll parameter.  One has to keep in mind here that large non-Gaussinities are generated by requiring the generalized speed of sound, $\sim \alpha_0+\sqrt{\beta_0}$, to be much smaller than unity and so the first of these inequalities seems somewhat less stringent than the others though there is no requirement on the $\bf{M}_n$'s to be all of the same order.\\
As anticipated a small $\bar M_6$ can in principle lead to a subleading contribution to the bispectrum signal thus rendering the corresponding flat shape-function less interesting. Borrowing the third order action of Eq.~(\ref{action}) and employing the estimates on the wavefunction we showed to hold at horizon crossing, one is able to derive the inequalities below:
\bea
\frac{\bar M_6^2 H^2}{M_2^4(\alpha_0+\sqrt{\beta_0})}>1\,;\qquad \frac{\bar M_6^2 H^2}{M_3^4(\alpha_0+\sqrt{\beta_0})^2}>1\,.   \label{ineq2}
\eea
This time it is necessary that both the inequalities in Eq.~(\ref{ineq2}) are satisfied. The coefficient $M_4$ is not found in the equation above as it first appears in theory Lagrangian at fourth order, so it is not involved in the tree level bispectrum calculations. The further  freedom on $M_4$ (and other coeffcients) that results from this simple fact can be used to study models of inflation which present a relatively small bispectrum together with a larger trispectrum signal \cite{t,4pt}. The inequalities given in  Eq.~(\ref{ineq1}) and (\ref{ineq2}) above are indeed compatible for values of $(\alpha_0+\sqrt{\beta_0})$ smaller than unity,  an assumption which is generally made when looking for models that can produce large NG, as shown in \cite{b,t} and detailed in \textit{Section 2} and \textit{3}. We see then that there is a whole, large window of values for the coefficient $\bar M_6$ that would allow for a running of $f_{NL}$ dominated by $\epsilon_6$.  This effect is generated purely by third order terms and specifically by extrinsic curvature-generated interaction terms driven by coefficients which first appear in the Lagrangian at third order.\\
Let us consider a specific example with realistic values for the generalized speed of sound. For simplicity, we take $\alpha_0=10^{-2}\,; \beta_0=0$: this corresponds to a generalized speed of sound $\sim \sqrt{\alpha_0}$ of about $1/10$. If we want the corresponding value of the $\bar M_6$ contribution to $f_{NL}$ to be well within the WMAP7 \cite{kom} limits for say, $f_{NL}^{{\rm orthogonal}}$ * , then one automatically obtains from the first column of the Table above Eq.~(\ref{gam6}) the inequality $\gamma_6 \leq 2\times 10^{-2}$. Our first inequality in Eq.(\ref{ineq1}) is easily satisfied in the case at hand and for the first relation in  Eq.(\ref{ineq2}) one obtains 
\bea
\frac{\bar M_6^2 H^2}{M_2^4(\alpha_0+\sqrt{\beta_0})}\sim \frac{2\times 10^{-2}}{10^{-2}}\geq 1\,\,\, .
\eea
\noindent So that, indeed, there is room for a leading $\epsilon_6$ in the running of $f_{NL}$ also after the bounds on $f_{NL}$ have been duly set into place.\\
Adding the results of this section to the analysis of \cite{b,t} one can safely say that, with respect to all observables one is ultimately interested in, the analysis of these curvature terms has shown they can have leading effects on all quantities and must therefore always be included in a thorough analysis of non-Gaussianities.\\
\vspace{9cm}

{\small *Implementing the bounds such as  $-410\leq f^{{\rm orthogonal}}_{NL}\leq 6$ is indeed the best we can require on the $\bar M_6$-driven contribution to $f_{NL}$ as the shape-function generated by $\bar M_6$ is certainly closer to the orthogonal shape than to the equilateral or local one.}

\newpage

\section{The Hamiltonian up to fourth order}
Before we proceed to write down the quartic Lagrangian, let us tress that the cubic Lagrangian is going to be just as relevant for the four-point function calculation as it contributes to the so called \textit{scalar exchange} diagram we will shortly calculate below. Adding the trispectrum analysis to the bispectrum data we gained in the previous section, here the hope is to be able to identify distinctive features for as many as possible different combinations of the ${\bf M}_n$'s in the form of specific patterns they produce in the shapes of the various correlators of curvature perturbations. The degeneracies among the results for different inflationary mechanisms that will inevitably arise might be removed by a joint analysis of the different n-point functions, starting with the bispectrum, the trispectrum, loop corrections to the power spectrum and so on.
Let us briefly go through some of the main features of the third order effective action above. All the comments can be straightforwardly extended to the fourth-order expression as well. In deriving the fourth-order Lagrangian we use the same algorithm that was used at third order so there is no need to reproduce it here. We do a small exeption for the following estimates we showed to hold at horizon crossing:

\be
\dot\pi \sim H \pi, \qquad \nabla \pi \sim    \frac{H}{\sqrt{\alpha_0 + \sqrt{\alpha_0^2 + 8 \beta_0}}}\,\, \pi \label{est} \equiv H/ \tilde{c}_s \, \, \pi,
\ee

\noindent because they will turn out to be, once again, very useful. The most general fourth-order action in the usual set up is:

\[
\fl \mathcal{S}_4=\int d^4 x \sqrt{-g}\left[ 
 \frac{1}{2!}M_2(t)^4 \frac{(\partial_i \pi)^4}{a^4} +2 M_3(t)^4 \frac{{\dot\pi}^2 (\partial_i \pi)^2}{a^2} 
+\frac{2}{3}M_4(t)^4 {\dot\pi}^4 
\right.
\]

\[ \left.  
\fl   - \frac{\bar M_1(t)^3}{4}\left(\frac{H(\partial_i \pi)^4}{a^4}-\frac{2 \dot \pi(\partial_i \pi)^2\partial_j^2 \pi}{a^4}   \right)
-\frac{\bar M_2(t)^2}{2} \left(\frac{(\partial_j \pi)^2(\partial_i^2 \pi)^2}{a^6} +\frac {2  \partial_k^2 \pi \partial_i \pi \partial_{ij} \pi \partial_j\pi}{a^6} \right) 
\right. 
\]

\[ 
\left.
\fl -\frac{\bar M_3(t)^2}{2} \left(  \frac{ (\partial_{ij}\pi)^2 (\partial_k \pi)^2}{a^6}+\frac{2 \partial_{i} \pi \partial_{ij} \pi\partial_{jk} \pi\partial_{k} \pi
}{a^6} \right)+\frac{2}{3}\bar M_4(t)^3 \,\, \frac{ \dot\pi (\partial_i \pi)^2 \partial_j^2 \pi}{a^4}
\right.
\]

\[
\left.
\fl -\frac{\bar M_6(t)^2}{3!} \frac{(\partial_k \pi)^2(\partial_{ij} \pi)^2}{a^6} -\frac{\bar M_7(t)}{3!}\left(\frac{3}{2}\frac{(\partial_i^2 \pi)^2 H(\partial_j \pi)^2}{a^6} +\frac{6\,\dot\pi \partial_k^2 \pi (\partial_j\partial_i^2 \pi) \partial_j\pi}{a^6}   \right)
\right.
\]

\bea
\left. 
\fl -\frac{\bar M_8(t)}{3!} \left( \frac{H (\partial_i \pi)^2 (\partial_j^2 \pi)^2}{a^6}+ \frac{H (\partial_i \pi)^2 (\partial_{jk} \pi)^2 }{2a^6}- \frac{2\, H \partial_k^2 \pi \partial_i \pi \partial_{ij} \pi \partial_j \pi}{a^6} + \frac{2\dot \pi \partial_k^2 \pi \partial_i^2 \partial_j \pi \partial_j \pi}{a^6}
\right. \right. \nonumber
\eea
\bea
\left. \left.
 \qquad \qquad + \frac{2\dot \pi \partial_k^2 \partial_i \pi \partial_{ij}\pi \partial_j \pi}{a^6}  +  \frac{2\dot \pi \partial_{ij} \partial_{ijk} \partial_k \pi}{a^6} \right) - \frac{\bar M_5(t)^2}{3!}  \frac{(\partial_i \pi)^2(\partial_j^2 \pi)^2}{a^6}
\right. \nonumber
\eea
\bea
\fl -\frac{\bar M_9(t)}{2}\left( {\frac{ H \partial_k^2 \pi (\partial_{ij}\pi)^2}{2 a^6}-\frac{ H \partial_{i} \pi \partial_{ij}\pi \partial_{jk}\pi \partial_{k}\pi }{a^6}+\frac{ \dot \pi  \partial_{ij}\pi \partial_{ijk} \pi \partial_{k}\pi }{a^6}+\frac{ \dot \pi \partial_i^2 \partial_j \pi \partial_{jk} \pi \partial_k \pi }{a^6}}\right)\nonumber
\eea

\[
\left.
\fl +\frac{\bar M_{10}^{3}(t)}{3}\frac{{\dot\pi}^3\partial_i^2 \pi}{a^2} 
-\frac{\bar M_{11}^{2}(t)}{3!}\frac{{\dot\pi}^2 (\partial_i^2 \pi)^2}{a^4} 
-\frac{\bar M_{12}^{2}(t)}{3!}\frac{{\dot\pi}^2 (\partial_{ij}\pi)^2 }{a^4} +\frac{\bar M_{13}(t)}{4!} \frac{2\,\dot\pi}{a^6}(\partial_i^2\pi)^3 
\right.
\]
\[ \left.  
\fl +\frac{\bar M_{14}(t)}{4!} \frac{2\, \dot\pi \partial_k^2\pi(\partial_{ij}\pi)^2}{a^6}  +\frac{\bar M_{15}(t)}{4!}\frac{2\, \dot \pi  \partial_{ij}\pi \partial_{jk}\pi \partial_{ki}\pi }{a^6} -\frac{\bar N_{1}(t)}{4!}\frac{(\partial_i^2 \pi)^4}{a^8} -\frac{\bar N_{2}(t)}{4!}\frac{(\partial_k^2 \pi)^2(\partial_{ij}\pi)^2}{a^8} 
\right.
\]
\bea
\left.
\fl  -\frac{\bar N_{3}(t)}{4!}\frac{\partial_{\rho}^2\pi   \partial_{ij}\pi \partial_{jk}\pi \partial_{ki}\pi}{a^8} -\frac{\bar N_{4}(t)}{4!}\frac{ (\partial_{ij} \pi)^4 }{a^8} -\frac{\bar N_{5}(t)}{4!} \frac{\partial_{ij}\pi \partial_{jk}\pi \partial_{k \rho}\pi \partial_{\rho i}\pi}{a^8}\right] \label{l4}
\eea

\noindent Note that, as pointed out in \cite{lh}, starting at fourth order in perturbations, one cannot immediately read off the Hamiltonian from the expression of the Lagrangian, in other words $H = - L$ does not hold here. We use the results one obtains by adopting the correct procedure which was outlined in detail in \cite{lh}.\\
Let us split the interaction Hamiltonian we will be concerned with as $H_{int}= H_3 + H_4$; one can prove that the overall interaction Hamiltonian is then:
\bea
\fl H_{int}= - L_3 - L_4 + \int d^3 x \sqrt{-g}\Big[ \frac{1}{2 M_2^4+H^2 \epsilon  M_P^2-3 H \bar{M}_1^3} \left( \frac{(\partial_i \pi)^4 M_2^8}{a^4} +\frac{4 {\dot \pi}^2 (\partial_i \pi)^2 M_2^4 M_3^4}{a^2} \right. \nonumber \\
\fl \left. +4 {\dot \pi}^4 M_3^8+\frac{(\partial_k \pi)^2 \partial_{i}^2 \partial_j \pi  \partial_j \pi M_2^4 \bar{M}_2^2}{a^6 }+\frac{2 {\dot \pi}^2 \partial_{i}^2 \partial_j \pi  \partial_j \pi M_3^4 \bar{M}_2^2}{a^4} +\frac{(\partial_{i}^2 \partial_j \pi  \partial_j \pi)^2  \bar{M}_2^4}{4 a^8} \right.  \nonumber \\
\fl \left. +\frac{(\partial_k \pi)^2 \partial_i^2 \partial_j \pi \partial_j \pi M_2^4 \bar{M}_3^2}{a^6 }+\frac{2 {\dot \pi}^2  \partial_i^2 \partial_j \pi \partial_j \pi M_3^4 \bar{M}_3^2}{a^4 } +\frac{( \partial_i^2 \partial_j \pi \partial_j \pi)^2 \bar{M}_2^2 \bar{M}_3^2}{2 a^8 } +\frac{(\partial_i^2 \partial_j \pi \partial_j \pi)^2 \bar{M}_3^4}{4 a^8} \right.\nonumber \\
\fl\left.  +\frac{4 {\dot \pi} (\partial_k \pi)^2 \partial_i^2 \pi M_2^4 \bar{M}_4^3}{3 a^4}  +\frac{8 {\dot \pi}^3 {\partial_i^2 \pi} M_3^4 \bar{M}_4^3}{3 a^2}  
+\frac{2 {\dot \pi} {\partial_k^2 \pi} {\partial_i^2 \partial_j \pi} {\partial_j \pi} \bar{M}_2^2 \bar{M}_4^3}{3 a^6} +\frac{2 {\dot \pi} {\partial_k^2 \pi}  {\partial_i^2 \partial_j \pi} {\partial_j \pi} \bar{M}_3^2 \bar{M}_4^3}{3 a^6 }
\right. \nonumber \\
\left. \fl +\frac{4 {\dot \pi}^2 (\partial_k^2 \pi)^2 \bar{M}_4^6}{9 a^4 }-\frac{(\partial_i \pi)^2 (\partial_k^2 \pi)^2 M_2^4 \bar{M}_5^2}{3 a^6 }-\frac{2 {\dot \pi}^2 (\partial_i^2 \pi)^2 M_3^4 \bar{M}_5^2}{3 a^4 } -\frac{(\partial_k^2 \pi)^2  \partial_i^2 \partial_j \pi \partial_j \pi \bar{M}_2^2 \bar{M}_5^2}{6 a^8} \right.  \nonumber \\
\fl \left.  -\frac{(\partial_k^2 \pi)^2 \partial_j \pi \partial_j \pi \bar{M}_3^2 \bar{M}_5^2}{6 a^8 }-\frac{2 {\dot \pi} (\partial_k^2 \pi)^3 \bar{M}_4^3 \bar{M}_5^2}{9 a^6 }  +\frac{(\partial_k^2 \pi)^4 \bar{M}_5^4}{36 a^8}-\frac{(\partial_k \pi)^2 (\partial_{ij}\pi)^2 M_2^4 \bar{M}_6^2}{3 a^6 } \right.  \nonumber \\
\fl \left. -\frac{2 {\dot \pi}^2 (\partial_{ij}\pi)^2 M_3^4 \bar{M}_6^2}{3 a^4 } -\frac{ (\partial_{kl}\pi)^2 \partial_i^2 \partial_j \pi \partial_j \pi \bar{M}_2^2 \bar{M}_6^2}{6 a^8 \left(2 M_2^4+H^2 \epsilon  M_P^2-3 H \bar{M}_1^3\right)}-\frac{(\partial_{kl}\pi)^2 \partial_i^2 \partial_j \pi \partial_j \pi \bar{M}_3^2 \bar{M}_6^2}{6 a^8 \left(2 M_2^4+H^2 \epsilon  M_P^2-3 H \bar{M}_1^3\right)} \right. \nonumber \\
\fl \left. -\frac{2 \dot \pi {\partial_k^2 \pi} (\partial_{ij}\pi)^2 \bar{M}_4^3 \bar{M}_6^2}{9 a^6}  +\frac{(\partial_k^2 \pi)^2 (\partial_{ij}\pi)^2 \bar{M}_5^2 \bar{M}_6^2}{18 a^8 }+\frac{(\partial_{ij}\pi)^4 \bar{M}_6^4}{36 a^8 } \right) \Big]  \nonumber \\
\label{h4}
\eea 
where the above terms besides $-(L_3 + L_{4})$ are all at fourth order in perturbations.\\
\section{Symmetries}
Having written the complete Hamiltonian, we now proceed to calculate the four-point function contributions arising from interaction terms at third and fourth order. We employ here the IN-IN formalism \cite{in-in1, in-in2, in-in3, w-qccc} and conveniently split the contributions to the four-point function as the ones arising from terms that make up the contact interaction diagram and the ones that generate the scalar exchange diagram as in the figure below.


\begin{center}
\fcolorbox{white}{white}{
  \begin{picture}(370,66) (47,-47)
    \SetWidth{1.0}
    \SetColor{Black}
    \Line(48,-46)(80,-14)
    \Line(80,-14)(48,18)
    \Line(80,-14)(128,-14)
    \Line(128,-14)(160,18)
    \Line(128,-14)(160,-46)
    \Line(352,-46)(416,18)
    \Line(352,18)(416,-46)
 \Text(8,-65)[lb]{\small{\Black{\textbf{Figure A}: On the left, the scalar exchange diagram. Contact interaction diagram on the right.}}}
  \end{picture}
}
\end{center}

\vspace{4mm}
\noindent It is useful at this stage to offer some comments on the calculations we are going to present. As mentioned, the literature already contains a thorough analysis of the trispectra for \textit{general single-field inflation} models, see for example \cite{chen-tris}. Work on the four-point function for ghost inflationary models has recently been presented \cite{huang, muko}. Our starting point, being based on a comprehensive effective theory, clearly encompasses all these models. Working with the effective Hamiltonian above translates into many immediate advantages as listed before but, on the other hand, in calculating the resulting four point function, one faces a substantial number of terms and it is therefore natural to look for some ordering principle which would single out some contributions to the trispectrum as the leading ones and allow us to concentrate on them only. In this context employing a symmetry for the whole theory can prove very useful. Indeed in \cite{4pt,muko} the authors consider only those allowed by a particular (approximate in \cite{4pt}) symmetry of the action, respectively:
\be
{\bf S1}:\,\,\, \pi \rightarrow -\pi\,\, ; \qquad \qquad {\bf S2}:\,\,\, \pi \rightarrow - \pi\quad\,\,\textit{and} \,\,  \quad t \rightarrow -t\,\, . \label{symm}
\ee
We plan here to employ our general effective theory to show that, allowing some freedom on the ${\bf M}_n$ coefficients that modulate the various terms in the third and fourth order action, within each one of the two distinct and quite restrictive symmetry requirements above there are novel curvature-generated terms in the action that should not be disregarded as negligible and that, furthermore, show some distinctive features in the shapes of the trispectrum. We will also describe terms allowed by both the symmetries in Eq.~(\ref{symm}) combined. Of course, one need not employ  symmetries to switch on or off any specific operator in the action. Most of the contributions are indeed freely adjustable by the correspondent $M_n$ coefficient, a procedure which is, in principle, legitimate since the underlying theory is unknown. We choose here to restrict ourselves to considering only symmetry-abiding terms. Let us comment on each one of the symmetries.\\ {\bf S1} is built upon the following considerations. Often the same ${\bf M}_n$ coefficients multiply terms of different perturbative orders; consequently the amplitude  $f_{NL}$ of the 3-point function will be related to the amplitude of higher order correlators, notably to $\tau_{NL}$, the amplitude for the four-point function. Whenever the leading part of the trispectrum is generated by these types of ${\bf M}_n$'s one can estimate that for its effect to be observable $\tau_{NL}$ has to be five orders of magnitude larger than $f_{NL}$ \cite{4pt}, which leaves little room for feasible models. On the other hand, one quickly realizes those ${\bf M}_n$'s whose first term starts only at the fourth perturbative order ($M_4, \bar M_{10}...$in Eq.~(\ref{l4})) are not plagued by this problem. This then represents a natural way to obtain inflationary models which allow a large, detectable trispectrum untied to the interactions which make up the bispectrum (which might well be small now)*.
Indeed, in \cite{4pt} the authors investigate on the size of all the interactions driven by the $M_4, \bar M_{11}, \bar N_1$** 
coefficients in Eq.~(\ref{l4}) and show that the leading interactions driven by these parameters  are all consistent with the $\pi \rightarrow - \pi$ prescription and are expected to give a comparable signal ***.
 By construction then, the terms in the interaction Hamiltonian that are going to contribute to the trispectrum and be consistent with the reasoning that inspired the {\bf S1} symmetry are only some of the ones that will make up the contact interaction diagram, namely those whose lowest order interaction is already at fourth order. This limits us to the contributions regulated by the following coefficients: $M_4, \bar M_{10}..\bar M_{15}, \bar N_{1}..\bar N_{5}$.\\
{\bf S2} symmetry, on the other hand, does not prohibit third order interactions, indeed in \cite{muko} the interaction $\dot \pi (\nabla \pi)^2$ is considered and, by inspection of Eq.~(\ref{action}), one can see that also other terms are allowed, the one regulated by $\bar M_5$ and, notably, the $\bar M_6\, \dot \pi (\partial_{ij}\pi)^2/a^4$ term. The $\bar M_6$-driven term is particularly interesting because its contribution to the bispectrum calculations of \cite{b} generates an interesting flat shape. The scalar exchange diagram will then be built out of the third order {\bf S2}-obeying terms in the action. In particular, inspired by previous findings, we are going to give a detailed account of the $\bar M_6$ contribution.\\
If both {\bf S1} \textit{and} {\bf S2} are to be enforced one must also exclude from the list of {\bf S1}-abiding interactions the ones multiplied by $\bar M_{10},\bar M_{13},\bar M_{14},\bar M_{15}$. A more clear picture of the situation concerning the various symmetries is presented in \textit{Table 1} below.\\

\begin{center}
\textbf{Table 1}\\
\vspace{5mm}
\begin{tabular}{| l || l | l | l | l | l| l | l| l| l| l| l|l|l| }
\hline			
       Coefficients   & $M_2$ & $M_3$ & $M_4$ & $\bar M_1$ & $\bar M_2$ & $\bar M_3$ & $\bar M_4$ & $\bar M_5$ & $\bar M_6$ & $\bar M_7$ & $\bar M_8$ & $\bar M_9$  \\ \hline 
  ${\bf S1}$ & X & X & \checkmark & X & X & X & X & X & X & X & X &X    \\ \hline
  ${\bf S2}$  & \checkmark &\checkmark  & \checkmark & X & X & X & X & \checkmark  & \checkmark & X & X   & X \\\hline
  Coefficients & $\bar M_{10}$ & $\bar M_{11}$ & $\bar M_{12}$ & $\bar M_{13}$ & $\bar M_{14}$ & $\bar M_{15}$ & $\bar N_1$ & $\bar N_2$ & $\bar N_3$ & $\bar N_4$ & $\bar N_5$ & $/$    \\ \hline 
  ${\bf S1}$  & \checkmark  & \checkmark & \checkmark & \checkmark & \checkmark & \checkmark & \checkmark & \checkmark & \checkmark & \checkmark &\checkmark &    \\ \hline
  ${\bf S2}$  & X & \checkmark  & \checkmark  & X & X & X & \checkmark & \checkmark & \checkmark & \checkmark & \checkmark  &  \\\hline
    
\end{tabular}
\end{center}
{\small { \bf  The Coefficients marked with `` \checkmark '' in correspondence of a given symmetry {\bf S} are {\bf S}-invariant, those marked with ``X'' violate the {\bf S} symmetry.  }}
\\
\vspace{4mm}

\noindent {\small {\bf *}One needs also to check that the interactions driven by coefficients that multiply also third  order fluctuations do not become important in the form radiative corrections to the bispectrum. This check is done in \cite{4pt} and ensures that loop corrections of those terms are not relevant.\\
{\bf **}In the same spirit of the analysis done in \cite{b} for all curvature-generated terms at third order, the authors of \cite{4pt} consider in the v2 of their paper some extrinsic-curvature terms generated at fourth order. They also comment on their importance in near de Sitter limit and their conclusions apply to our $\bar M_{11}, \bar N_1$ parameters.\\
{\bf ***}It would be interesting to understand to what kind of models, in terms of the fundamental scalar field, the simple resulting effective Lagrangian corresponds in this case.}

\noindent Note that each ${\bf M}_n$ coefficient might multiply many interactions at each perturbative orders and therefore we mark the coefficient as invariant under a symmetry when all the \textit{leading} interactions it multiplies are invariant under {\bf S1} or {\bf S2}. Determing the properties of the coefficients in the second row requires no effort, as one can easily verify these ${\bf M}_n$'s first appear in the action as multipliers of fourth-order terms. Things are less linear with the coefficients in the first row (except for $M_4$) as they appear at fourth order both multiplying bare interaction terms and multiplying other coefficients as well as interaction terms (for an example of the latter case see the terms written explicitly in Eq.~(\ref{h4})). They also appear at third and some also at second order in perturbations. One then must carefully check that, given a particular coefficient ${\bf M}_n$, in none of the interactions it multiplies at any order the leading terms violate the symmetry. For $M_2, M_3, \bar M_5, \bar M_6$ in the first row one can verify after some checks that these terms all parametrize indeed approximately invariant interactions upon requiring the coefficient $\bar M_4^3$ to be much smaller than the typical ${\bf M}_n$ such as  $M_2..\bar M_6$. This is because in the fourth-order Hamiltonian in Eq.~(\ref{h4}) there are terms of the form

\be
\propto \frac{1}{M^4}\,\, \bar M_4^3\times \{M_2^4, M_3^4, \bar M_5^2, \bar M_6^2 \}\times ({\bf S2}-\textit{violating\,\, interaction})
\ee
which one then assumes to be subleading. We stress this point because it emerges clearly and naturally in the effective theory approach.\\

\section{Trispectrum (amplitudes and shapes)}
\subsection{IN-IN Formalism}
We are going to employ the IN-IN formalism to calculate the four point function of curvature perturbation. The most general and compact expression for such a quantity is:

\bea
\fl \langle \Omega|\zeta_{k1}\zeta_{k2}\zeta_{k3}\zeta_{k4}(t)|\Omega \rangle = \langle 0|\bar T \{e^{i\int_{-\infty}^{t_0} d^3 x dt^{'} \mathcal{H}(x)}\} \zeta_{k1}\zeta_{k2}\zeta_{k3}\zeta_{k4}(t)\,  T\{ e^{-i\int_{-\infty}^{t_0} d^3 x^{'} dt^{''} \mathcal{H}(x)} \}|0\rangle \nonumber\\ , \label{ttb}
\eea
where $\bar T$ and $T$ indicate respectively anti-time order and time order operations,  $|0\rangle$ and $|\Omega\rangle$ stand for the vacuum of the free and interacting theory.\\
Expanding both the exponentials in Eq.~(\ref{ttb}), we single out the first non vanishing terms that will contribute to the scalar exchange and contact interaction diagrams.
\bea
 \langle \Omega|\zeta_{k1}\zeta_{k2}\zeta_{k3}\zeta_{k4}(t)|\Omega\rangle = \nonumber\\ \langle 0|\bar T \{i\int_{-\infty}^{t_0} d^3 x dt^{'} \mathcal{H}_3(x)\} \zeta_{k1}\zeta_{k2}\zeta_{k3}\zeta_{k4}(t)\,  T\{ -i\int_{-\infty}^{t_0} d^3 x^{'} dt^{''} \mathcal{H}_3(x^{'}) \}|0\rangle \nonumber\\
 +\langle 0|\bar T \{\frac{i^2}{2}\int{ \int{ d^3 x \,dt^{'}\, d^3 x^{'}\, dt^{''} \mathcal{H}_3(x) \mathcal{H}_3(x^{'})}}\} \zeta_{k1}\zeta_{k2}\zeta_{k3}\zeta_{k4}(t)|0 \rangle \nonumber\\
 +\langle 0|\zeta_{k1} \zeta_{k2}\zeta_{k3}\zeta_{k4}(t) T \{\frac{(-i)^2}{2}\int_{-\infty}^{t_0}{ \int_{-\infty}^{t_0}{ d^3 x \,dt^{'}\, d^3 x^{'}\, dt^{''} \mathcal{H}_3(x) \mathcal{H}_3(x^{'})}}\}|0\rangle
\nonumber\\
 +\langle 0|\bar T \{i\int_{-\infty}^{t_0}{d^3 x \,dt\, \mathcal{H}_4(x)}\} \zeta_{k1}\zeta_{k2}\zeta_{k3}\zeta_{k4}(t)|0\rangle \nonumber\\
 +\langle 0|\zeta_{k1} \zeta_{k2}\zeta_{k3}\zeta_{k4}(t) T \{-i\int_{-\infty}^{t_0}{ d^3 x^{'}\, dt^{''}  \mathcal{H}_4(x^{'})}\}|0\rangle +...\label{z4}
\eea
where  $\mathcal{H}_3,\mathcal{H}_4 $ are the third and fourth-order Hamiltonian in the interaction picture. The latter two terms make up the contact interaction diagram, the rest is responsible for the scalar exchange. Let us also remind the reader that the gauge invariant observable $\zeta$ is, at first approximation, linearly related to the scalar $\pi$ via $\zeta=-H \pi$. Also, already at this stage one can see that the result of the four point function is going to depend on  six variables. All wavefunctions, once in Fourier space, depend only on the magnitude of their momenta. There are at most ten fields involved in the contractions, eight of which will always depend on the magnitude of the four external momenta ($k_1,k_2,k_3,k_4$). We are left with one last contraction between two fields depending on the magnitude of one vector which, by construction, is going to be the sum of two external momenta. It turns out that, employing the overall momentum conservation, two variables are sufficient to describe any of these linear combinations, we choose $(k_{12} \equiv |\vec k_1+ \vec k_2|,k_{14} \equiv |\vec k_1+ \vec k_4|)$, giving  a total of six variables. As clear from above, the $\bar M_6$-driven third-order interaction we are going to consider further depends on scalar products between the various momenta but, as one can easily verify, these can all be fully specified by using the six variables introduced above. All the variables we will employ are represented in the figure below.

\begin{center}
\fcolorbox{white}{white}{
  \begin{picture}(276,196) (191,-43)
    \SetWidth{1.0}
    \SetColor{Black}
    \Line[arrow,arrowpos=0.5,arrowlength=5,arrowwidth=2,arrowinset=0.2](192,52)(400,4)
    \Line[arrow,arrowpos=0.5,arrowlength=5,arrowwidth=2,arrowinset=0.2](400,4)(432,84)
    \Line[arrow,arrowpos=0.5,arrowlength=5,arrowwidth=2,arrowinset=0.2](432,84)(256,148)
    \Line[arrow,arrowpos=0.5,arrowlength=5,arrowwidth=2,arrowinset=0.2](256,148)(192,52)
    \Line[arrow,arrowpos=0.5,arrowlength=5,arrowwidth=2,arrowinset=0.2](256,148)(400,4)
    \Line[dash,dashsize=2,arrow,arrowpos=0.5,arrowlength=5,arrowwidth=2,arrowinset=0.2](192,52)(432,84)
    \Text(272,14)[lb]{\Large{\Black{$k_1$}}}
    \Text(428,36)[lb]{\Large{\Black{$k_2$}}}
    \Text(360,122)[lb]{\Large{\Black{$k_3$}}}
    \Text(208,116)[lb]{\Large{\Black{$k_4$}}}
    \Text(276,94)[lb]{\Large{\Black{$k_{14}$}}}
    \Text(368,57)[lb]{\Large{\Black{$k_{12}$}}}


    \Arc[clock](366.455,20.679)(15.531,-158.551,-279.431)
    \Text(348,32)[lb]{\small{\Black{$\alpha$}}}
    \Arc[clock](397.214,19.929)(13.258,175.365,27.255)
    \Text(397,22)[lb]{\small{\Black{$\beta$}}}
    \Arc[dash,dashsize=10,clock](392.783,7.705)(33.382,169.13,54.852)
    \Text(366,16)[lb]{\small{\Black{$\gamma$}}}
  \Text(108,-12)[lb]{\small {\bf Figure B:} the regular tetrahedron described by the four external momenta and $k_{12}, k_{14}$}
   \end{picture}
}
\end{center}
\noindent In order to get a tetrahedron as the one in Fig~B one must enforce the following inequalities:
\be
 cos(\alpha -\beta) \ge cos(\gamma)\ge cos(\alpha+\beta)
\ee
with 
\be
\fl cos(\alpha)= \frac{k_1^2+k_{14}^2-k_4^2}{2k_1 k_{14}}\,; \quad cos(\beta)= \frac{k_2^2+k_{14}^2-k_3^2}{2k_2 k_{14}}\,;\quad cos(\gamma)= \frac{k_1^2+k_{2}^2-k_{12}^2}{2k_1 k_{2}}  .
\ee
From $-1 \le cos(\alpha,\beta,\gamma) \le 1$ one also obtains the usual  triangles inequalities. Here we single out some of the inequalities which we are going to use in what follows: 
\bea
k_{12}^2 + k_{14}^2 \le 4\, ; \quad \sqrt{1+k_4^2 -2k_4} \le k_{14}\le \sqrt{1+k_4^2 +2k_4}. \label{ineq}
\eea
In order to have a visual intuition and understanding of  the result, once the calculation of the several contributions to the trispectrum is performed  one needs to set up a number of configurations in which four out of the six variables are held fixed. Having more than one configuration also increases one's ability to distinguish the signatures of different interactions. Following \cite{chen-tris}, we adopt the set up described below:
\begin{itemize}
\item \textit{Equilateral configuration}: all the external momenta have the same magnitude $k=k_1=k_2=k_3=k_4$; the two variables left are plotted as $k_{12}/k, k_{14}/k$. Note that when plotting in this configuration we will use the first inequality in Eq.~(\ref{ineq}). Incidentally, this is the only configuration for which exact calculations for the trispectrum in ghost inflation have been presented (see \cite{muko}) so far. Note also that for the equilateral as well as for the other configurations, one conveniently plots the result of the calculations in Eq.~(\ref{z4}) for any specific interaction term multiplied by a factor of $\prod_{i=1}^{4}k_i^3$. It is done also because this factor is generally common to all the contributions and so removing it sharpens the differences between the plots of each interaction term.
\item \textit{Folded configuration}: here one has $k_{12}\rightarrow 0$ as well as $k_1 = k_2$ and $k_3 = k_4$. The second and third inequalities in Eq.~(\ref{ineq}) must be enforced in this case. The variables $k_{14}$ and $k_4$ are the ones plotted in this configuration.
\item \textit{Specialized planar limit configuration}: in this case we have $k_1=k_3=k_{14}$ as well as:
\be
k_{12}=\Big[k_1^2+\frac{k_2 k_4}{2\,k_1^2}\Big(k_2 k_4 + \sqrt{(4k_1^2 -k_2^2)(4k_1^2 -k_4^2)} \,\,   \Big)     \Big]^{1/2}.
\ee
The variables plotted are going to be $k_2/k_1$ and $k_4/k_1$.
\item \textit{Near double squeezed limit configuration}: the tetrahedron is now a planar quadrangle and $k_3 = k_4 = k_{12}$. The region of interest is in particular the one for which $k_3,\,k_4,\,\, k_{12}\rightarrow 0$ where the following relation holds:
\bea
\fl k_2=\frac{\sqrt{k_1^2(-k_{12}^2 +k_3^2+k_4^2)- k_{s1}^2 k_{s2}^2+k_{12}^2 k_{14}^2+k_{12}^2 k_{4}^2+k_{14}^2 k_{4}^2-k_{14}^2 k_{3}^2-k_4^4 +k_{3}^2 k_{4}^2}}{\sqrt{2}k_4} \nonumber\\
\eea

with
\bea
\fl  k_{s1}^2=2 \sqrt{(k_1 k_4 + {\bf k}_1 \cdot {\bf k}_4)(k_1 k_4 - {\bf k}_1 \cdot {\bf k}_4)} \quad  k_{s2}^2= \sqrt{(k_3 k_4 + {\bf k}_3 \cdot {\bf k}_4)(k_3 k_4 - {\bf k}_3 \cdot {\bf k}_4)}. \nonumber \\
\eea
In this case as well the last two inequalities of Eq.~(\ref{ineq}) will be imposed on the variables $k_{14}/k_1$ and $k_4/k_1$. Note that, only in this configuration, what one actually plots is the result of Eq.~(\ref{z4}) times  $\prod_{i=1}^{4}k_i^2$, instead of $\prod_{i=1}^{4}k_i^3$. This is once again done in order to better appreciate the difference among the many interaction terms. 
\end{itemize}
 We now consider  the result for the scalar exchange contribution focusing in particular on an interaction term (the one proportional to $\bar M_6$ in Eq.~\ref{action} ) case which proved very interesting in plotting the shape of the bispectrum \cite{b} as seen in \textit{Section 4}.\\
In all the calculations that follow we use a simplifying assumption which has been verified to hold for 3-point functions and is expected to hold for higher correlators as well \cite{b}. Instead of using the generalized wavefunction which comprises the \textit{general single-field inflation} solution, the Ghost inflation one, etc. as its simplified limits, we employ the usual Hankel function $H_{3/2}(k \tilde{c_s}\tau)$ as a solution to the equation of motion for the quadratic action. The rationale for such a simplification is that, as one can readily verify, the main contribution to higher order correlators comes as usual from the horizon-crossing region and precisely in that region the behaviour of the general solution of \textit{Section 2} resembles very closely the one of the simpler specific DBI wavefunction. We elaborate further on this fact in \textit{Appendix C} where some examples and comparisons of explicit calculations are provided.\\
\noindent Before moving to the detailed analysis of the shape-functions for several interactions terms in various configurations, let us comment briefly on the amplitudes generically associated with these interaction terms. As noted before, building on the freedom on the ${\bf M}_n$ coefficients allowed by the theory and on the possibility of employing a small speed of sound, $c_s^2 \ll 1$ (the same holds for the parameters which represent the generalization of $c_s^2$, i.e. $\alpha_0$ and $\beta_0$), one can obtain large values for the amplitude associated to each one of the curvature-generated terms we are going to study. This has been quantitatively verified for all the terms of Eq.~(\ref{action}) in \cite{b}. As an example, consider the $\bar M_6$-driven fourthorder interaction term. To estimate the size of the amplitude associated with a given interaction term one considers its ratio with the quadratic terms of the theory at freezing \cite{eft08}. Applying this prescription to our example one obtains:
\be
\frac{\mathcal{L}_{\bar M_6^2 (\nabla \pi )^2(\nabla^2 \pi)^2}}{\mathcal{L}_2}\sim \frac{\bar M_6^2 \,H^6\,\pi^4/\tilde{c_s}^6 }{M_2^4 \,H^2 \,{\pi}^2}\sim \frac{\bar M_6^2 \,H^2 }{M_2^4}\frac{\zeta^2}{\tilde{c_s}^6}
\ee 
 where the linear relation $\zeta = - H \pi $ has been used; taking $\zeta \sim 10^{-5}$ gives a rough estimate of the size of the non linear corrections.
A number of  useful consistency checks for trispectrum calculations have been outlined in the literature (see first and sixth references in \cite{trispectrum}). We leave the task of performing such checks to future work.   
\subsection{Scalar exchange diagram}
\begin{figure}[hp]
	\includegraphics[scale=0.50]{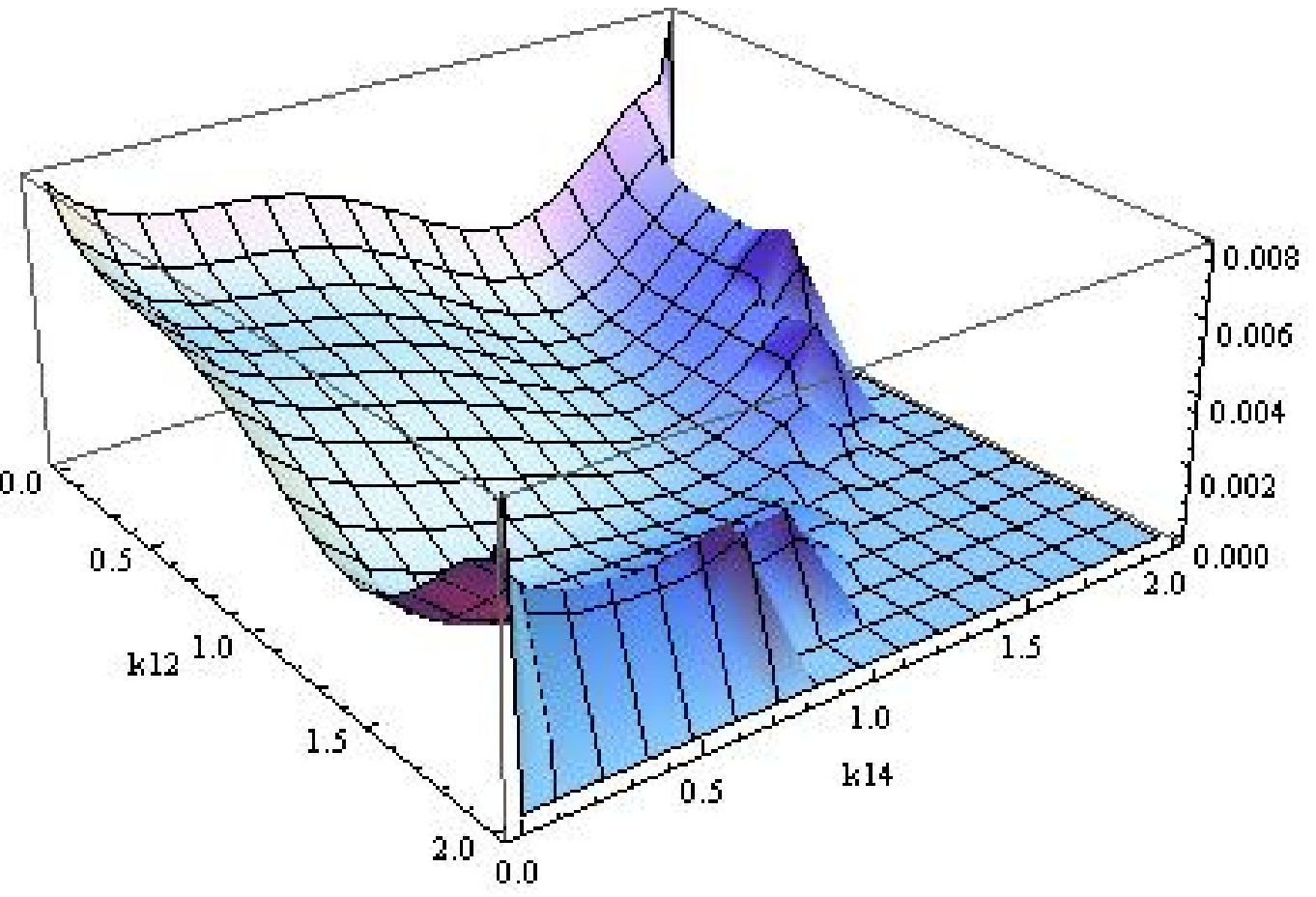}
	\hspace{10mm}
		\includegraphics[scale=0.50]{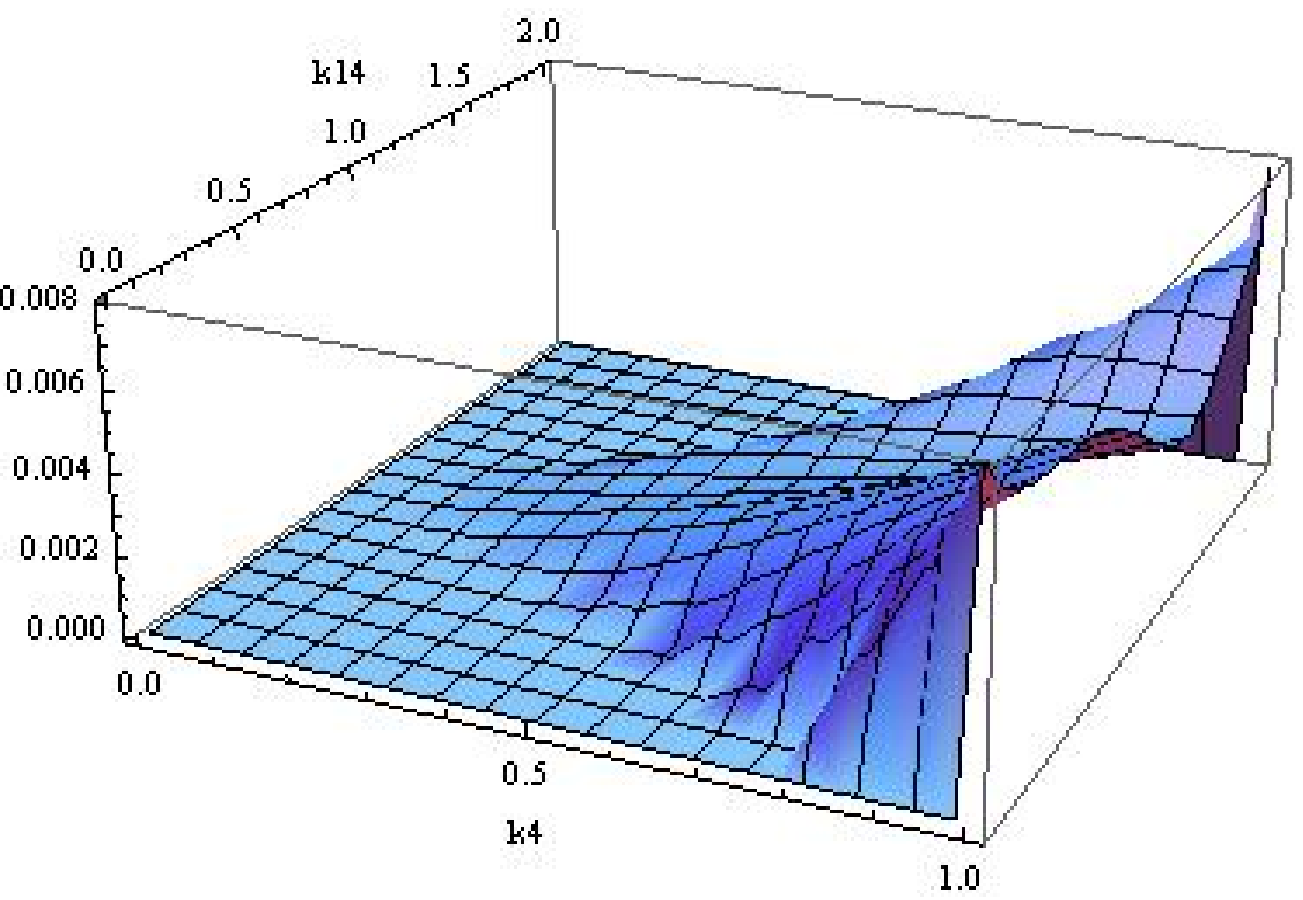}
\caption{The \textit{equilateral} configuration shape is presented on the left from the scalar exchange contribution of the $\bar M_6$-driven interaction. The shape-function is different from the plots presented in \cite{chen-tris}. On the other hand, once a necessary change of variables has been performed, it is qualitatively similar to the shape for the contact interaction diagram which arises from the ghost inflation $(\nabla \pi)^4$ interaction term in \cite{muko}.\\ On the right we plotted our findings for the $\bar M_6$-generated interaction in the \textit{folded} configuration. It very much resembles the ones obtained in \cite{chen-tris} for the scalar exchange diagrams from DBI-like terms, especially from the interaction $\dot \pi (\nabla \pi)^2$ . In all the pictures above and below $k_1$ has ben set equal to unity without loss of generality.}
\label{m61}
\end{figure}

\begin{figure}[hp]
	\includegraphics[scale=0.54]{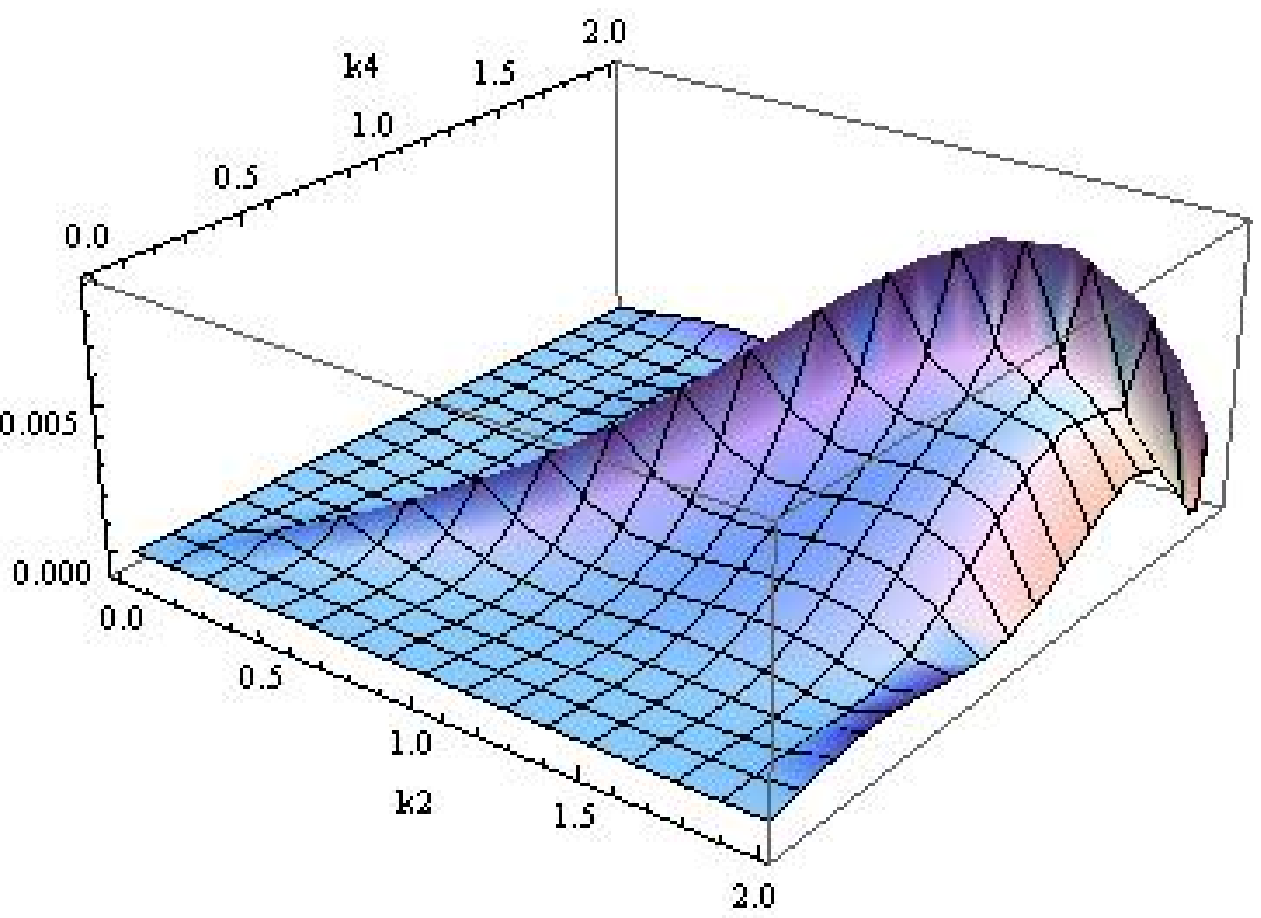}
	\hspace{20mm}
		\includegraphics[scale=0.43]{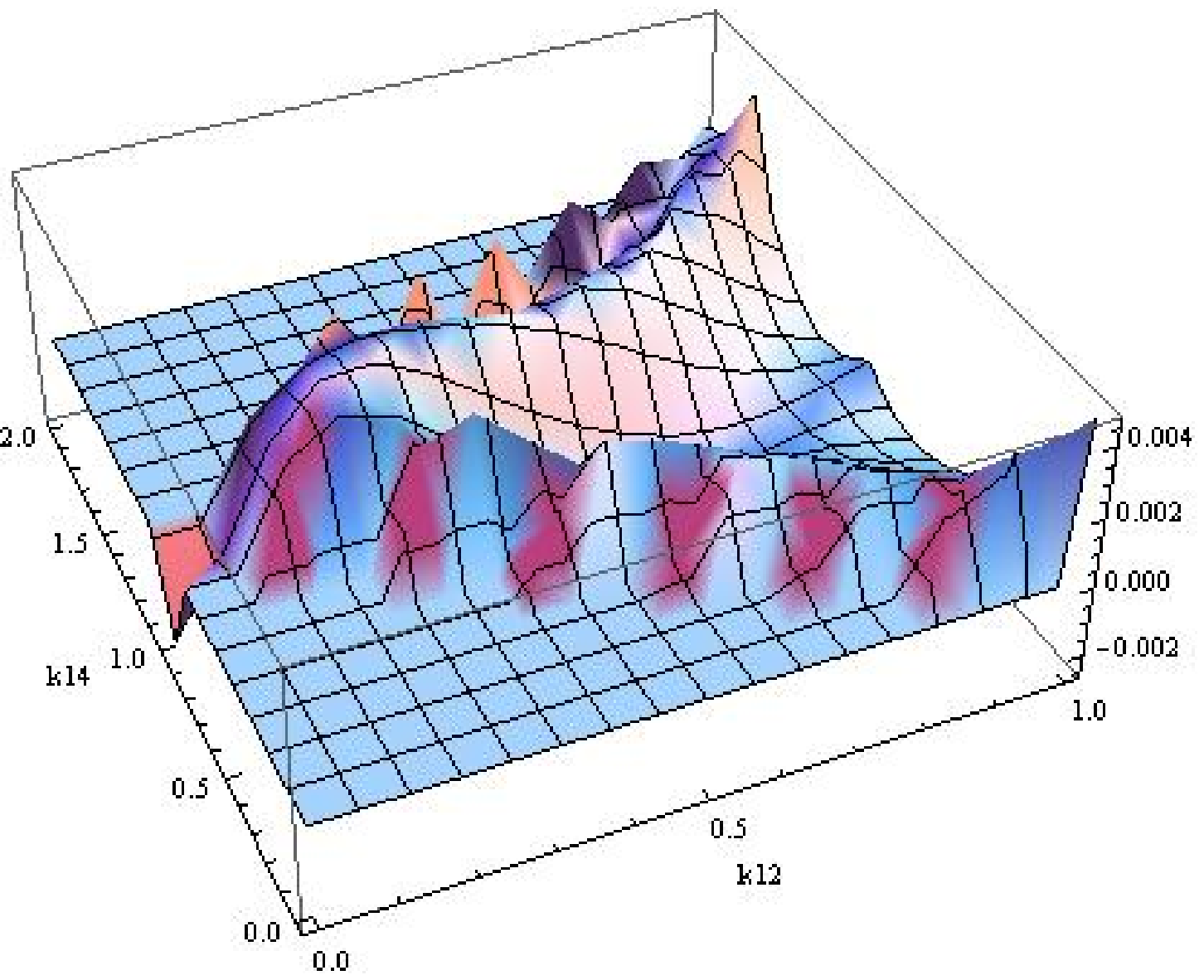}
\caption{On the left the plot obtained for the $\bar M_6$ interaction in the \textit{planar} configuration plot. Note  here some of the interesting features: as $k_2,k_4 \rightarrow 0$ the shape function goes to zero, much like it happens for DBI-generated interactions. As $k_2,k_4 \rightarrow 2$ the shape function reaches values that are negative, albeit slightly so. Looking at the $k_2 = k_4$ line we see that it is convex, rather than concave as found for other interaction types in \cite{chen-tris}\\ On the right the \textit{planar limit double squeezed configuration} is plotted. In the region of interest, namely for $k_{12} \rightarrow 0$, the shape function is non-zero, finite and negative; this again is different than what found in \cite{chen-tris} for a variety of DBI-originated terms.  }
\label{m62}
\end{figure}

Here we are going to consider the interaction term $\bar M_6\,\, \dot \pi (\partial_{ij}\pi)^2/a^4$. Note that, as opposed to the $\bar M_8, \bar M_9$-regulated terms in Eq.~(\ref{action}) which give a flat shape for the bispectrum much like the $\bar M_6$-driven interaction, this term is actually invariant under the symmetry {\bf S2} while for it to be (approximately) invariant under {\bf S1} one needs to require its $\bar M_6$ coefficient to be much smaller than $M_4$ which would in turn make its signal undetectable.
We now write more explicitly the contribution of the $\bar M_6$-driven third order interaction to the scalar exchange diagram. For all the details of the calculation, including contractions, we refer the reader to \textit{Appendix B}. Consider here just one particular contraction of the fields, the sample contribution we are after looks like the following: 
\bea
\fl <\pi_{k_1} \pi_{k_2}\pi_{k_3}\pi_{k_4}>^{s.e.}_{\bar M_6}= \nonumber \sum_{{ \textit{all contractions}}}
\eea
\bea
\fl \pi^{*}_{k_1}\pi^{*}_{k_2}\pi_{k_3}\pi_{k_4}(0)\,\int_{-\infty}^{t\rightarrow \,0}{dt_1  a^3 \dot \pi_{k_{12}} \pi_{k_1}\pi_{k_2}({\bf k}_1\cdot {\bf k}_2)^2 /a^4}\int_{-\infty}^{t\rightarrow \,0}{dt_2  a^3 \dot \pi^{*}_{k_{12}} \pi^{*}_{k_3}\pi^{*}_{k_4} ({\bf k}_3\cdot {\bf k}_4)^2 /a^4}  \nonumber\\
\eea
\bea
\fl - 2\mathcal{R}_e \Big[ \pi^{*}_{k_1}\pi^{*}_{k_2}\pi^{*}_{k_3}\pi^{*}_{k_4}\,\int_{-\infty}^{t\rightarrow \,0}{dt_1   \frac{a^3}{a^4}  \dot \pi^{*}_{k_{12}} \pi_{k_1}\pi_{k_2}(t_1)({\bf k}_1\cdot {\bf k}_2)^2 }\int_{-\infty}^{t_1}{dt_2  \frac{a^3}{a^4} \dot \pi_{k_{12}} \pi_{k_3}\pi_{k_4}(t_2) ({\bf k}_3\cdot {\bf k}_4)^2 }   \Big], \nonumber \\ \label{calc}
\eea
where an overall momentum conservation delta and a factor of $(2\pi)^3$ have been omitted for simplicity.\\In Fig. \ref{m61}, \ref{m62} below we plot what one obtains by summing over all contractions, accounting for the symmetry factors of the vertices and plotting the result.\\ For the sake of comparison we often make reference to the shapes obtained in \cite{chen-tris} and \cite{muko}. In the former work the so called local trispectrum is also plotted and compared with findings for \textit{general single-field inflation} models; in the latter one shapes for the trispectrum of ghost inflation are presented in the equilateral configuration only. As mentioned, our starting Lagrangian comprises both these inflationary models; we decided to concentrate on plotting the novel curvature-generated terms that are invariant under {\bf S1}, {\bf S2} or both. 

\newpage

Overall we see that performing the shape analysis for this {\bf S2}-abiding term we are able to find some distinctive features with respect to the DBI-generated contributions.  This is quite interesting also given the fact that the very same $\bar M_6$-modulated interaction term gives rise to a flat shape for the three point function \cite{b} which contributes to enlarge the allowed classification of bispectra shape-functions for single-field models of inflation (with Bunch-Davies vacuum).\\ 
\subsection{Contact interaction diagram}
We now turn to the calculation of various terms that contribute to the contact interaction diagrams. The terms driven by $M_3, M_4$ are DBI-generated and have been calculated in a number of papers, notably \cite{trispectrum,chen-tris}. $M_2$ is found in both DBI and Ghost inflationary theories \cite{chen-tris, muko}. If we are to preserve both symmetries then we need to focus on $\bar M_{11},\bar M_{12}, \bar N_{1}..\bar N_{5}$ as one can easily check from \textit{Table 1}.\\ Given any interaction term at fourth order, its contribution to the contact interaction diagram can be written as:

\bea
\fl \langle \pi_{k_1}\pi_{k_2}\pi_{k_3}\pi_{k_4}(t\rightarrow 0)\rangle ^{c.i.}_{H_4} = 
-2 \mathcal{I}_m \Big[ \langle 0|\bar T \{ i\int_{-\infty}^{t\rightarrow 0}{ d^3 x \,\, dt^{'} \mathcal{H}_4(x) \pi^{*}_{k1}\pi^{*}_{k2}\pi^{*}_{k3}\pi^{*}_{k4}(t\rightarrow 0)}\}|0\rangle \Big]  \nonumber\\
\eea

\noindent where $\mathcal{I}_m$ stands for taking the imaginary part; a delta enforcing momentum conservation and unimportant numerical factors have been omitted.\\
We have applied the above formula to a number of fourth-order interaction terms providing some examples of notable {\bf S1} \textit{and} {\bf S2}-abiding terms, {\bf S1} \textit{ or} {\bf S2} invariant contributions and, finally, terms that do not respect any of the symmetries above. We start with the {\bf S2}-invariant interaction term $\sim M_2^4 (\nabla \pi)^4$. This contribution is present in both DBI-like and Ghost inflationary models (e.g. in DBI one simply has  $1/c_s^2 = 1-2 M_2^4/M_P^2 \dot H$). Although already written down in \cite{muko, chen-tris}, this term has not been plotted in all four configurations described in section 4 we employ here. This is because in DBI-like theories it is expected to be subdomimant with respect to the $M_4^4 {\dot \pi}^4$ term. \\

$\bullet$  {\bf $ {\cal O}_{1}=1/2\,\,   M_2^4 \, (\partial_{i} \pi)^4 \,\,/ a^4 $}\\

\noindent For the shape function of the operator $(\nabla \pi)^4$ we see that a number of interesting issues arise. First, the plot in the equilateral configuration does not resemble any of those plotted in \cite{chen-tris} *. 
Then, as we mentioned in Fig.~\ref{m62}, in the double squeezed configuration the $k_{12} \rightarrow 0$ limit gives a non-zero finite shape function. This is important because, up to the results in \cite{chen-tris}, this limit was thought as very useful to distinguish the leading contributions coming from interactions at third order in perturbations from the ones at fourth order in fluctuations.

\begin{figure}[hp]
	\includegraphics[scale=0.53]{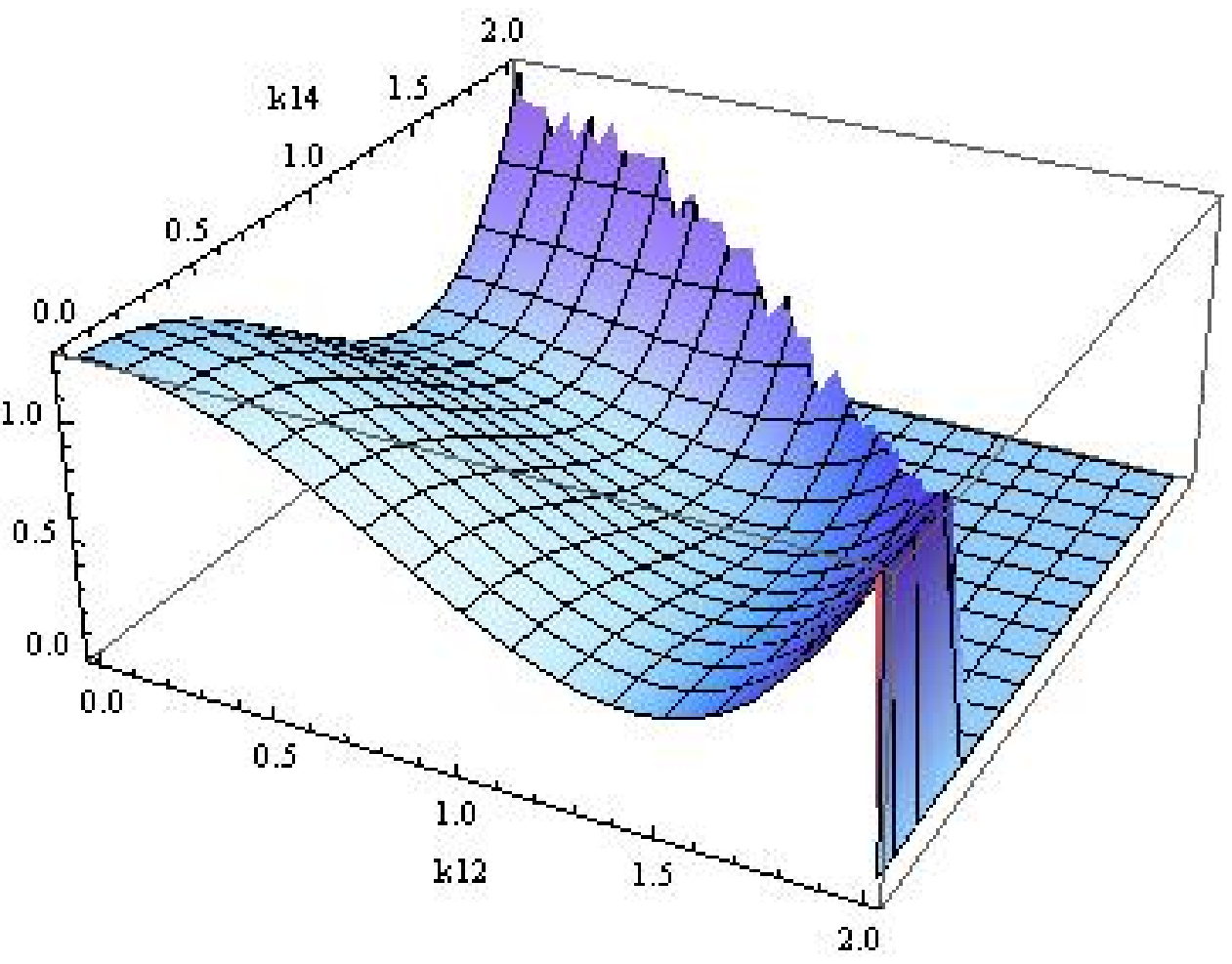}
	\hspace{10mm}
		\includegraphics[scale=0.57]{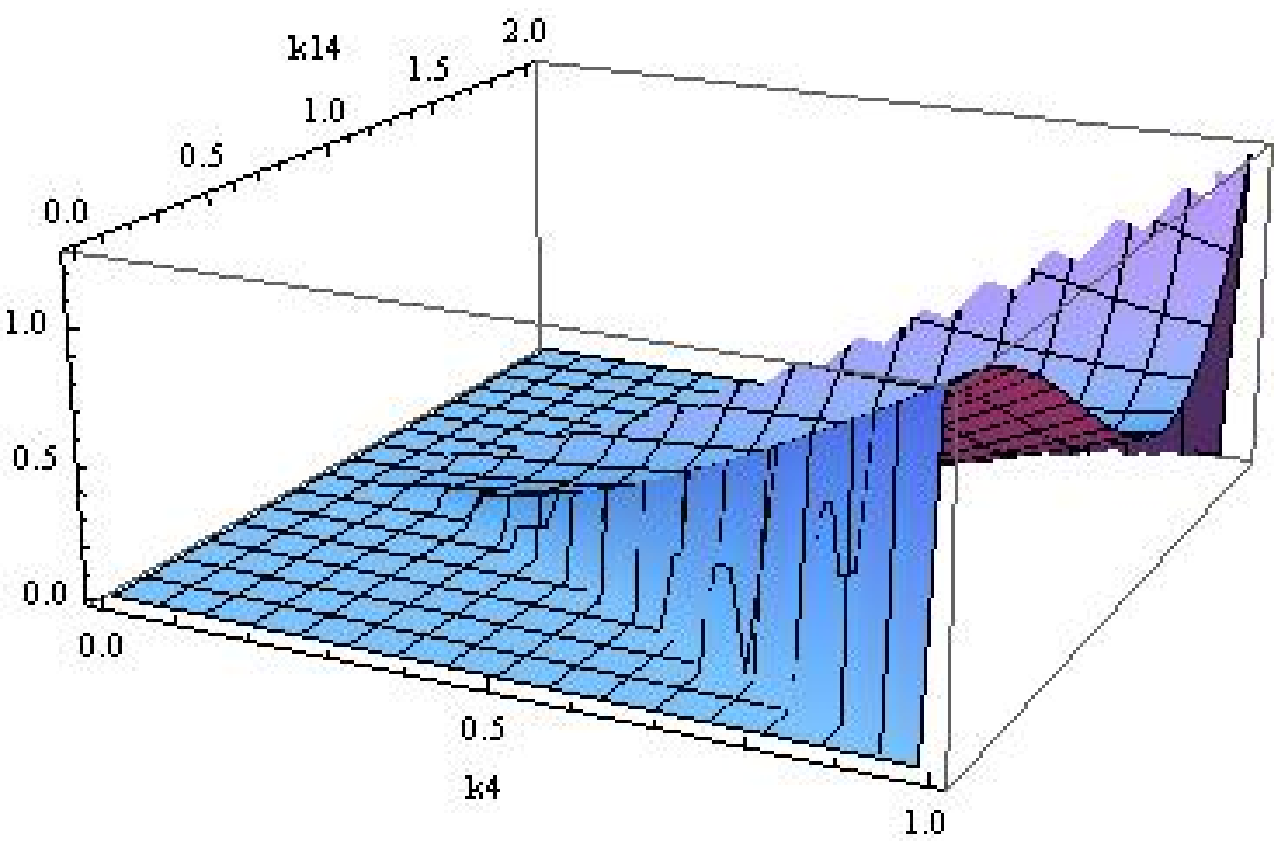}
\caption{The \textit{equilateral} configuration shape for the ${\cal O}_{1}$ operator is presented on the left. It is different from the results in \cite{chen-tris} and very much resembles the plot we obtained for the scalar exchange calculation. Notice that upon performing a change of variables, it is basically identical to the $(\nabla \pi)^4$ interaction term plotted in \cite{muko} (We elaborate further on this point in \textit{Appendix C}).\\ On the right we plotted our findings for the ${\cal O}_{1}$ interaction term in the \textit{folded} configuration. As it will be for the other interactions, this configurations provides no particularly distinctive features that would allow to single out the constributions from the different interaction operators.}
\label{M41}
\end{figure}
\noindent  In fact, all of the terms contributing to the scalar exchange diagram in \cite{chen-tris} give a shape function which in the $k_{12}\rightarrow 0$ is finite. On the contrary, the leading  contact interaction diagram contributions analyzed in \cite{chen-tris} do vanish in this limit.\\
\vspace{1mm}
\noindent \textit{{\small  {\bf *} See \textit{App. C} for a detailed account of the plot of this term first done in \cite{muko} with different variables and results essentially identical to ours despite a simplyfing assumption on our part.}}\\
Note also that if one is to relax the assumption of a Bunch-Davies vacuum, the authors of \cite{chen-tris} showed that this is not true anymore. 

\begin{figure}[hp]
	\includegraphics[scale=0.55]{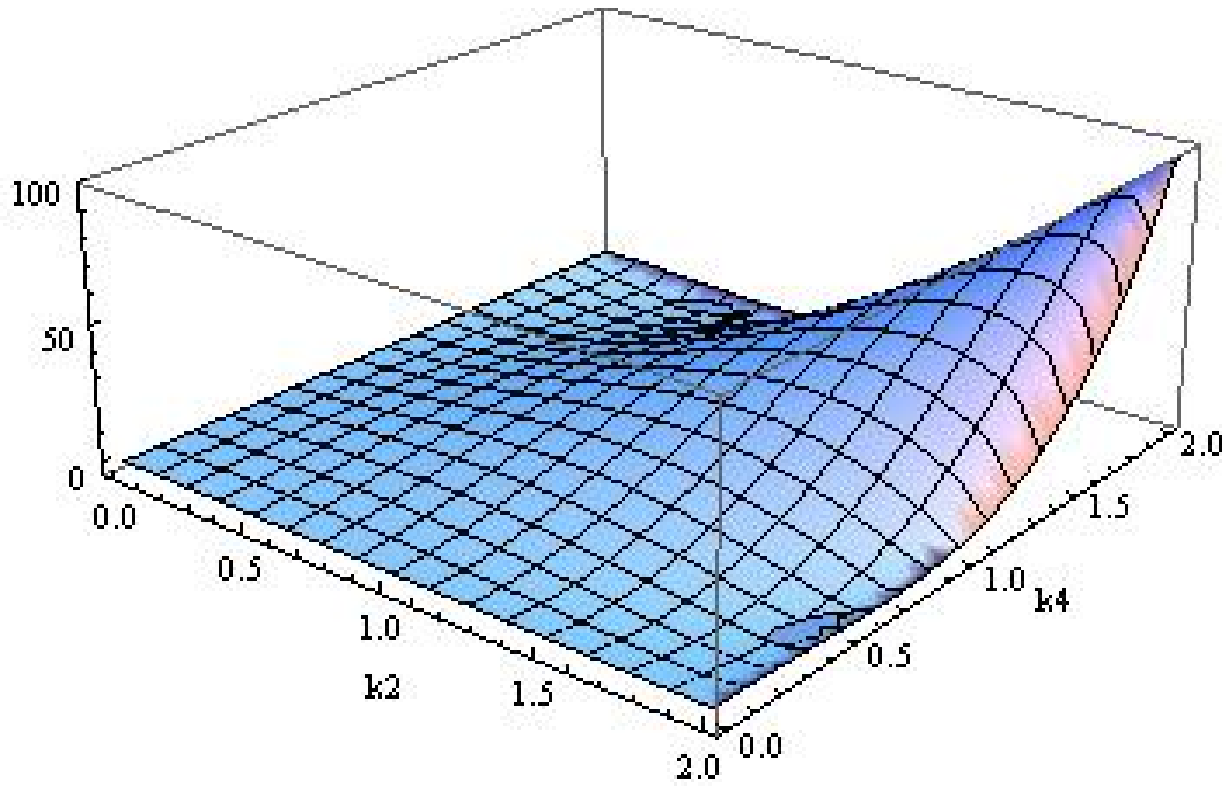}
	\hspace{10mm}
		\includegraphics[scale=0.48]{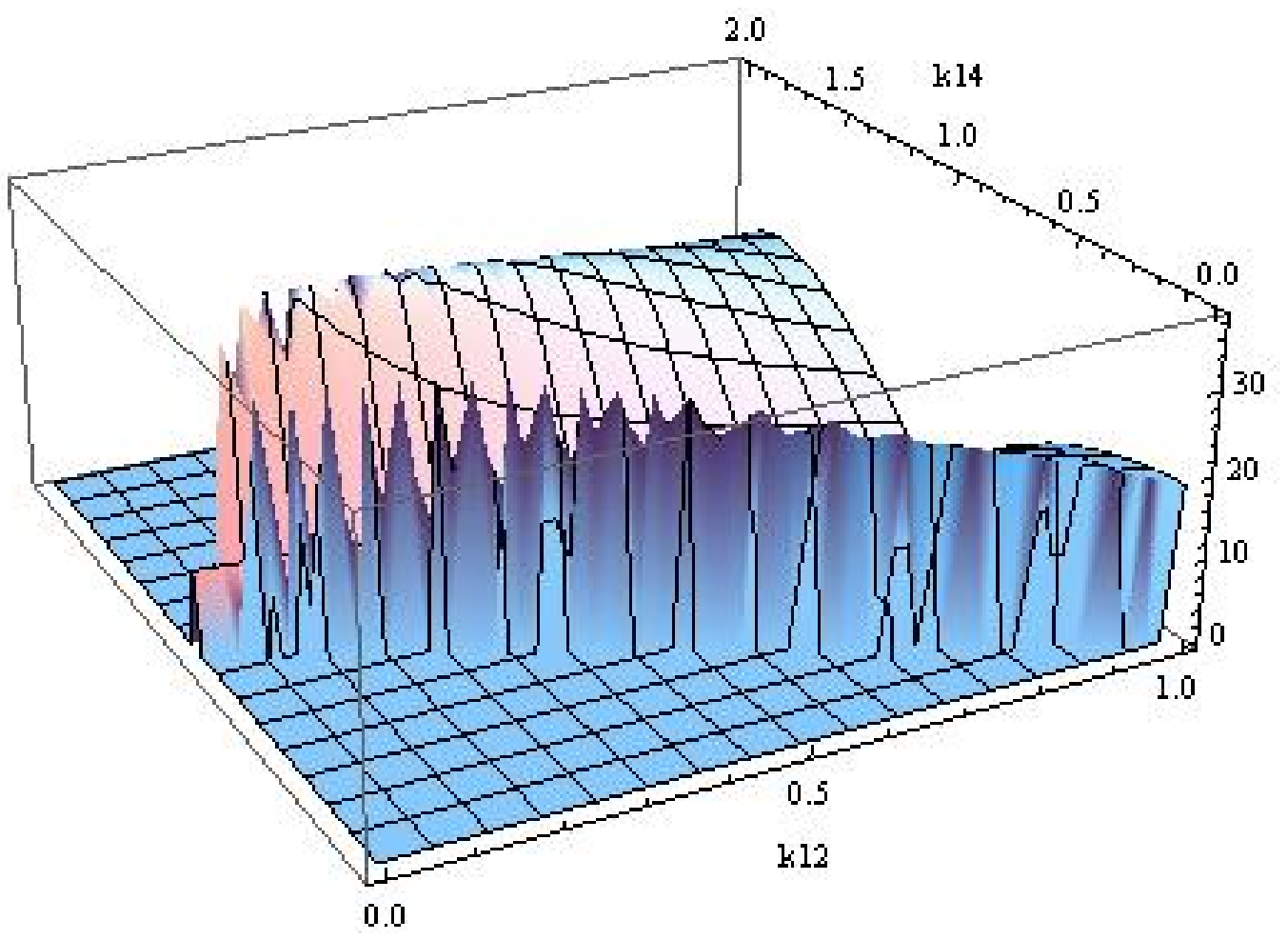}
\caption{The ${\cal O}_{1}$ interaction  \textit{planar} configuration shape  on the left is not exceedingly different from the ones presented in \cite{chen-tris}: it vanishes for $k_2, k_4 \rightarrow 0$, it is peaked for $k_2 = 2 = k_4$. On the $k_2=k_4$ line the shapefunction is convex, rather than concave as for the contact interaction term plotted in \cite{chen-tris}.\\ On the right we plotted the ${\cal O}_{1}$ interaction shape function in the \textit{planar limit double squeezed configuration}.  Here we immediately note an interesting feature: despite this being a contribution to the contact interaction diagram, in the $k_4=k_{12} \rightarrow	0$ it gives a finite, non zero shape function. We comment more on this fact in the text.}
\label{M42}
\end{figure}

\noindent Next, we continue keeping our attention focused on terms which are {\bf S1} \textit{and} {\bf S2} invariant. These include, in terms of their free coefficient, $\bar M_{11}, \bar M_{12}, \bar N_1.. \bar N_5 $. Since they generate shape functions which are qualitatively very similar, we chose to plot just two representative terms in this list. \\

$\bullet$  {\bf $ {\cal O}_{2}=1/6\,\,  \bar M_{11}^2 \, {\dot \pi}^2 \,(\partial_{i}^2 \pi)^2 \,\,/ a^4 $}\\

In the plots below we see that the $\bar M_{11}$-driven term,  generates shapes which are very similar to the ones plotted in \cite{chen-tris} for the DBI-generated term $\sim {\dot \pi}^4$.
\newpage
\begin{figure}[hp]
	\includegraphics[scale=0.53]{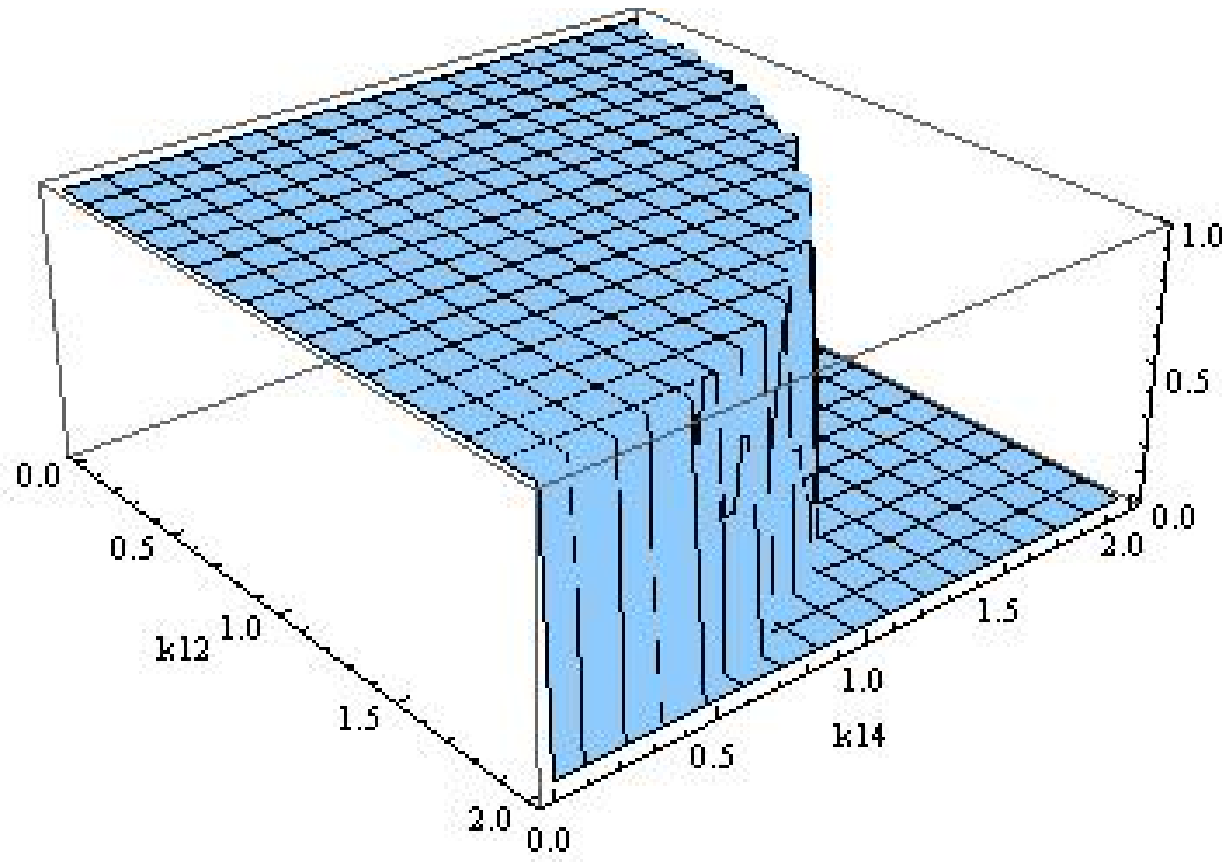}
	\hspace{10mm}
		\includegraphics[scale=0.61]{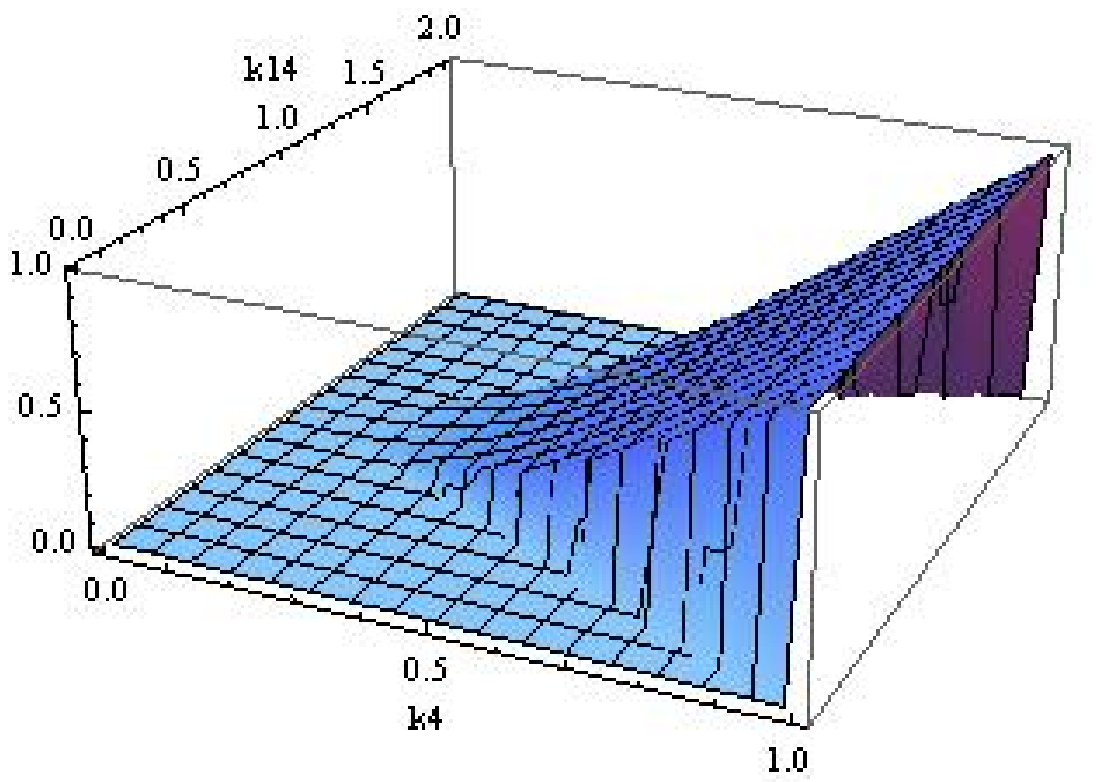}
\caption{The \textit{equilateral} configuration shape is presented on the left for the $ {\cal O}_{2}$ operator. Because of the way the space derivatives are written in Fourier space there's no $k_{12}, k_{14}$ dependence and so one gets a plateau. On the right our findings for $ {\cal O}_{2}$ in the \textit{folded} configuration.}
\label{M111}
\end{figure}

\begin{figure}[hp]
	\includegraphics[scale=0.61]{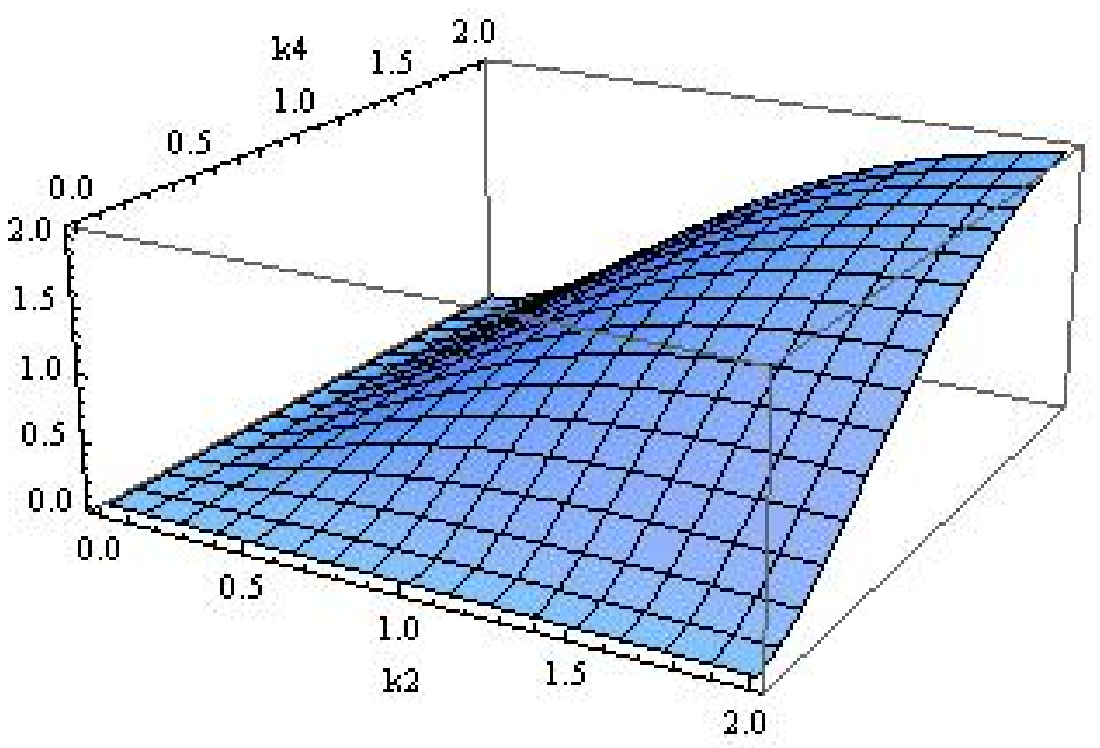}
	\hspace{10mm}
		\includegraphics[scale=0.47]{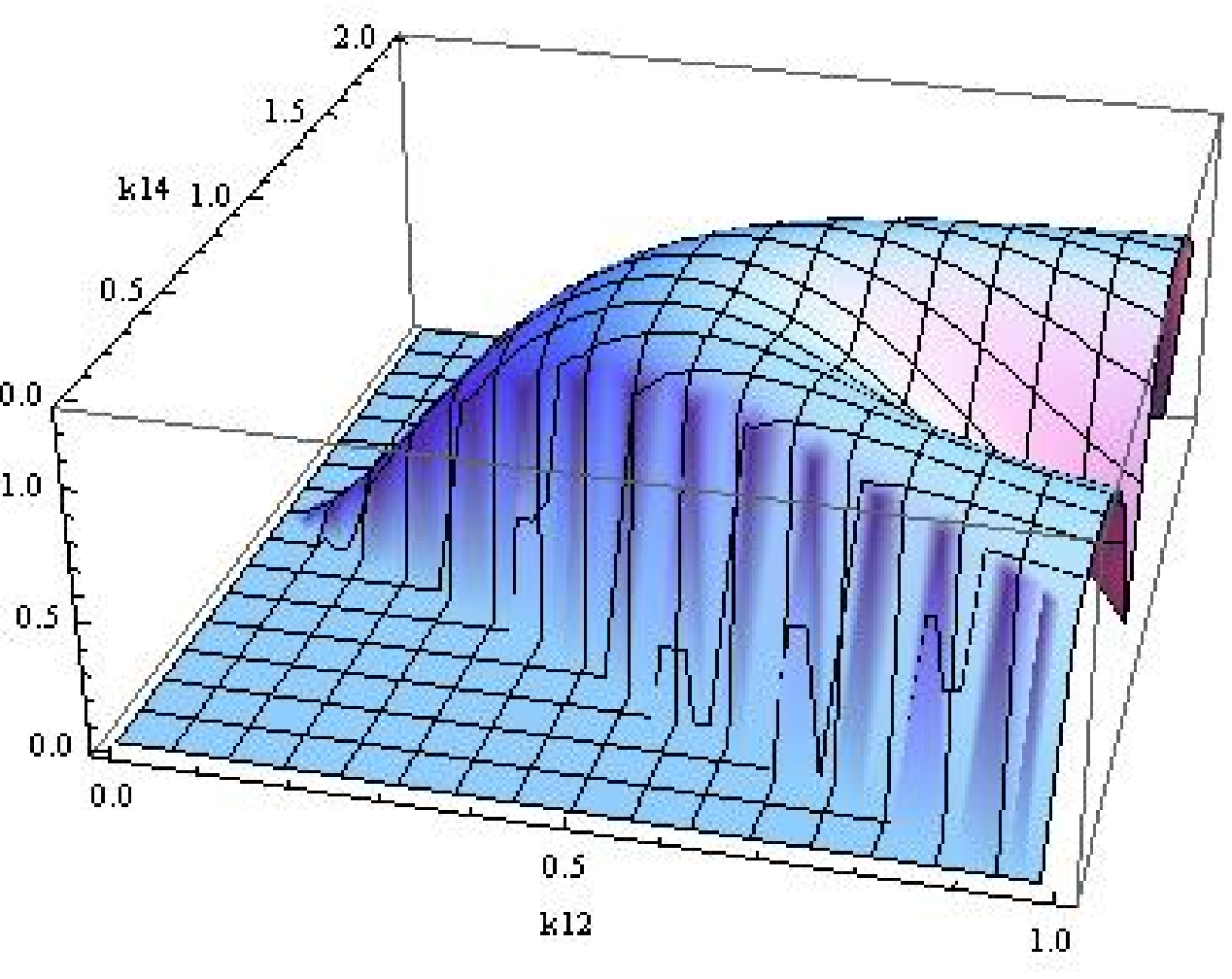}
\caption{The $ {\cal O}_{2}$ \textit{planar} configuration shape  on the left: it vanishes for $k_2, k_4 \rightarrow 0$, it is peaked for $k_2 = 2 = k_4$. On the $k_2=k_4$ line the shapefunction is now concave, just like one would get for the ${\dot \pi}^4$ interaction. \\ On the right we plotted the shape function in the \textit{planar limit double squeezed configuration} associated to the $ {\cal O}_{2}$ interaction term.}
\label{M112}
\end{figure}

\noindent We proceed with the other representative term:\\

$\bullet$  {\bf $ {\cal O}_{3}=1/4!\,\,  \bar N_3 \, \partial_{\rho}^2 \pi \,  \partial_{ij} \pi \, \partial_{jk} \pi \, \partial_{ki} \pi\,\,/ a^8 $}\\

The differences with respect to the $\bar M_{11}$-driven interaction shapes are to be find in the first and the third configuration: in the first configuration they are due to the $k$-dependence of the interaction, on the third configuration $N_3$ gives a plot similar to the one tuned by the $M_2$ coefficient(see Fig.~ \ref{M42}).
\newpage
\begin{figure}[hp]
	\includegraphics[scale=0.62]{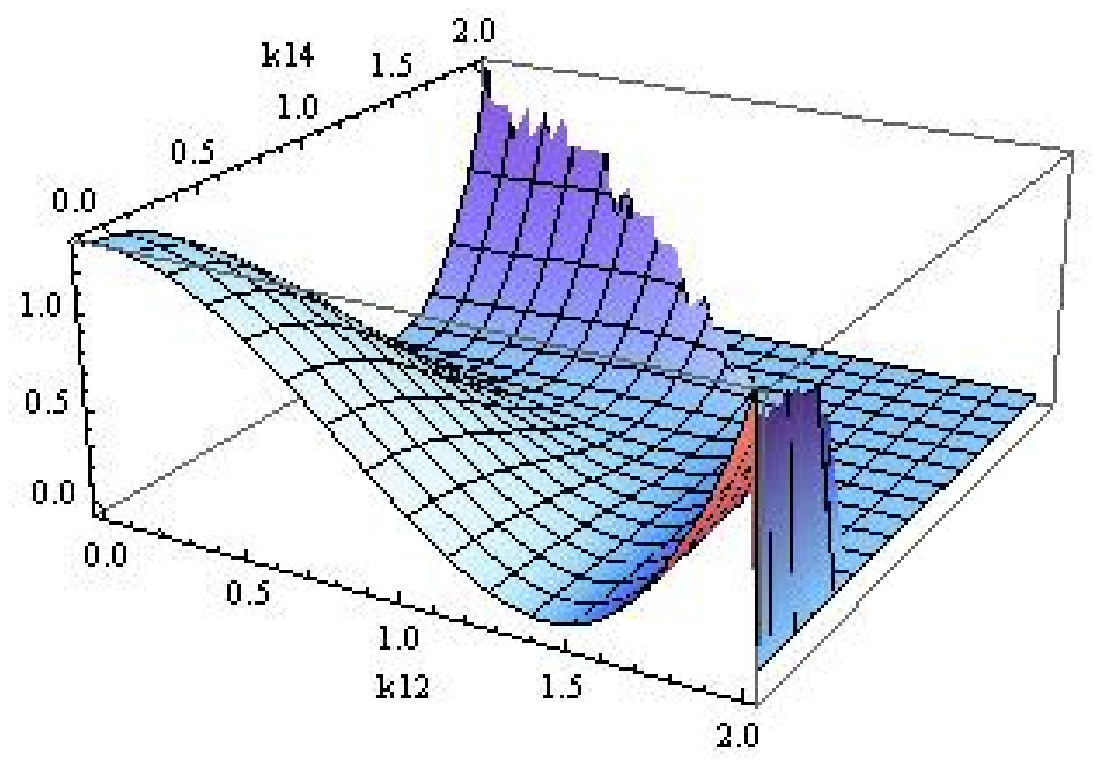}
	\hspace{10mm}
		\includegraphics[scale=0.55]{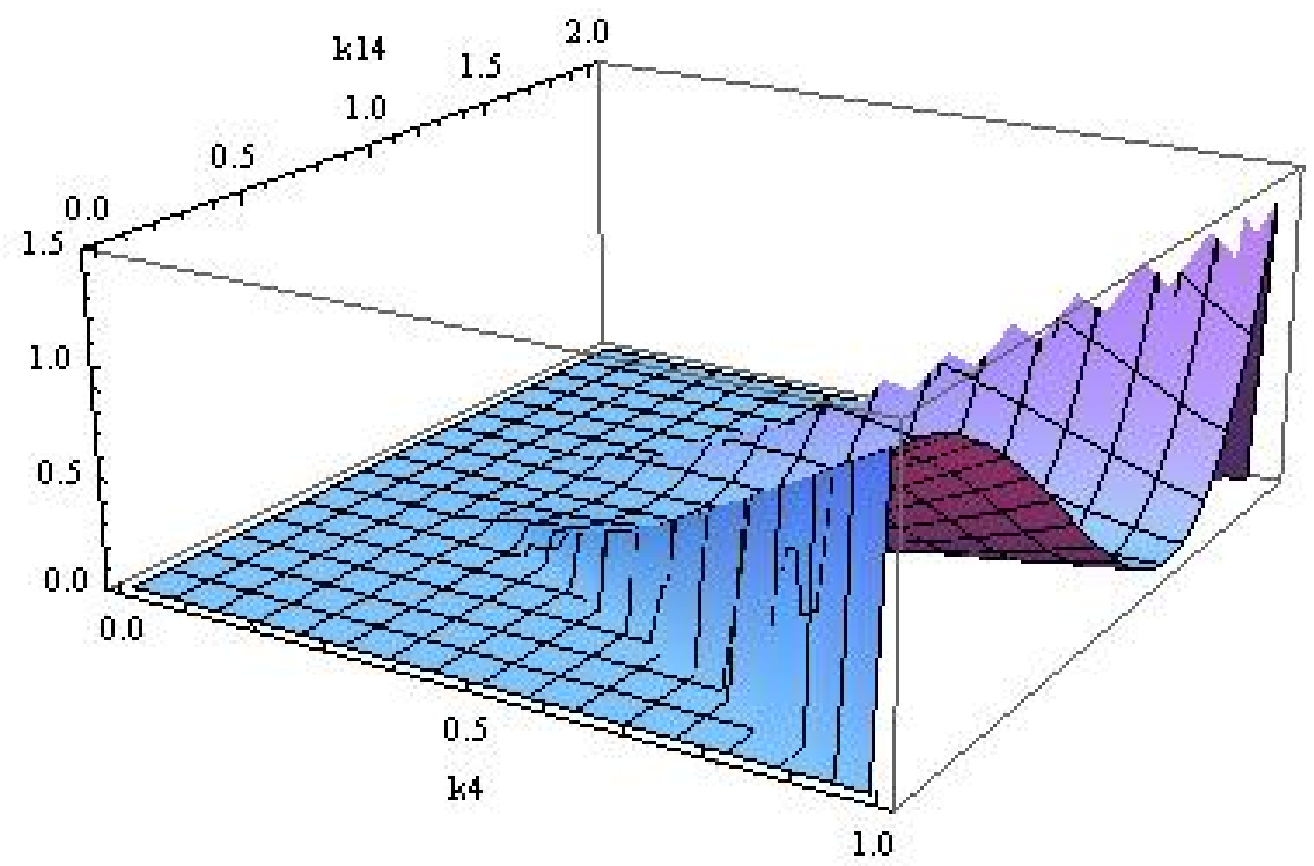}
\caption{The \textit{equilateral} configuration shape for $ {\cal O}_{3}$ is presented on the left. On the right our findings for the $ {\cal O}_{3}$ operator in the \textit{folded} configuration.}
\label{N31}
\end{figure}

\begin{figure}[hp]
	\includegraphics[scale=0.59]{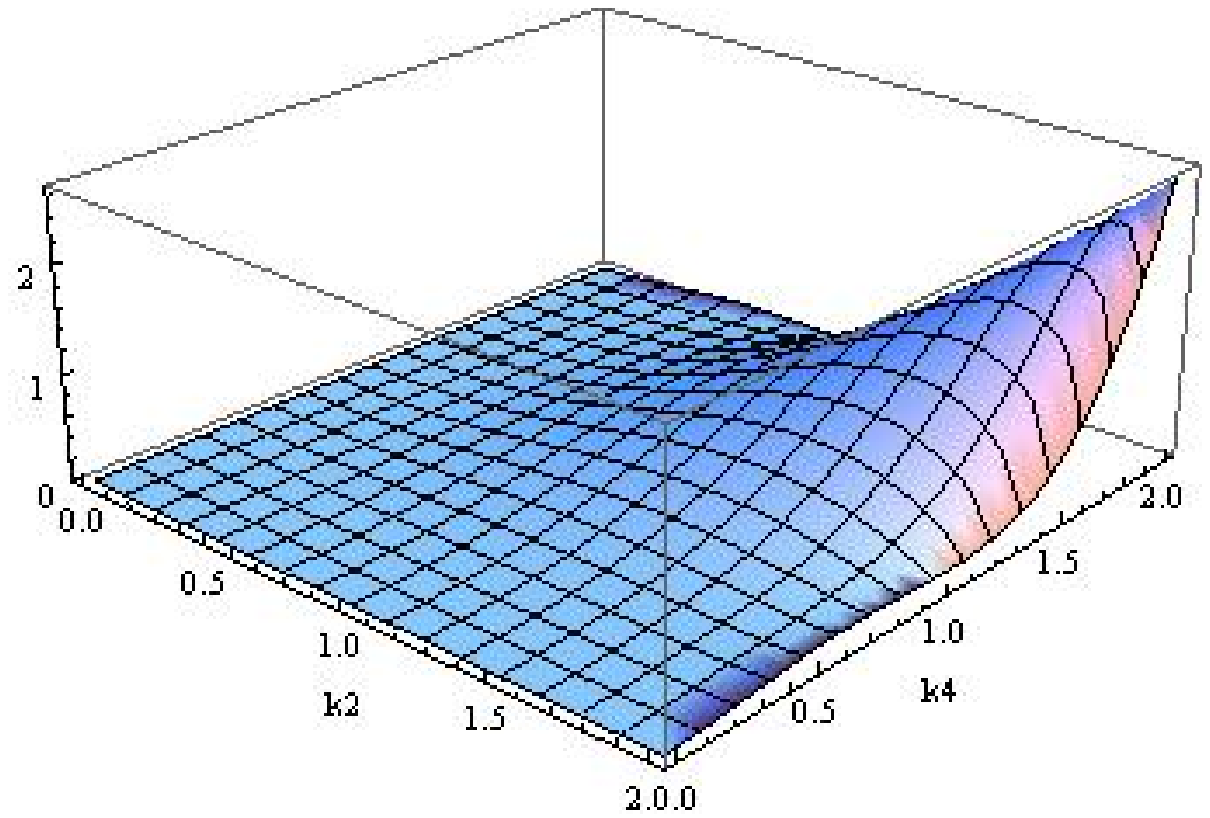}
	\hspace{10mm}
		\includegraphics[scale=0.57]{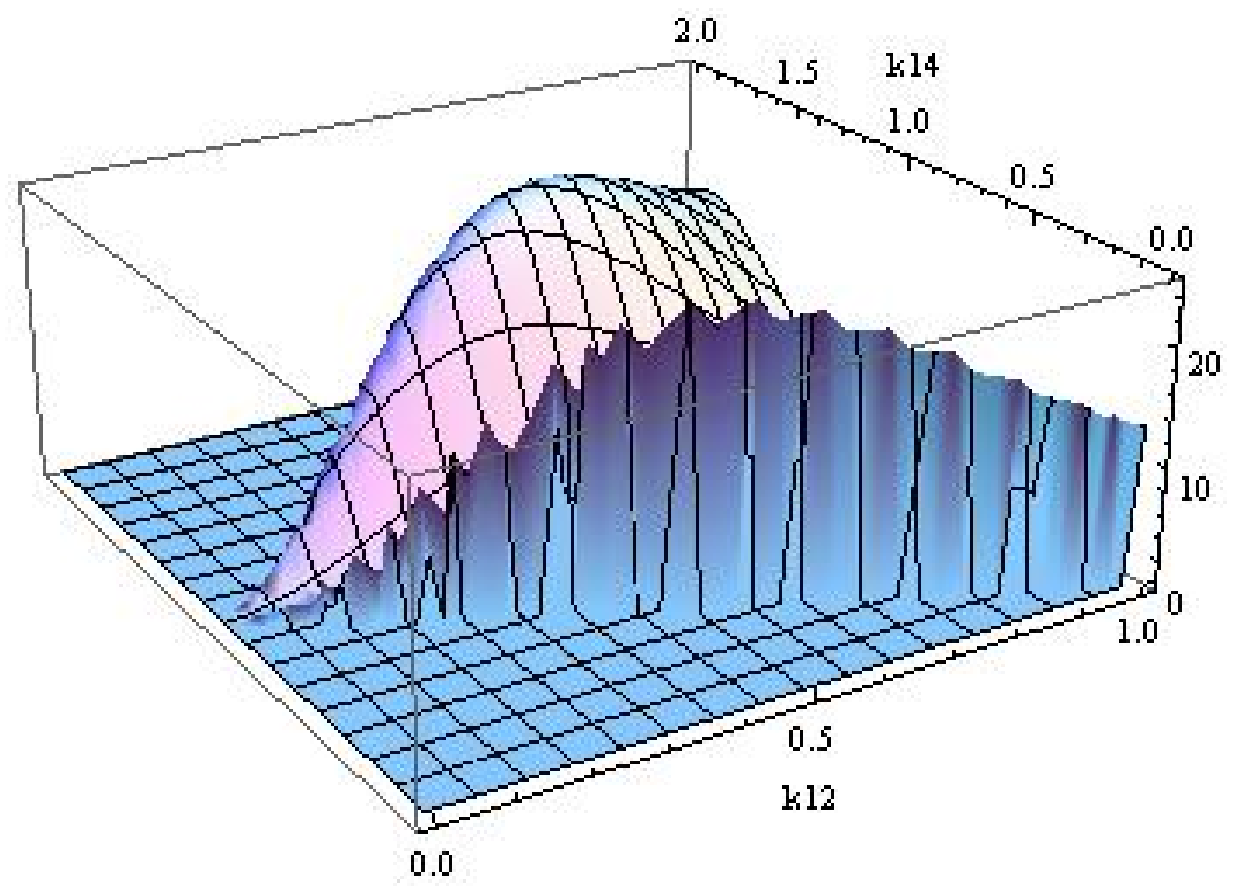}
\caption{The \textit{planar} configuration shape  on the left for $ {\cal O}_{3}$. On the right we plotted the shape function of the $ {\cal O}_{3}$ interaction term in the \textit{planar limit double squeezed configuration}.}
\label{N32}
\end{figure}

\noindent We now turn our attention onto terms which violate one of the symmetries, {\bf S1} in this case. Indeed, we analyze the interaction $ \sim M_2^4 M_3^4 {\dot \pi}^2 (\partial_{i} \pi)^2\,/a^2$\\

$\bullet$  {\bf $ {\cal O}_{4}= (M_2^4 M_3^4)/(2M_2^4 +M_{P}^2 \epsilon H^2 -3\bar M_1^3 H)\,\times  {\dot \pi}^2 \, ( \partial_{i} \pi)^2\, / a^2 $}\\
\newpage
\begin{figure}[hp]
	\includegraphics[scale=0.60]{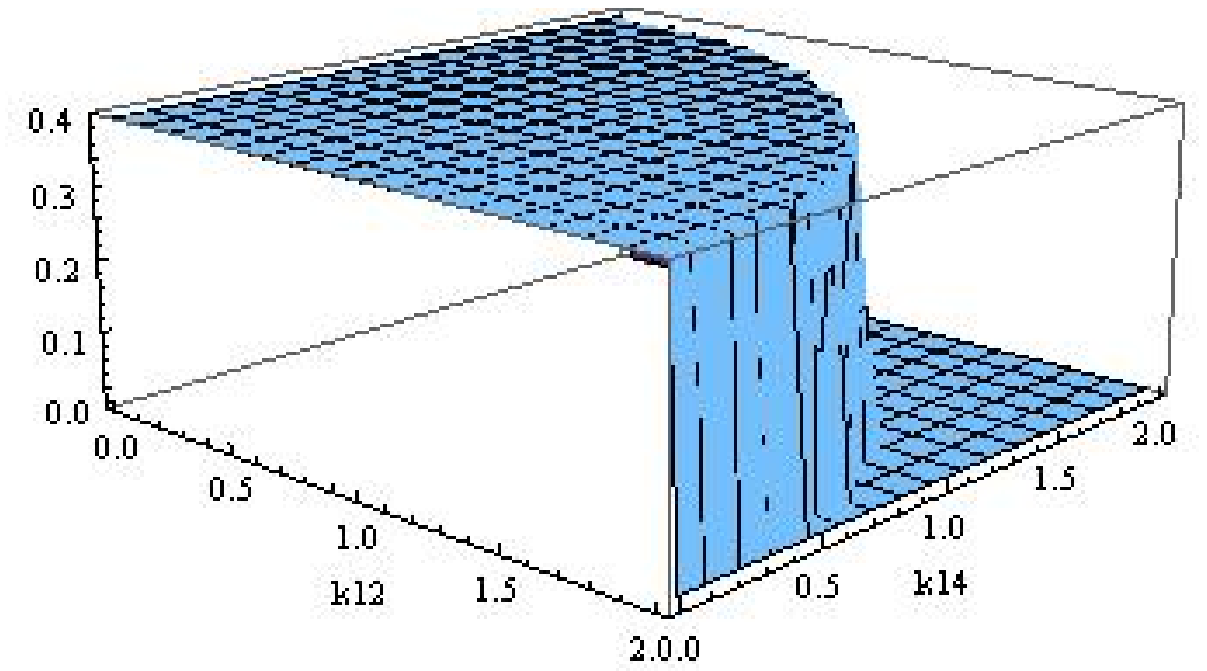}
	\hspace{10mm}
		\includegraphics[scale=0.56]{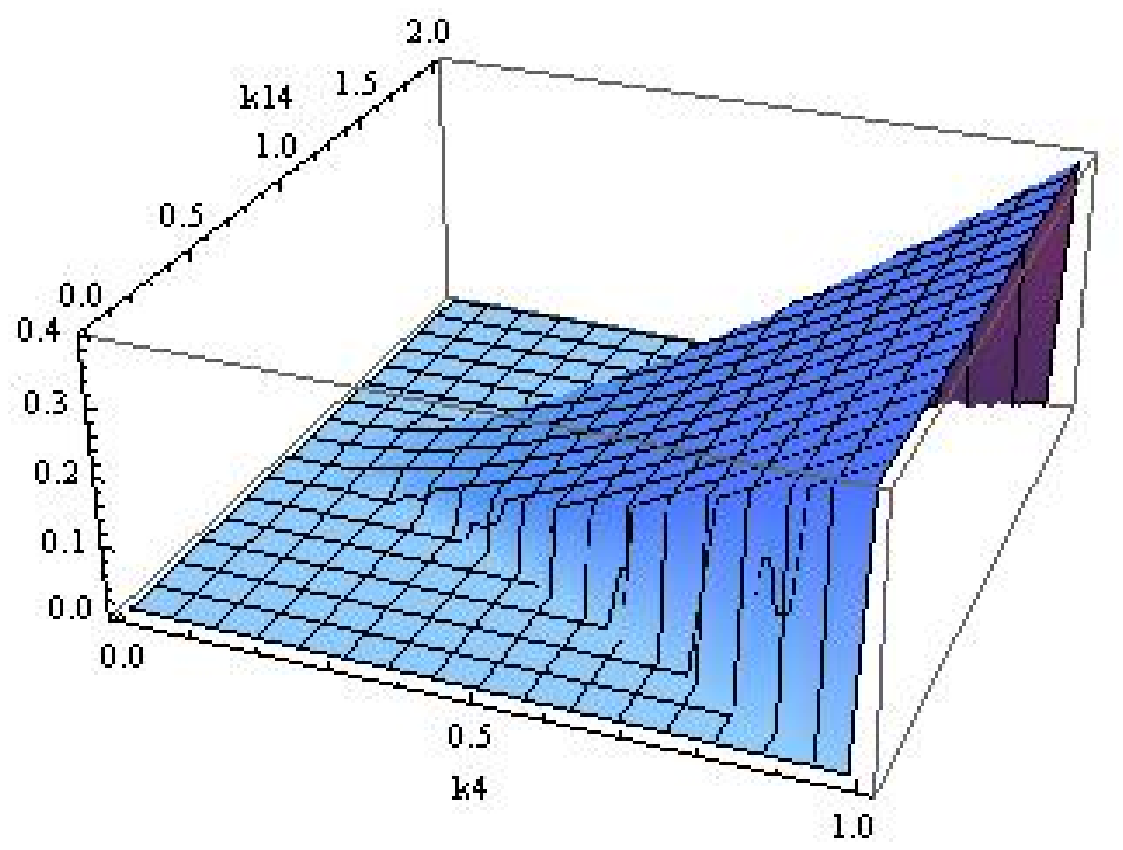}
\caption{The \textit{equilateral} configuration shape for $ {\cal O}_{4}$ is presented on the left. On the right our findings for the \textit{folded} configuration associated to $ {\cal O}_{4}$. Both are very similar to we obtained for the $\bar M_{11}$ in Fig.~\ref{M111} and to what was found for the $M_4^4$-driven interaction in the literature \cite{chen-tris}.}
\label{M231}
\end{figure}

For this interaction term we see the interesting feature presents itself in the fourth configuration where the $k_{12} \rightarrow 0$ limit gives a finite shape function.

\begin{figure}[hp]
	\includegraphics[scale=0.60]{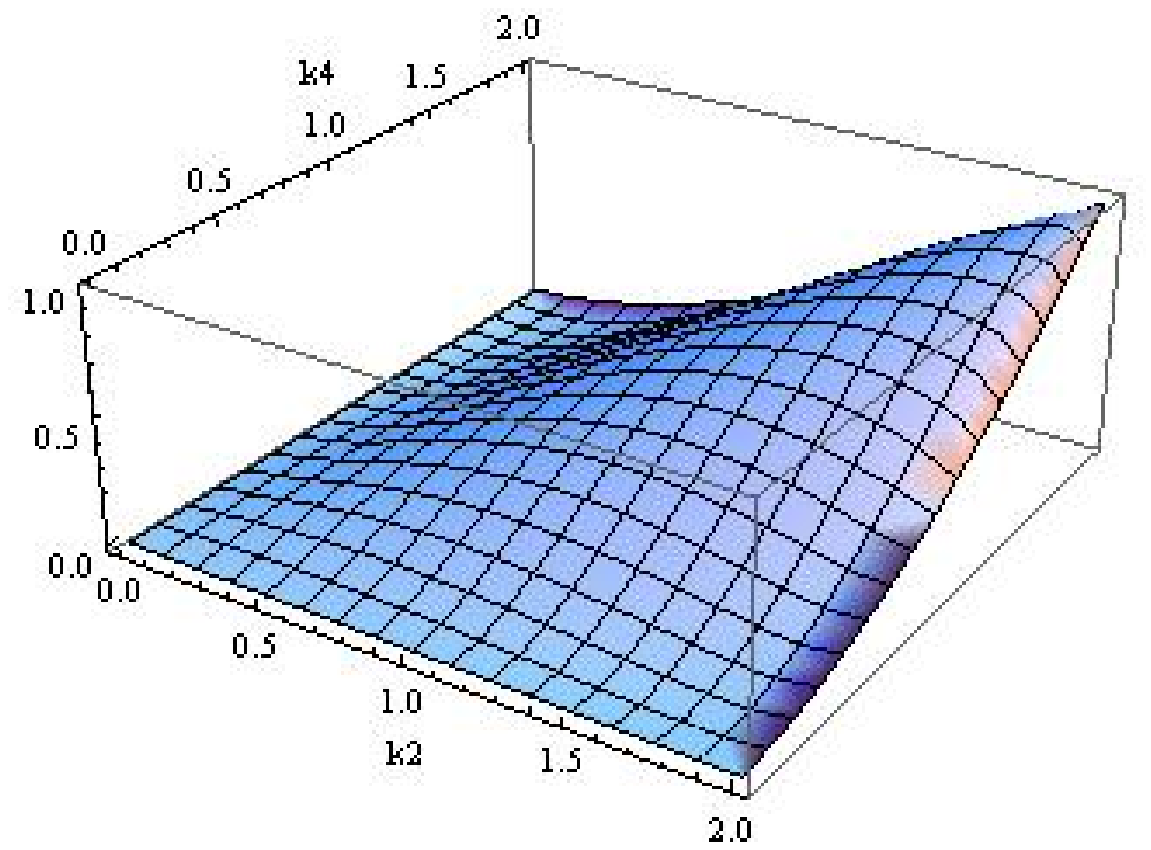}
	\hspace{10mm}
		\includegraphics[scale=0.52]{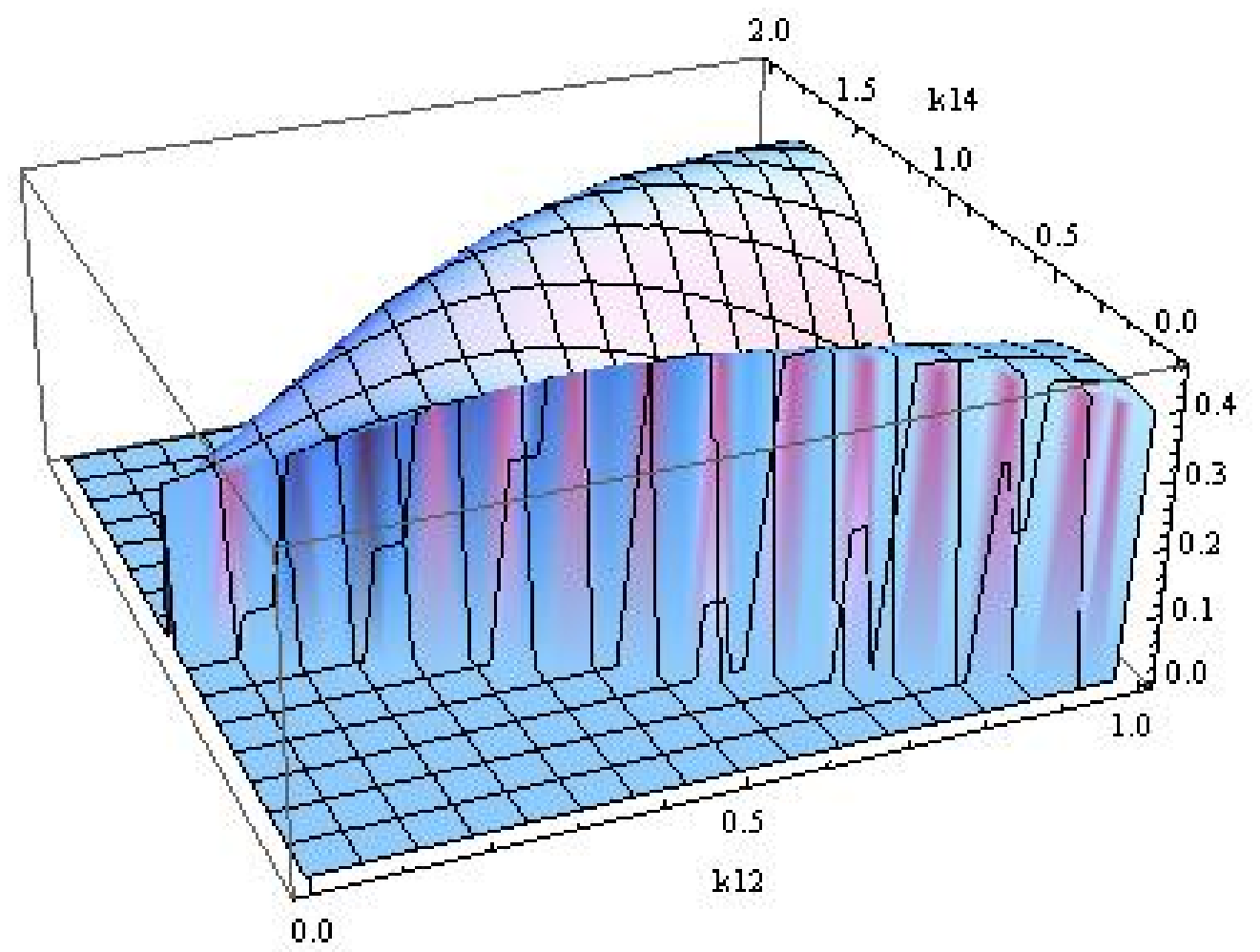}
\caption{The \textit{planar} configuration shape for $ {\cal O}_{4}$  on the left. On the right we plotted the $ {\cal O}_{4}$-generated shape function in the \textit{planar limit double squeezed configuration} which gives again a non-zero and finite shape function even for fourth-order interactions such as the one under scrutiny here.}
\label{M232}
\end{figure}

We now proceed to plot our findings for one more term, precisely the leading fourth-order interaction term among the ones driven by $\bar M_6$. It is clear that whenever this term gives a leading third order contribution (something one can achieve given the freedom on most ${\bf M}_n$'s ), it violates {\bf S1}. {\bf S2} however, is preserved by the leading terms associated to this coefficient at third and fourth order as is clear from Eq.~(\ref{h4}) and \textit{Table 1}.\\

$\bullet$  {\bf $ {\cal O}_{5}=1/6\,\, \bar M_6   (\partial_k \pi)^2 \, ( \partial_{ij} \pi)^2\, / a^6 $}\\
\newpage
\begin{figure}[hp]
	\includegraphics[scale=0.54]{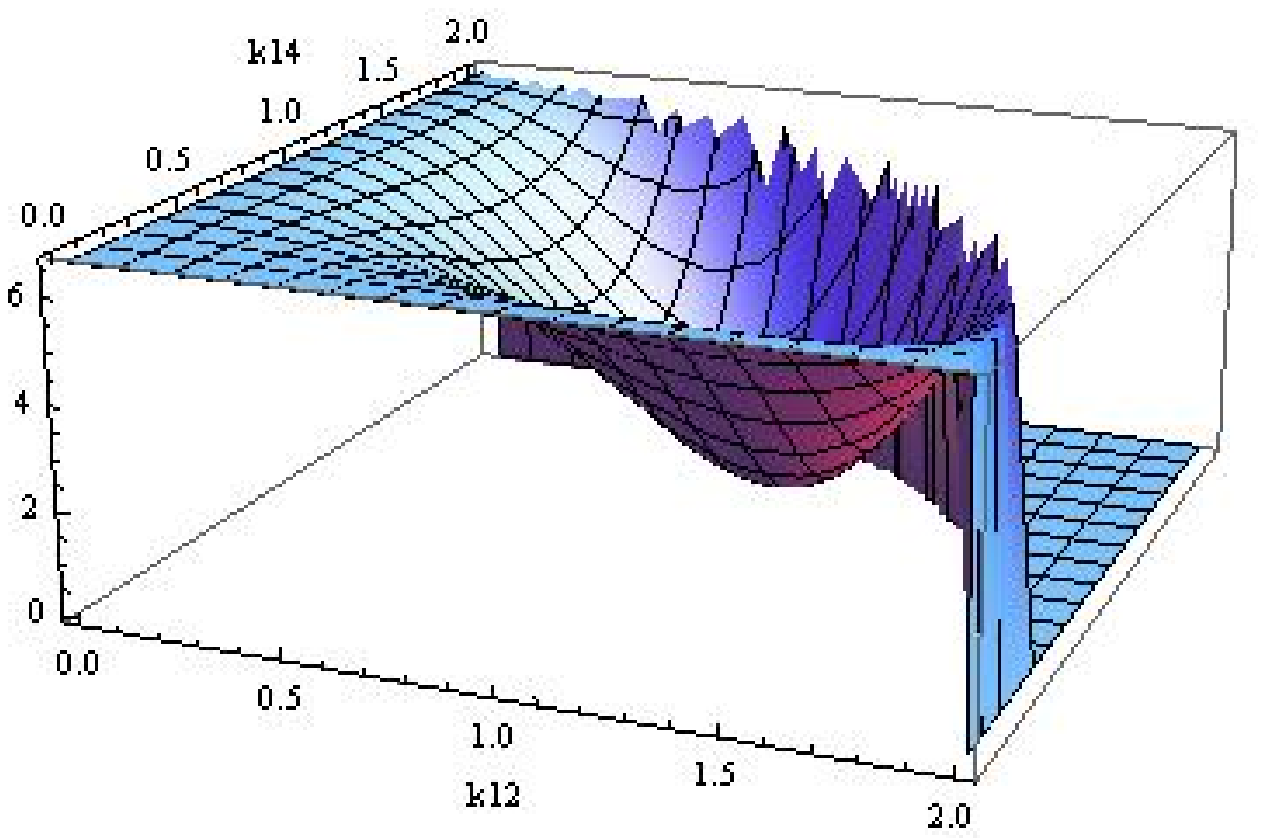}
	\hspace{10mm}
		\includegraphics[scale=0.60]{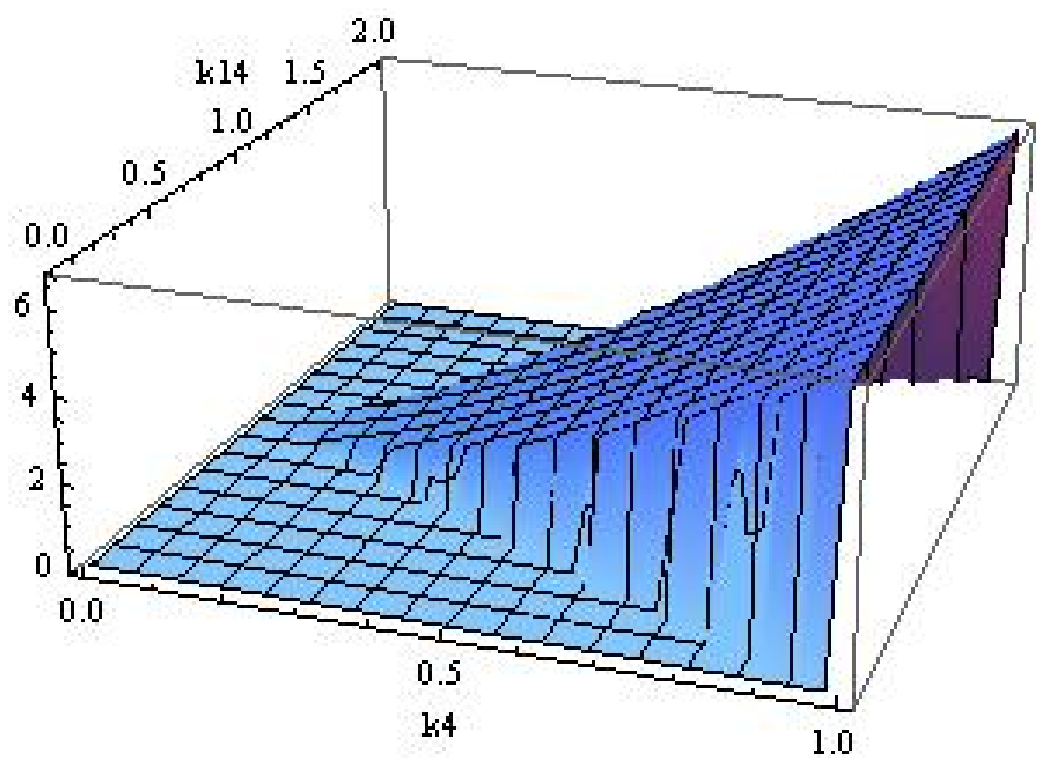}
\caption{The \textit{equilateral} configuration shape for the operator $ {\cal O}_{5}$ is presented on the left: this shape for the plot function has not been seen before in the equilateral configuration.\\ On the right our findings for the $ {\cal O}_{5}$ interaction term in the \textit{folded} configuration.}
\label{M641}
\end{figure}

\begin{figure}[hp]
	\includegraphics[scale=0.64]{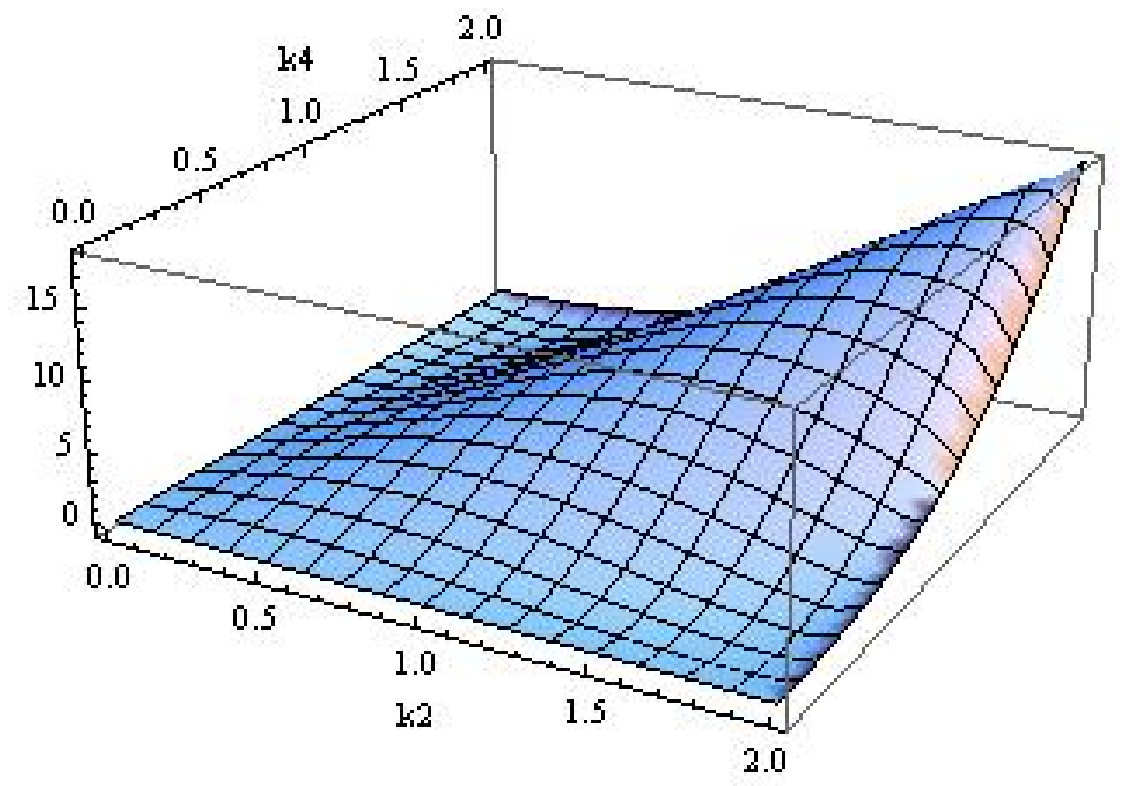}
	\hspace{10mm}
		\includegraphics[scale=0.51]{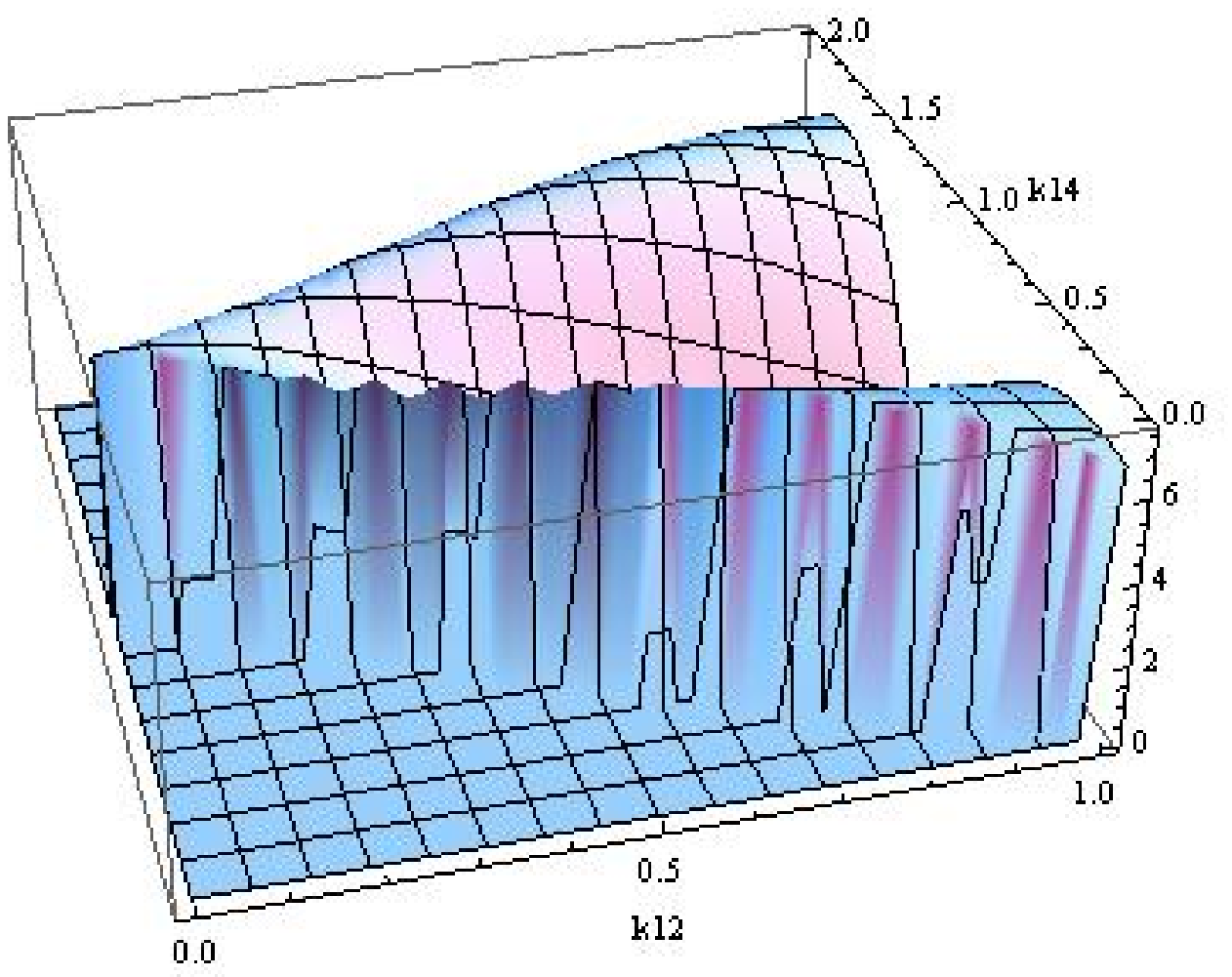}
\caption{The \textit{planar} configuration shape for $ {\cal O}_{5}$  on the left. On the right we plotted the $ {\cal O}_{5}$-generated shape function in the \textit{planar limit double squeezed configuration} }
\label{M642}
\end{figure}

As one can see from Fig.~\ref{M641}, the plot in the equilateral configuration has no analogue in the shapes of \cite{chen-tris, muko} for this configuration. It is somewhat reminiscent of the shape obtained for the ${\dot \pi}^4$ of \cite{chen-tris} but again, we stress it was obtained in a different configuration. The results plotted in Fig.~\ref{M642} show once again that it is not safe in theories more general than DBI to attribuite to the \textit{planar limit double squeezed} configuration the role to provide a distinctive signature in the $k_{12} \rightarrow 0$ limit that would enable one to distinguish between third and fourth-order interaction contributions (see the discussion in \cite{chen-tris}).

\newpage

\section{Conclusions}
In this thesis we have summarized the work done in \cite{b,t,rn}. The raison d'$\hat {e}$tre of these investigations was to study as general as possible a theory of single-field inflation and characterize the various inflationary mechanisms it encompasses mainly according to their non-Gaussian properties.\\
To that aim, we employed the effective field theory approach of \cite{eft08}, which turned out to be a very powerful tool at our disposal. As detailed in \textit{Section 2}, we first solved the equation of motion for the effective Lagrangian and obtained a wavefunction that interpolates between the classical solution of known inflationary models such as $P(X,\phi)$-models and Ghost inflation. Our solution is actually more general, it does not just interpolate between known models. From the wavefunction one easily obtains the power spectrum of curvature perturbations, its tilt and its running. As expected, these quantities now depend on more variables than the usual three slow-roll parameters, there are actually five, generalized slow-roll parameters. We stress here that the two additional degrees of freedom are generated by extrinsic curvature terms in the quadratic Lagrangian; these can be significant in specific cases such as near de Sitter limit, or for small values of the generalized speed of sound or a combination thereof.\\
The natural way to proceed is to tackle non-Gaussianities. We did so for the bispectrum in \textit{Section 3} and \textit{4}. Here we showed that there exist a whole region of the parameters where extrinsic curvature-generated interaction terms play an important, possibly leading, role in determining the amplitude $f_{NL}$. Even more interestingly, a number of such operators independently generate a shape-function which peaks in the \textit{flat} configuration: a feature which is quite uncommon for single-field models of inflation. Prompted by these findings, we also calculated the running of the contribution to $f_{NL}$ given by one of such curvature operators, the one driven by the $\bar M_6$ coefficient: it turns out that the running depends, as it should, on linear combinations of the generalized slow-roll parameters but also on a parameter, $\epsilon_6$, which is a function of the coefficient $\bar M_6$ and its time derivative. We showed that, upon imposing suitable and mild bounds on the coefficients driving the various interactions, the running of $f_{NL}$ is dominated by $\epsilon_6$ without spoiling any of the interesting features on the three and four-point function generated by $\bar M_6$ and without affecting the leading value of the power spectrum. In particular, we showed that the running $n_{NG}$ can be such that $n_{NG} \gg \mathcal{O}(\epsilon,\eta,s)$.\\
Mimicking  the bispectrum analysis, and guided by the requirement of some additional symmetries on the action as
an ordering principle \cite{muko,4pt}, in \textit{Section 7} and \textit{8} we singled out the distinctive features of
the trispectrum one obtains when considering extrinsic curvature-generated terms of a very
general fourth-order Hamiltonian obtained in \textit{Section 6}.  It is important to note that all of these intereactions allow, by construction, for a large trispectrum. Some of them present features which also emerge in DBI-inflation and Ghost inflation \cite{chen-tris, muko}.
We have focused mainly on third and fourth-order interaction terms which have not been analyzed before and whose analysis reveals novel interesting effects. We were able to show that, unlike what happens in $P(X,\phi)$ models, the analysis of the double squeezed configuration cannot give a clear cut clue as to what kind of signal comes from leading third-order terms as opposed to fourth-order terms in perturbations. 
We found many interactions generating a shape in the equilateral configuration that mimics the behaviour of the ghost interaction term ( i.e. $(\nabla \pi)^4$ ) shape function first plotted in \cite{muko}, which is quite different from the shapes of the DBI model (we also extended the Ghost inflation plots of \cite{muko} to three other configurations).
Triggered by the fact several interactions generated an interesting flat shape for the bispectrum, we considered the effect of one of those terms (for consistency we chose again the one controlled by the $\bar M_6$ coefficient) for the trispectrum. We calculated and plotted the contributions of this term to the scalar exchange and contact interaction diagram: a shape-function which has not been found before emerged in the equilateral configuration for the contact interaction contribution.\\ Both at cubic and quartic order in perturbations we again and again came accross the realization of the following feature: a shape function which in general single field inflation models is only obtained either by employing a linear combination of operators (as far as the Bispectrum
is concerned) or relaxing the Bunch-Davies vacuum requirement for the theory, quite naturally (Bunch-Davies vacuum, no linear combinations) arises in more general setups as the one employed here. Furthermore, it does so when considering several and independent interaction terms.\\
All the above results clearly point to at least two important facts: first, the effective field theory approach has proven very fruitful in addressing the need to describe inflationary models from a unifying general perspective. Second, in the quest for predictions on important cosmological observables it is crucial to consider the effect of extrinsic curvature-generated interaction terms in the inflationary action.\\
From here one might proceed in several directions. It is true that extrinsic curvature terms have shown several interesting and distinctive features. Is there a UV-safe theory that comprises a phase of expansion of the universe described by higher derivative interactions? Something of this sort is described for example in recent literature \cite{galileon1,galileon2}.
All the effective field theory machinery has been used here assuming a Bunch-Davies vacuum for the theory. One might well ask what would happen if we were to start from an excited state, much in the spirit of \cite{holman}. Another possibility is to expand on the effective approach employed here: removing the shift symmetry requirement, it should be possible to describe from a very general perspective the so called \textit{resonance models} which have generated quite an interest$^{*}$ 
in the recent literature \cite{chen-res}-\cite{pajer2}. 

\vspace{2.3cm}
\noindent {\small {\bf *} These studies have revealed a somewhat unexpected and intriguing fact: small periodic features in an inflationary model can have important consequences on the non-Gaussianities of such model.}

\newpage

\newpage

\section{Appendix A}
\subsection*{Explicit expression for the slow-roll parameter $\epsilon_{\Gamma}$}
We give here an explicit expression for the time dependence of the slow-roll parameters which we called $\epsilon_{\Gamma}$. This quantity if first written in terms of the $M, \bar M$ parameters:
\bea
\fl \Gamma[t]= \gamma \left[\frac{5}{4}+\frac{H^2 M_P^2 \epsilon +\frac{1}{2} M_1{}^4}{\left(H^2 M_P^2 \epsilon +2 M_2{}^4+3 M_1{}^4\right) \left(\frac{4 \left(H^2 M_P^2 \epsilon +\frac{1}{2} M_1{}^4\right)}{H^2 M_P^2 \epsilon +2 M_2{}^4+3 M_1{}^4}-4 i \sqrt{\frac{M_0{}^4}{2 \left(H^2 M_P^2 \epsilon +2 M_2{}^4+3 M_1{}^4\right)}}\right)}\right]\times\nonumber\\
\fl \gamma\left[\frac{5}{4}+\frac{H^2 M_P^2 \epsilon +\frac{1}{2} M_1{}^4}{\left(H^2 M_P^2 \epsilon +2 M_2{}^4+3 M_1{}^4\right) \left(\frac{4 \left(H^2 M_P^2 \epsilon +\frac{1}{2} M_1{}^4\right)}{H^2 M_P^2 \epsilon +2 M_2{}^4+3 M_1{}^4}+4 i \sqrt{\frac{M_0{}^4}{2 \left(H^2 M_P^2 \epsilon +2 M_2{}^4+3 M_1{}^4\right)}}\right)}\right],
\eea
where $\gamma[t]$ is the Euler function. From here one calculates the quantity $\epsilon_{\Gamma}=-\frac{\dot \Gamma}{H \Gamma }$, obtaining:
\bea
\fl \frac{\dot \Gamma}{H \Gamma}= \left(i \left(Poly\Gamma\left[0,\frac{5}{4}+\left(4-\frac{8 i \left(H^2 M_P^2 \epsilon +2 M_2{}^4+3 M_1{}^4\right) \sqrt{\frac{M_0{}^4}{2 H^2 M_P^2 \epsilon +4 M_2{}^4+6 M_1{}^4}}}{2 H^2 M_P^2 \epsilon +M_1{}^4}\right)^{-1}\right]\times \right. \right.\nonumber \\
\fl \left. \left.  \left(i \left(2 H^2 M_P^2 \epsilon +M_1{}^4\right)-2 \left(H^2 M_P^2 \epsilon +2 M_2{}^4+3 M_1{}^4\right) \sqrt{\frac{M_0{}^4}{2 H^2 M_P^2 \epsilon +4 M_2{}^4+6 M_1{}^4}}\right){}^2 \right. \right. \nonumber\\
\fl -Poly\Gamma\left[0,\frac{5}{4}+\left(4+\frac{8 i \left(H^2 M_P^2 \epsilon +2 M_2{}^4+3 M_1{}^4\right) \sqrt{\frac{M_0{}^4}{2 H^2 M_P^2 \epsilon +4 M_2{}^4+6 M_1{}^4}}}{2 H^2 M_P^2 \epsilon +M_1{}^4}\right)^{-1}\right]\times \nonumber\\
\fl \left. \left(i \left(2 H^2 M_P^2 \epsilon +M_1{}^4\right)+2 \left(H^2 M_P^2 \epsilon +2 M_2{}^4+3 M_1{}^4\right) \sqrt{\frac{M_0{}^4}{2 H^2 M_P^2 \epsilon +4 M_2{}^4+6 M_1{}^4}}\right){}^2\right)\times \nonumber\\
\fl \left(2 H^4 M_P^4 \epsilon ^2 \frac{d}{dt}\left(\frac{M_0{}^4}{2 H^2 M_P^2 \epsilon +4 M_2{}^4+6 M_1{}^4}\right)+4 H^2 M_P^2 \epsilon  M_2{}^4  \frac{d}{dt}\left(\frac{M_0{}^4}{2 H^2 M_P^2 \epsilon +4 M_2{}^4+6 M_1{}^4}\right) \right.\nonumber \\
\fl +7 H^2 M_P^2 \epsilon  M_1{}^4 \frac{d}{dt}\left(\frac{M_0{}^4}{2 H^2 M_P^2 \epsilon +4 M_2{}^4+6 M_1{}^4}\right) +2 M_2{}^4 M_1{}^4 \frac{d}{dt}\left(\frac{M_0{}^4}{2 H^2 M_P^2 \epsilon +4 M_2{}^4+6 M_1{}^4}\right)\nonumber\\ 
\fl +3 M_1{}^8 \frac{d}{dt} \left(\frac{M_0{}^4}{2 H^2 M_P^2 \epsilon +4 M_2{}^4+6 M_1{}^4}\right)-16 H M_P^2 \epsilon  M_2{}^4 \dot H \frac{M_0{}^4}{2 H^2 M_P^2 \epsilon +4 M_2{}^4+6 M_1{}^4}\nonumber \\
\fl -20 H M_P^2 \epsilon  M_1{}^4 \dot H \frac{M_0{}^4}{2 H^2 M_P^2 \epsilon +4 M_2{}^4+6 M_1{}^4}-8 H^2 M_P^2 M_2{}^4 \dot  \epsilon  \frac{M_0{}^4}{2 H^2 M_P^2 \epsilon +4 M_2{}^4+6 M_1{}^4}   \nonumber \\
\fl -10 H^2 M_P^2 M_1{}^4 \dot \epsilon  \frac{M_0{}^4}{2 H^2 M_P^2 \epsilon +4 M_2{}^4+6 M_1{}^4}+32 H^2 M_P^2 \epsilon  M_2{}^3 \dot M_2 \frac{M_0{}^4}{2 H^2 M_P^2 \epsilon +4 M_2{}^4+6 M_1{}^4} \nonumber \\
\fl +16 M_2{}^3 M_1{}^4 \dot M_2 \frac{M_0{}^4}{2 H^2 M_P^2 \epsilon +4 M_2{}^4+6 M_1{}^4}+40 H^2 M_P^2 \epsilon  M_1{}^3 \dot M_1 \frac{M_0{}^4}{2 H^2 M_P^2 \epsilon +4 M_2{}^4+6 M_1{}^4}      \nonumber \\
\fl -16 M_2{}^4 M_1{}^3 \dot M_1 \frac{M_0{}^4}{2 H^2 M_P^2 \epsilon +4 M_2{}^4+6 M_1{}^4}  +\left(2 H^2 M_P^2 \epsilon +M_1{}^4\right) \left(H^2 M_P^2 \epsilon +2 M_2{}^4+3 M_1{}^4\right)\times \nonumber \\
\fl \left. \left. \frac{M_0{}^3 \left(4 \left(H^2 M_P^2 \epsilon +2 M_2{}^4+3 M_1{}^4\right) \dot M_0-M_0 \left(H M_P^2 \left(2 \epsilon  \dot H+H \dot \epsilon \right)+8 M_2{}^3 \dot M_2+12 M_1{}^3 \dot M_1\right)\right)}{2 \left(H^2 M_P^2 \epsilon +2 M_2{}^4+3 M_1{}^4\right){}^2}\right)\right) / \nonumber\\
\fl \left( H \sqrt{\frac{2 M_0{}^4}{2 H^2 M_P^2 \epsilon +4 M_2{}^4+6 M_1{}^4}} \left(\left(2 H^2 M_P^2 \epsilon +M_1{}^4\right){}^4 +8 \left(2 H^2 M_P^2 \epsilon +M_1{}^4\right){}^2\times \right. \right.\nonumber\\
\fl \left(H^2 M_P^2 \epsilon +2 M_2{}^4+3 M_1{}^4\right){}^2 \frac{M_0{}^4}{2 H^2 M_P^2 \epsilon +4 M_2{}^4+6 M_1{}^4}+16 \left(H^2 M_P^2 \epsilon +2 M_2{}^4+3 M_1{}^4\right){}^4 \times \nonumber \\
\fl \left. \left. \frac{M_0{}^4}{2 H^2 M_P^2 \epsilon +4 M_2{}^4+6 M_1{}^4}{}^2\right)\right).
\eea

\subsection*{Explicit expression for the $\Theta$ functions of Eq.~(\ref{run}).}

\bea
\fl \Theta= \frac{7 H^2 M_P^2 \epsilon+8 \left(2 M_2{}^4+3 M_1{}^4\right) }{2  \left(H^2 M_P^2 \epsilon+2 M_2{}^4+3 M_1{}^4\right)}+\nonumber\\ 
\fl  \frac{8 H^4 M_P^4 \epsilon^2-16 H^2 M_P^2 \epsilon \left(2 M_2{}^4+3 M_1{}^4\right)+H^2 M_P^2 \epsilon  \left(-14 H^2 M_P^2 \epsilon +8 M_2{}^4+15 M_1{}^4\right) }{2  \left(H^2 M_P^2 \epsilon+2 M_2{}^4+3 M_1{}^4\right) \left(2 H^2 M_P^2 \epsilon+M_1{}^4+M_0{}^2 \sqrt{2 H^2 M_P^2 \epsilon +4 M_2{}^4+6 M_1{}^4}\right)};\nonumber\\
\fl \Theta_{\eta}=  \frac{ H^2 M_P^2  \epsilon }{4 \left(H^2 M_P^2 \epsilon +2 M_2{}^4+3 M_1{}^4\right)}+\nonumber\\
\fl \frac{ \left(6 H^2 M_P^2 \epsilon +24 M_2{}^4+33 M_1{}^4\right)H^2 M_P^2 \epsilon }{4 \left(H^2 M_P^2 \epsilon +2 M_2{}^4+3 M_1{}^4\right) \left(2 H^2 M_P^2 \epsilon +M_1{}^4+M_0{}^2 \sqrt{2 H^2 M_P^2 \epsilon +4 M_2{}^4+6 M_1{}^4}\right)};\nonumber\\
\fl\Theta_{2}= \frac{ 2 M_2{}^4}{ \left(H^2 M_P^2 \epsilon +2 M_2{}^4+3 M_1{}^4\right)}+\nonumber\\
\fl- \frac{ \left(6 H^2 M_P^2 \epsilon +3 M_1{}^4\right)2 M_2{}^4 }{ \left(H^2 M_P^2 \epsilon +2 M_2{}^4+3 M_1{}^4\right) \left(2 H^2 M_P^2 \epsilon +M_1{}^4+M_0{}^2 \sqrt{2 H^2 M_P^2 \epsilon +4 M_2{}^4+6 M_1{}^4}\right)};\nonumber \\
\fl \Theta_{1}=  \frac{3 M_1{}^4 }{ \left(H^2 M_P^2 \epsilon +2 M_2{}^4+3 M_1{}^4\right) }+\nonumber\\
\fl \frac{\left(+ 3M_1{}^4-4 H^2 M_P^2 \epsilon +4 M_2{}^4\right)3 M_1{}^4 }{ \left(H^2 M_P^2 \epsilon +2 M_2{}^4+3 M_1{}^4\right) \left(2 H^2 M_P^2 \epsilon +M_1{}^4+M_0{}^2 \sqrt{2 H^2 M_P^2 \epsilon +4 M_2{}^4+6 M_1{}^4}\right)}; \nonumber \\ 
\fl \Theta_{0}=  \frac{6 M_0^2 }{ \left(2 M_0{}^2+\sqrt{2} \left(2 H^2 M_P^2 \epsilon +M_1{}^4\right) \sqrt{\frac{1}{H^2 M_P^2 \epsilon +2 M_2{}^4+3 M_1{}^4}}\right)}. 
\eea
\subsection*{Explicit expression for variables  in Eq.~(\ref{eoma}),(\ref{eomb}).}
\bea
\fl \frac{f^{''}}{f} = 2 a^2 H^2 \left(1-\frac{\epsilon}{2} +\frac{6 M_2{}^3 \dot M_2}{H \left(2 M_2{}^4+3 M_1{}^4-M_P^2 \dot H\right)}+\frac{9 M_1{}^3 \dot M_1}{H \left(-2 M_2{}^4-3 M_1{}^4+M_P^2 \dot H\right)} \right.\nonumber\\
\fl \left.\qquad \qquad \qquad +\frac{3 M_P^2 \ddot H}{4 H \left(-2 M_2{}^4-3 M_1{}^4+M_P^2 \dot H\right)}       \right) \, .
\eea
\bea
\tilde \alpha_0 =\alpha_0(1+\frac{\dot \alpha_0}{H \alpha_0}) \,;\qquad  \tilde \beta_0=\beta_0(1+\frac{\dot \beta_0}{H \beta_0}) \,.
\eea
In going from  Eq.~(\ref{eoma}) to Eq.~(\ref{eomb}) it was safely assumed that $a(\tau)\simeq -\frac{1}{H \tau(1-\epsilon)}$, which in turn means that the cmplete and most general expression for the parameter $x_0$ is, to first order in generalized slow-roll parameters, given by:
\bea
\fl x_0 = \left(-\frac{\epsilon}{2} +\frac{6 M_2{}^3 \dot M_2}{H \left(2 M_2{}^4-3 M_1{}^4-M_P^2 \dot H\right)}+\frac{9 M_1{}^3 \dot M_1}{H \left(-2 M_2{}^4+3 M_1{}^4+M_P^2 \dot H\right)} \right.\nonumber\\
\fl \left.\qquad \qquad \qquad +\frac{3 M_P^2 \ddot H}{4 H \left(-2 M_2{}^4+3 M_1{}^4+M_P^2 \dot H\right)} +2\epsilon  +\frac{3}{2}\epsilon \frac{ M_P^2 {\dot H(t)}}{2M_2^4-M_P^2 \dot H +3 M_1^4 }   \right)\, .
\eea

\newpage

\section{Appendix B}
 A more detailed presentation of the scalar exchange calculation for the $\bar M_6$-driven term is presented here. We start from the scalar exchange part of  Eq.~(\ref{z4}):
 \bea
 \langle \Omega|\zeta_{k_1}\zeta_{k_2}\zeta_{k_3}\zeta_{k_4}(t)|\Omega\rangle_{s.e.}= \nonumber\\ \langle 0|\bar T \{i\int_{-\infty}^{t_0} d^3 x dt^{'} \mathcal{H}_3(x)\} \zeta_{k_1}\zeta_{k_2}\zeta_{k_3}\zeta_{k_4}(t)\,  T\{ -i\int_{-\infty}^{t_0} d^3 x^{'} dt^{''} \mathcal{H}_3(x) \}|0\rangle \nonumber\\
 +\langle 0|\bar T \{\frac{i^2}{2}\int{ \int{ d^3 x \,dt\, d^3 x^{'}\, dt^{''} \mathcal{H}_3(x) \mathcal{H}_3(x^{'})}}\} \zeta_{k_1}\zeta_{k_2}\zeta_{k_3}\zeta_{k_4}(t)|0\rangle \nonumber\\
 +\langle 0|\zeta_{k_1} \zeta_{k_2}\zeta_{k_3}\zeta_{k_4}(t) T \{\frac{(-i)^2}{2}\int_{-\infty}^{t_0}{ \int_{-\infty}^{t_0}{ d^3 x \,dt\, d^3 x^{'}\, dt^{''} \mathcal{H}_3(x) \mathcal{H}_3(x^{'})}}\}|0\rangle .
\nonumber\\ \label{sed}
\eea
Using Wick contraction on a generic operator  $\phi$, one has:
\bea
\fl T\{\phi(t_1)\phi(t_2)\phi(t_3)\phi(t_4)\}=N\{\phi(t_1)\phi(t_2)\phi(t_3)\phi(t_4)+\textit{all \,\, contractions}  \},
\eea
 where \textit{N} is the normal ordering operator.
Note also that, being our $\pi \sim \zeta$ operators squeezed between two vacua of the free theory, this reduces to considering only terms which are writeable as fully contracted contributions. For the anti-time order operator the same formula holds, only one needs to define contractions differently. We show below this difference:
\bea
\fl \contraction{\phi(}{\vec x_1, t_1) \phi( }{ \vec x_2, t_2},
\phi(\vec x_1 , t_1)\,\,\,\, \phi(\vec x_2, t_2)_{\bf{T}} =[\phi^{+}(\vec x_1 ,t_1), \phi^{-}(\vec x_2,t_2)] \theta(t_1-t_2)+[\phi^{+}(\vec x_2 ,t_2), \phi^{-}(\vec x_1,t_1)] \theta(t_2-t_1) \nonumber\\
\eea

\bea
\fl \contraction{\phi(}{\vec x_1, t_1) \phi( }{ \vec x_2, t_2},
\phi(\vec x_1 , t_1)\,\,\,\, \phi(\vec x_2, t_2)_{{\bf \bar T}} =[\phi^{+}(\vec x_2 ,t_2), \phi^{-}(\vec x_1,t_1)] \theta(t_1-t_2)+[\phi^{+}(\vec x_1 ,t_1), \phi^{-}(\vec x_2,t_2)] \theta(t_2-t_1) \nonumber\\
\eea

where 
\be
\fl \phi^{+}(\vec x_2 ,t_2)= \int{\frac{d^3 k}{(2\pi)^3} \phi(\vec k,t_2)\, {\bf a_k}e^{i \vec{k} \cdot \vec x_2}}; \quad  \phi^{-}(\vec x_2 ,t_2)= \int{\frac{d^3 k}{(2\pi)^3} \phi^{*}(\vec{k},t_2)\, {\bf a^{\dagger}_{k}}e^{-i \vec{k} \cdot \vec x_2}}.
\ee

\noindent Using the definitions above one gets several different contributions from Eq.~(\ref{sed}). Note also that, using time and anti-time order definitions, the last two lines of Eq.~(\ref{sed})  are just each other's conjugate and can therefore be grouped together. We now procede to write an explicit expression for the four point function generated by the $\bar M_6$-driven contribution to the scalar exchange diagram:
\bea
\fl \langle \pi_{k1} \pi_{k2}\pi_{k3}\pi_{k4}\rangle^{s.e.}_{\bar M_6}\propto
\eea
\bea
\fl  \frac{M_2^8}{3}\left[ 4\cdot \left(\pi^{*}_{k1}\pi^{*}_{k2}\pi_{k3}\pi_{k4}(0)\,\int_{-\infty}^{0}{dt_1  \frac{a^3}{a^4} \dot \pi_{k12} \pi_{k1}\pi_{k2}({\bf k}_1\cdot {\bf k}_2)^2 }\int_{-\infty}^{0}{dt_2 \frac{a^3}{a^4} \dot \pi^{*}_{k12} \pi^{*}_{k3}\pi^{*}_{k4}({\bf k}_3\cdot {\bf k}_4)^2} \right. \right. \nonumber \\
\fl \left. \left. +\textit{ 5}\,\,\textit{ permutations} \right)+\right. \nonumber \\
\fl \left. \left( 2 \cdot 2 \cdot\,  \pi^{*}_{k1}\pi^{*}_{k2}\pi_{k3}\pi_{k4}(0)\,\int_{-\infty}^{0}{dt_1  \frac{a^3}{a^4} \dot \pi_{k1} \pi_{k12}\pi_{k2}(-{\bf k}_{12}\cdot {\bf k}_2)^2} \int_{-\infty}^{0}{dt_2 \frac{a^3}{a^4} \dot \pi^{*}_{k12} \pi^{*}_{k3}\pi^{*}_{k4} ({\bf k}_3\cdot {\bf k}_4)^2 } \right. \right. \nonumber \\
\left. \left. \fl +\textit{11}\,\,\textit{ permutations}  \right) \right.\nonumber \\
\left.\fl    \left( 2 \cdot 2 \cdot\,  \pi^{*}_{k1}\pi^{*}_{k2}\pi_{k3}\pi_{k4}(0)\,\int_{-\infty}^{0}{dt_1  \frac{a^3}{a^4} \dot \pi_{k12} \pi_{k1}\pi_{k2}({\bf k}_{1}\cdot {\bf k}_2)^2} \int_{-\infty}^{0}{dt_2 \frac{a^3}{a^4} \dot \pi^{*}_{k3} \pi^{*}_{k12}\pi^{*}_{k4} ({\bf k}_{12}\cdot {\bf k}_4)^2 } \right.     \right.\nonumber\\
\left. \left. \fl  +\textit{11}\,\,\textit{ permutations}  \right)    \right. \nonumber \\
\left. \fl   \left( 4 \cdot   \pi^{*}_{k1}\pi^{*}_{k2}\pi_{k3}\pi_{k4}(0)\,\int_{-\infty}^{0}{dt_1  \frac{a^3}{a^4} \dot \pi_{k1} \pi_{k12}\pi_{k2}(-{\bf k}_{12}\cdot {\bf k}_2)^2} \int_{-\infty}^{0}{dt_2 \frac{a^3}{a^4} \dot \pi^{*}_{k3} \pi^{*}_{k12}\pi^{*}_{k4} ({\bf k}_{12}\cdot {\bf k}_4)^2 } \right.    \right. \nonumber \\
\left. \left. \fl    +\textit{23}\,\,\textit{ permutations}  \right)  \right]+ \nonumber
\eea
\bea
\fl  -\frac{2\,M_2^8}{3} \mathcal{R}_e  \left[ 4\cdot \left(\pi^{*}_{k1}\pi^{*}_{k2}\pi^{*}_{k3}\pi^{*}_{k4}(0)\,\int_{-\infty}^{0}{dt_1  \frac{a^3}{a^4} \dot \pi^{*}_{k12} \pi_{k1}\pi_{k2}({\bf k}_1\cdot {\bf k}_2)^2 }\int_{-\infty}^{t_1}{dt_2 \frac{a^3}{a^4} \dot \pi^{}_{k12} \pi^{}_{k3}\pi^{}_{k4}({\bf k}_3\cdot {\bf k}_4)^2} \right. \right. \nonumber \\
\fl \left. \left. +\textit{ 5}\,\,\textit{ permutations} \right)+\right. \nonumber \\
\fl \left. \left( 2 \cdot 2 \cdot\,  \pi^{*}_{k1}\pi^{*}_{k2}\pi^{*}_{k3}\pi^{*}_{k4}(0)\,\int_{-\infty}^{0}{dt_1  \frac{a^3}{a^4} \dot \pi_{k1} \pi^{*}_{k12}\pi_{k2}(-{\bf k}_{12}\cdot {\bf k}_2)^2} \int_{-\infty}^{t_1}{dt_2 \frac{a^3}{a^4} \dot \pi^{}_{k12} \pi^{}_{k3}\pi^{}_{k4} ({\bf k}_3\cdot {\bf k}_4)^2 } \right. \right. \nonumber \\
\left. \left. \fl +\textit{11}\,\,\textit{ permutations}  \right) \right.\nonumber \\
\left.\fl    \left( 2 \cdot 2 \cdot\,  \pi^{*}_{k1}\pi^{*}_{k2}\pi^{*}_{k3}\pi^{*}_{k4}(0)\,\int_{-\infty}^{0}{dt_1  \frac{a^3}{a^4} \dot \pi^{*}_{k12} \pi_{k1}\pi_{k2}({\bf k}_{1}\cdot {\bf k}_2)^2} \int_{-\infty}^{t_1}{dt_2 \frac{a^3}{a^4} \dot \pi^{}_{k3} \pi^{}_{k12}\pi^{}_{k4} ({\bf k}_{12}\cdot {\bf k}_4)^2 } \right.     \right.\nonumber\\
\left. \left. \fl  +\textit{11}\,\,\textit{ permutations}  \right)    \right. \nonumber \\
\left. \fl   \left( 4 \cdot   \pi^{*}_{k1}\pi^{*}_{k2}\pi^{*}_{k3}\pi^{*}_{k4}(0)\,\int_{-\infty}^{0}{dt_1  \frac{a^3}{a^4} \dot \pi_{k1} \pi^{*}_{k12}\pi_{k2}(-{\bf k}_{12}\cdot {\bf k}_2)^2} \int_{-\infty}^{t_1}{dt_2 \frac{a^3}{a^4} \dot \pi^{}_{k3} \pi^{}_{k12}\pi^{}_{k4} ({\bf k}_{12}\cdot {\bf k}_4)^2 } \right.    \right. \nonumber \\
\left. \left. \fl    +\textit{23}\,\,\textit{ permutations}  \right)  \right]. \label{appb}
\eea
One then performs these calculations and plots the results to obtain Fig.~\ref{m61},\ref{m62}. The situation for the contact interaction diagram contributions is considerably simpler as there is just one time intergral to be performed and two less fields to be taken into account.

\newpage 
\section{Appendix C}
We want here to show with an example what seems to be a general feature concerning the use of (reasonably) approximated wavefunctions in the calculation of higher order correlators. In \cite{b} we found that in  performing an exact calculation for correlators in a very general theory such as the one we employed in this paper, whenever  a given interaction term was producing a shape function for the trispectrum which one could qualitatively classify as, say, equilateral, so was the calculation  performed with a simplified wavefunction. This is due to two independent reasons. First, we start from the realization that, precisely in the horizon-crossing region, which is where one expects the main contribution to any n-point to come from, the exact general wavefunction \cite{b} and the usual one, $H^{(1)}_{3/2}(\tilde{c_s}k\tau)$, which in these theories is an approximated solution, behave very similarly. Secondly, in \cite{b} we concluded that most of the distinctive effects of the bispectrum where due not to the particular k-modes dependence of the result of the integrals like the one in Eq.~(\ref{calc}),  but on the fraction of that k-dependence that could be taken outside the integral, so on the part of the k-dependence not directly attached to the time behaviour of the wavefunction and which is common to the exact and approximated wavefunction.\\
We now compare the trispectrum shapefunction of a ghost inflation interaction term, $(\nabla \pi)^4$, performed with the exact ghost solution in \cite{muko} with the results we obtain employing the approximated DBI wavefunction, just what we used in obtaining all the shape functions presented here.\\
The interaction reads:
\be
\frac{M_2^4}{2} \sum_{i=1}^{3} \sum_{j=1}^{3} \frac{(\partial_i \pi)^2(\partial_j \pi)^2}{a^4}
\ee
Its trispectrum  shapefunction obtained through the approximated methods has been ploted in Fig. \ref{M41}. In order to compare it with the exact calculation of \cite{muko} we need to change variables and turn to:
\bea
\fl C_2= \hat{k}_1 \cdot \hat{k}_2;\,\,\, C_3= \hat{k}_1 \cdot \hat{k}_3;\,\,\, C_4=-1-C_2-C_3; \quad k_{12}= \sqrt{2(1+C_2)}\quad k_{14}=  \sqrt{2-C_2 -C_3)} \nonumber \\
\eea
We now show the plots obtained by performing this change of variable on our approximated result alongside the plot obtained with the exact ghost wavefunction taken directly from \cite{muko}.
\vspace{10cm}
\begin{figure}[hp]
	\includegraphics[scale=0.58]{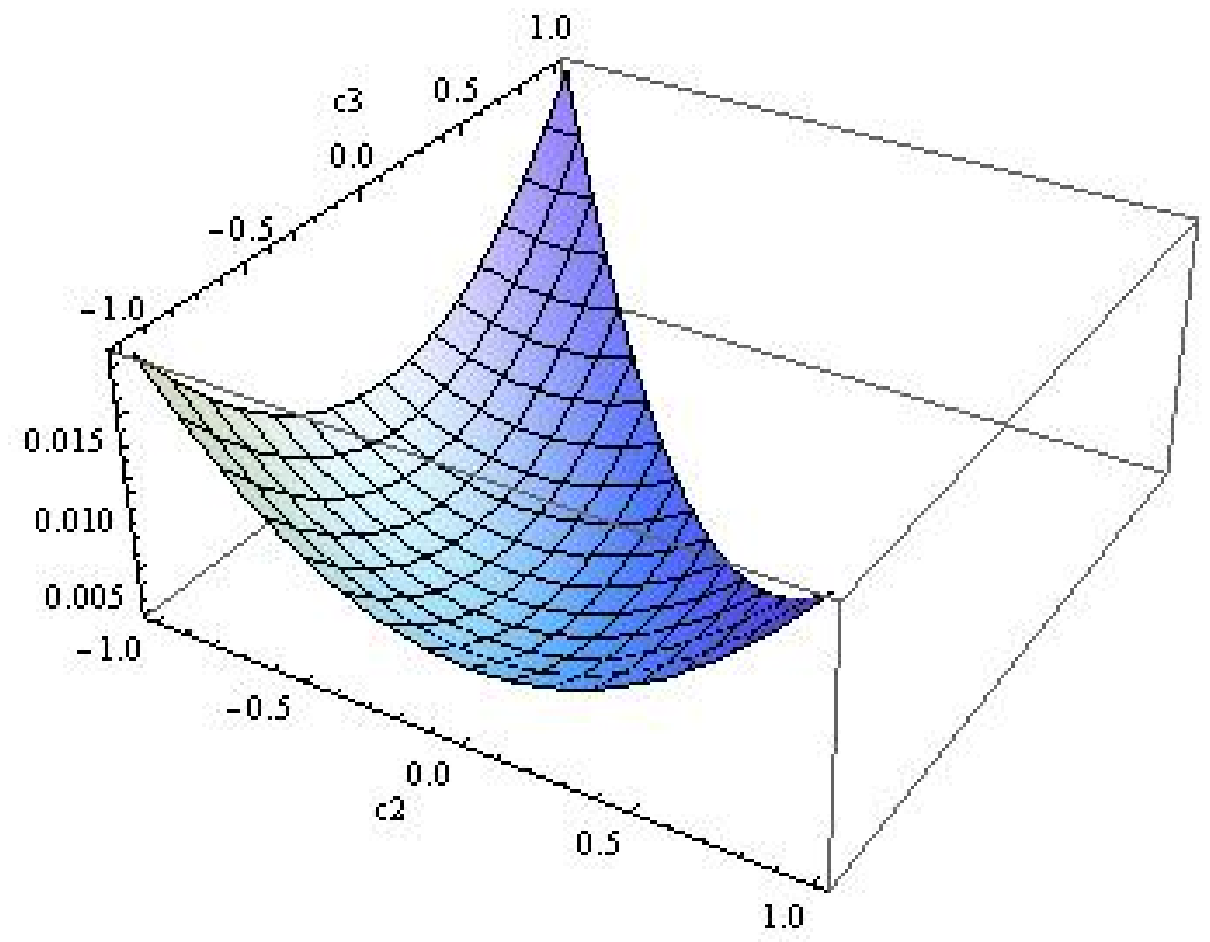}
	\hspace{10mm}
		\includegraphics[scale=0.72]{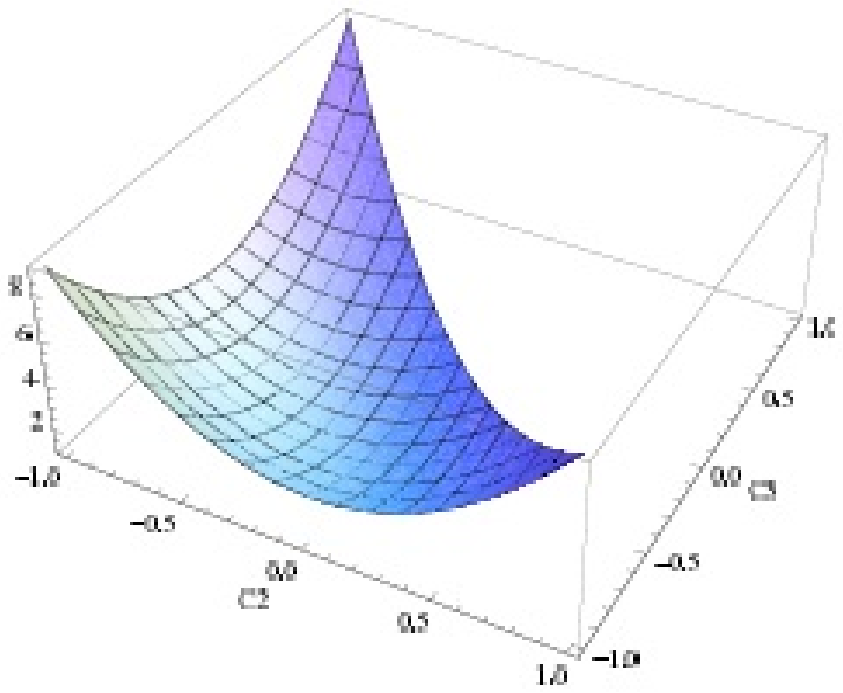}
\caption{On the left the approximated result. The two shapefunctions are qualitatively identical barring an unimportant numerical coefficient due to a different normalization.}
\label{Mghost}
\end{figure}

\section{Appendix D: IN-IN formalism}
See \textit{Section 9.1}
\newpage
\vspace{20cm}
\newpage

\section{References}
\bibliographystyle{JHEP}

\end{document}